\documentclass[12pt]{iopart}

\pdfoutput=1

\usepackage{graphicx,sidecap}
\usepackage{bm}
\usepackage{braket}
\usepackage{amssymb}
\usepackage{cite}
\usepackage{mathrsfs}
\usepackage{amsmath}
\usepackage{xcolor}
\usepackage{hyperref}
\usepackage{enumerate}
\usepackage[toc,title]{appendix}
\usepackage{verbatim}
\usepackage[margin=30pt, bf, font=footnotesize, center, justification=justified]{caption}[2004/07/16]

\hypersetup{colorlinks,bookmarksopen,bookmarksnumbered,
citecolor=red,
linkcolor=blue,
pdfstartview=green,
urlcolor=teal}

\catcode`@=11 
\renewcommand\tableofcontents{%
  \section*{\contentsname}%
  \@starttoc{toc}%
}
\catcode`@=12

\def\be{\begin{equation}}
\def\ee{\end{equation}}

\def\bea{\begin{eqnarray}}
\def\eea{\end{eqnarray}}

\def\Tr{{\rm Tr}}

\begin{document}

\title[Entanglement hamiltonians in 1D free lattice models after a global quantum quench
]
{Entanglement hamiltonians in 1D free lattice models after a global quantum quench
}

\vspace{.5cm}

\author{Giuseppe Di Giulio, Ra\'ul Arias and Erik Tonni}
\address{SISSA and INFN Sezione di Trieste, via Bonomea 265, 34136 Trieste, Italy.}

\vspace{.5cm}

\begin{abstract}
We study the temporal evolution of the entanglement hamiltonian of an interval 
after a global quantum quench
in free lattice models in one spatial dimension. 
In a harmonic chain we explore a quench of the frequency parameter.
In a chain of free fermions at half filling we consider the evolution of the 
ground state of a fully dimerised chain through the homogeneous hamiltonian. 
We focus on critical evolution hamiltonians.
The temporal evolutions of the gaps in the entanglement spectrum are analysed.
The entanglement hamiltonians in these models 
are characterised by matrices that provide also contours for the entanglement entropies.
The temporal evolution of these contours for the entanglement entropy 
is studied, also by employing existing conformal field theory results for the semi-infinite line
and the quasi-particle picture for the global quench.
\end{abstract}

\maketitle

\newpage

\tableofcontents

\newpage

\section{Introduction}
\label{sec:intro}

The entanglement associated to the state of a quantum system and to a bipartition of its Hilbert space 
has attracted a lot of attention during the last decade.
Many techniques have been developed to quantify entanglement in
quantum many-body systems, quantum field theories and quantum gravity \cite{ep-rev, ch-rev, cc-09-rev, fazio-review, cramer-plenio-rev, weedbrook, muk-tad-rev}.

Given a bipartition $\mathcal{H} = \mathcal{H}_A \otimes \mathcal{H}_B$ of the Hilbert space 
and a state characterised by the density matrix $\rho$, the reduced density matrix associated to $\mathcal{H}_A$ is 
$\rho_A \equiv \textrm{Tr}_{\mathcal{H}_B} \rho$.
In this manuscript we focus on bipartitions of Hilbert spaces that come from 
spatial bipartitions; hence $A$ denotes a spatial region and $B$ its complement. 
\cite{ee-initial-papers}.
The reduced density matrix is a positive semi-definite hermitian operator normalised 
to $\textrm{Tr}_{\mathcal{H}_A}  \rho_A  =1$
(hereafter the trace will be always over $\mathcal{H}_A$). 
These properties allow to write $\rho_A$ as 
\be
\label{rho_A}
\rho_A 
=
\frac{e^{-\widehat{K}_A}}{\mathcal{Z}_A }
\ee
where the hermitian operator $\widehat{K}_A$ is the entanglement hamiltonian 
(also called modular hamiltonian) and $ \mathcal{Z}_A  = \Tr (e^{-\widehat{K}_A})$.

The moments of the reduced density matrix $\Tr \rho_A^n$, that are parameterised by the integer power 
$n \geqslant 2$, provide the R\'enyi entropies $S_A^{(n)} \equiv \tfrac{1}{1-n} \log (\Tr \rho_A^n)$.
The entanglement entropy is defined as the Von Neumann entropy of the reduced density matrix 
$S_A = -\Tr(\rho_A \log \rho_A)$.
A very important property of the R\'enyi entropies is that $S_A^{(n)} \to S_A$ as $n \to 1$.
This limit can be very complicated because it requires to perform an analytic continuation in $n$.
The entanglement entropies are $S_A^{(n)} $ for $n \geqslant 1$, where $S_A^{(1)} \equiv S_A$ is assumed.
These scalar quantities depend only on the eigenvalues of $\rho_A$ (the entanglement spectrum)
\cite{ent-spectrum-top};
hence the entanglement hamiltonian contains more information about the entanglement of the bipartition 
with respect to the corresponding entanglement entropies.

In quantum field theories, Bisognano and Wichmann found 
the explicit expression of the entanglement hamiltonian of half space through the stress tensor of the model
\cite{bw}.
In conformal field theories (CFT), this formula has been employed as the starting point to obtain 
entanglement hamiltonians for other interesting configurations 
\cite{hislop-longo, chm, ch-09-eh-2int, klich-13, ct-16, trs-18-rainbow, ryu-ludwig-18-cft, sierra-alcaraz-18}.
Crucial tools in this analysis are the methods developed for CFT with boundary \cite{bcft}.
In quantum gravity models in the context of the gauge/gravity correspondence, elaborating on the holographic prescription 
for the entanglement entropy \cite{RT, HRT}, also the entanglement hamiltonian has been explored
\cite{JLMS}.

In many body quantum systems, entanglement hamiltonians 
have been studied for bosonic Gaussian states in harmonic lattices and for fermionic Gaussian states 
in lattice models of free fermions
\cite{
ep-rev, ch-rev, Peschel-Chung-oscillators, Peschel-Chung-00, peschel-03-modham, 
peschel-04-eh-fermions, eigenvector free fermions, banchi-pirandola-15, Arias-16, Arias-17, eisler-peschel-17, ep-18, Arias-18, etp-19}. 
The entanglement hamiltonian is a quadratic operator in these free models; 
hence it is fully characterised by a matrix.
The results available in the literature are mostly based on numerical analysis of this matrix. 
Analytic expressions for the entanglement hamiltonian in the thermodynamic limit have been obtained in \cite{eisler-peschel-17} for the static configuration 
where $A$ is a single block in the infinite chain of free fermions in the ground state.
The matrix characterising this entanglement hamiltonian at half filling 
has non vanishing odd diagonals at any allowed distance from the main diagonal. 
The continuum limit of this entanglement hamiltonian has been studied in \cite{etp-19} 
and the expected CFT result of  \cite{hislop-longo, chm} has been derived through this method. 
In spin chain systems, entanglement hamiltonians have been studied in  
\cite{peschel-truong}.

The entanglement spectrum is another very insightful quantity to consider in order to get 
insights about the bipartite entanglement.
It has been explored by employing both CFT methods  
\cite{ct-16, calabrese-lefevre} and numerical techniques 
\cite{lauchli-spectrum,assaad-ent-spectrum}.
The entanglement spectrum is not enough to construct the entanglement hamiltonian.

The contours for the entanglement entropies are simpler to obtain with respect to 
the entanglement hamiltonian and they are not fully determined by the entanglement spectrum as well. 
The contours for the entanglement entropies are positive functions on the subsystem $A$ 
such that their integral over $A$ provides the entanglement entropies.
The aim of these scalar quantities is to describe the contribution of the various sites in $A$ 
to the entanglement entropies of the bipartition 
and the entanglement spectrum is not enough to determine them.
In free lattice models, contours for the entanglement entropies have been constructed and 
numerical analysis have been performed \cite{br-04, chen-vidal, frerot-roschilde, cdt-17-contour, trs-18-rainbow}.
For some static configurations described by CFT, they are related to the corresponding entanglement hamiltonians \cite{cdt-17-contour}.
Density functions for the entanglement entropy have been explored also in \cite{hubeny-ee-density}.
The contours for the entanglement entropies are special density functions because further constraints are required for them \cite{chen-vidal, cdt-17-contour}.
Nonetheless, a complete list of properties that allows to select a unique contour for the entanglement entropies is not known.

The bipartite entanglement has been studied during the last decade also to explore
the out of equilibrium dynamics of quantum systems.
Global quantum quenches are insightful processes 
where the system is prepared in the ground state $| \psi_0 \rangle$ 
of a translationally invariant hamiltonian
and at $t=0$ a sudden change modifies the hamiltonian keeping the translation invariance. 
The non trivial dynamics is due to the fact that  the initial state $| \psi_0 \rangle$ is not
an eigenvector of the hamiltonian determining its unitary evolution. 
The temporal evolutions of various quantities after global quantum quenches
and also other kinds of quenches have been studied 
in quantum field theories, in many-body systems on the lattice 
\cite{cc-05-global quench, cc-06-quench-corr, cc-07-quench-extended, cardy-15, 
cc-07-local quench,  ep-07-local-quench, rigol-07-GGE, 
scc-boson-quench, sotiriadis-cardy, caux-essler-13, fagotti-essler-13,
coser-14-quench-neg, tagliacozzo-torlai,
caux-essler-prosen-15-GGE, myers-galante, mezei-muller-16-quench-scalar,
ac-18-qp-quench, surace-tagliacozzo-18}
and also in quantum gravity through the gauge/gravity correspondence 
\cite{HRT, holographic quenches}
(see the reviews \cite{cc-16-quench rev, fagotti-essler-review, vidmar-rigol-16-GGEreview}
for an extensive list of references).

In this manuscript we study the temporal evolution of the 
entanglement hamiltonian and of a contour for the entanglement entropy
after a global quantum quench in free lattice models.
For simplicity, we consider, in one spatial dimension,
a harmonic chain and a chain of free fermions.
In the harmonic chain, we explore a quench of the frequency parameter
such that the unitary evolution is governed by the massless hamiltonian.
In the chain of free fermions, we consider
the global quench introduced in \cite{ep-07-local-quench, ep-rev},
where the system is prepared in the ground state of a fully dimerised chain
while the evolution hamiltonian is fully homogeneous.

This manuscript is organised as follows. 
In \S\ref{sec:EHwilliamson_total} and \S\ref{sec-williamson-fermion}
we discuss the construction of the entanglement hamiltonians
and of the contours for the entanglement entropies in the models of interest.
In \S\ref{sec:cft-naive} we provide a naive formula for the contour function
obtained through CFT results. 
The numerical results for the harmonic chain 
and for the chain of free fermions are described 
in \S\ref{sec:HCnumerics} and \S\ref{sec:fermion-quench} respectively. 
In \S\ref{sec:conclusions} we draw some conclusions. 
The Appendices\;\ref{sec_app:EH derivation}, 
\ref{app:cholesky}, \ref{app:EH-block-diag} and \ref{app:larget-largel-derivation}
contain technical details about some issues discussed in the main text.

\section{Harmonic lattices}
\label{sec:EHwilliamson_total}

In this section we review the construction of the matrix characterising the entanglement hamiltonian
for Gaussian states in harmonic lattices. 
We focus on one dimensional chains, but the discussion can be easily extended
to harmonic lattices in higher dimensions.

The hamiltonian of the harmonic chain with nearest neighbour spring-like interaction reads
\be
\label{HC ham}
\widehat{H} = \sum_{i=0}^{L-1} \left(
\frac{1}{2m}\,\hat{p}_i^2+\frac{m\omega^2}{2}\,\hat{q}_i^2 +\frac{\kappa}{2}(\hat{q}_{i+1} -\hat{q}_i)^2
\right)
\ee
where $L$ is the number of lattice sites
and the hermitian operators $\hat{q}_i$ and $\hat{p}_i$ satisfy the canonical commutation relations 
$[\hat{q}_i, \hat{q}_j] = [\hat{p}_i, \hat{p}_j]  =0$ and $[\hat{q}_i, \hat{p}_j] = \textrm{i} \delta_{ij}$ 
(throughout this manuscript $\hbar =1$).
The boundary conditions, that are crucial to determine the expression of the correlators,
do not influence explicitly the following discussion.
By arranging the position and momentum operators into the vector 
$\boldsymbol{\hat{r}} \equiv (\hat{q}_1 , \dots , \hat{q}_L, \hat{p}_1, \dots, \hat{p}_L)^{\textrm{t}}$,
the canonical commutation relations can be written in the form $[\hat{r}_i, \hat{r}_j] = \textrm{i} J_{ij}$,
being  $J$ the standard symplectic matrix 
\be
\label{Jmat}
J \equiv
\bigg( \hspace{-.1cm} \begin{array}{cc}
 \boldsymbol{0} &  \boldsymbol{1} \\
 - \,\boldsymbol{1} &  \boldsymbol{0} \\
\end{array}  \hspace{-.05cm}  \bigg)
\ee
where $\boldsymbol{1}$ is the $L \times L$ identity matrix and $ \boldsymbol{0}$ is the $L \times L$ matrix made by zeros. 
Notice that $J^{\textrm{t}} = -J$ and $J^2 = - \boldsymbol{1}$. 
We also need that $J^{\textrm t}(a\oplus b) J = b\oplus a$.

The linear transformations $\boldsymbol{\hat{r}} \to \boldsymbol{\hat{r}}' = S \boldsymbol{\hat{r}} $
preserving the canonical commutation relations define the real symplectic group $\textrm{Sp}(L)$,
made by the real $2L \times 2L$ matrices $S$ that satisfy the relation $S J S^{\textrm t} = J$  \cite{deGosson}.
Given a symplectic matrix $S$, we have that 
$\textrm{det}(S) =  1$, $S^{\textrm{t}} \in \textrm{Sp}(L)$ and $S^{-1} = J S^{\textrm t} J^{-1}$.
The above observations imply that $S^{-\textrm{t}} = J^{\textrm t}  S J$, where we have adopted the notation $M^{-\textrm{t}} \equiv  ( M^{\textrm{t}} )^{-1}$.

By employing a canonical transformation, the hamiltonian (\ref{HC ham}) can be written 
as the hamiltonian of a free boson with mass $\omega$ discretised on a lattice with spacing $a=\sqrt{m/\kappa}$. 
This implies that we can set $m=\kappa=1$ without loss of generality.
The continuum limit of this model gives the free scalar boson with mass $\omega$ in two spacetime dimensions.
In the massless regime, this quantum field theory is a conformal field theory (CFT) with central charge $c=1$. 

In this manuscript we focus on Gaussian states of  (\ref{HC ham}), which
are completely characterised by the correlators $\langle \hat{r}_i \rangle$ (first moments) and $\langle \hat{r}_i \, \hat{r}_j\rangle$ (second moments).
Since a shift of the first moments corresponds to a unitary transformation that preserves the Gaussian nature of the state, 
let us consider Gaussian states having $\langle \hat{r}_i \rangle=0$.
Thus, the second moments fully describe these states and 
they can be collected into the $2L \times 2L$ covariance matrix 
$\gamma \equiv \textrm{Re} \langle \boldsymbol{\hat{r}}\, \boldsymbol{\hat{r}}^{\textrm{t}}\rangle$,
which is a real, symmetric and positive definite matrix 
\cite{cramer-plenio-rev, weedbrook, ee-hc-lattice-refs, br-04, deGosson}.
A canonical transformation $\boldsymbol{\hat{r}} \to \boldsymbol{\hat{r}}' = S \boldsymbol{\hat{r}} $ 
characterised by the symplectic matrix $S$ induces the transformation $\gamma \to \gamma' = S \gamma S^{\textrm t}$
on the covariance matrix.
The  covariance matrix of a pure Gaussian state satisfies the relation  $(\textrm{i} J \, \gamma)^2 = \tfrac{1}{4}\,\boldsymbol{1}$.

Given the harmonic chain (\ref{HC ham}) in a Gaussian state $\rho$ characterised by the covariance matrix $\gamma$,
let us introduce a bipartition of the Hilbert space $\mathcal{H} = \mathcal{H}_A \otimes \mathcal{H}_B$ 
corresponding to a spatial bipartition $A\cup B$.
The reduced density matrix $\rho_A \equiv \textrm{Tr}_{\mathcal{H}_B} \rho$ associated to the spatial subsystem $A$ 
characterises a mixed state also when the whole system $A \cup B$ is in a pure state.
For the harmonic chain (\ref{HC ham}) and its higher dimensional generalisations,  
$\rho_A$ remains Gaussian for any choice of $A$.
This implies that $\rho_A $ is fully described by the reduced covariance matrix
$\gamma_A \equiv \textrm{Re} \langle \boldsymbol{\hat{r}}\, \boldsymbol{\hat{r}}^{\textrm{t}}\rangle |_A$,
obtained by extracting from $\gamma$ of the entire system the rows and the columns 
corresponding to the lattice sites belonging to the subsystem $A$.
The reduced covariance matrix $\gamma_A$ is real, symmetric and positive definite.
In the numerical analysis of this manuscript we consider only the case where $A$ is an interval made by $\ell$ sites,
hence $\gamma_A$ is a $2\ell \times 2\ell$ matrix.
Nonetheless, the considerations reported in this section hold for a generic number of spatial dimensions, once $\ell$
is understood as the number of sites in the subsystem $A$.

\subsection{Entanglement hamiltonian}
\label{sec:EHwilliamson}

In the harmonic chain, the entanglement hamiltonian $\widehat{K}_A$ corresponding to a region $A$ introduced in (\ref{rho_A}) 
is a quadratic hermitian  operator, hence it can be written as 
\cite{Peschel-Chung-oscillators, ch-rev}
\be
\label{ent-ham HC}
\widehat{K}_A
= \frac{1}{2}\, \boldsymbol{\hat{r}}^{\textrm t} H_A\, \boldsymbol{\hat{r}}
\qquad
\boldsymbol{\hat{r}} = 
\bigg( \hspace{-.05cm} 
\begin{array}{c}
\boldsymbol{\hat q} \\  \boldsymbol{\hat p}
\end{array} \hspace{-.05cm}  \bigg)
\ee
where the $2\ell$ dimensional vector $\boldsymbol{\hat{r}}$ collects the position and the momentum operators 
$\hat{q}_i$ and $\hat{p}_i$ with $i \in A$.
The matrix $H_A$, that fully determines the entanglement hamiltonian in (\ref{ent-ham HC}), 
is real, symmetric and positive definite.
This guarantees that 
$\widehat{K}_A$ is hermitian, like the operators $\hat{q}_i$ and $\hat{p}_i$.
In the following we often call $H_A$ the entanglement hamiltonian matrix.
 
 A very important tool employed to quantify the bipartite entanglement in harmonic lattices
 is the Williamson's theorem \cite{Williamson, williamson-other proofs, Simon99 proofWilliamson}.
It holds for any real, symmetric and positive matrix having even order,
but in our analysis we employ it for the $2\ell \times 2\ell$ matrices $\gamma_A$ and $H_A$.
Considering the reduced covariance matrix $\gamma_A$ first, the Williamson's theorem guarantees that
we can construct a symplectic matrix $W\in \textrm{Sp}(\ell)$  such that
\be
\label{williamson th gammaA}
\gamma_A = W^{\textrm t} \,\mathcal{D}_\textrm{\tiny d} \,W
\ee
where $\mathcal{D}_\textrm{d} \equiv \mathcal{D} \oplus \mathcal{D}$
and the diagonal matrix $\mathcal{D}=\textrm{diag} (\sigma_1 , \dots , \sigma_\ell)$ 
collects the symplectic eigenvalues $\sigma_k > 0$ of $\gamma_A$.
The symplectic eigenvalues  are uniquely determined up to permutation and constitute the symplectic spectrum.
We will refer to the r.h.s. of (\ref{williamson th gammaA}) as the Williamson's decomposition of $\gamma_A $.
We will choose a decreasing ordering for the symplectic eigenvalues.

The symplectic spectrum of $\gamma_A$ provides the entanglement entropy $S_A$ and the R\'enyi entropies $S_A^{(n)}$ as follows \cite{cramer-plenio-rev, br-04, ee-hc-lattice-refs, weedbrook, peschel-03-modham, ep-rev}
\be
\label{SA and SAn from nuk}
S_A = \sum_{k=1}^\ell s(\sigma_k)
\qquad
S_A^{(n)} = \sum_{k=1}^\ell s_n(\sigma_k)
\ee
where $s(y)$ and $s_n(y)$ are the analytic functions given by 
\be
\label{sx def}
s(y) \equiv (y+1/2)\log(y+1/2) - (y-1/2)\log(y-1/2)
\ee
and 
\be
\label{snx def}
s_n(y) \equiv \frac{1}{n-1} \, \log\big[(y+1/2)^n - (y-1/2)^n\big] .
\ee
The parameter $n$ is an integer $n \geqslant 2$. 
Performing an analytic continuation of this integer parameter,
the entanglement entropy $S_A$ can be obtained as
$S_A^{(1)} \equiv \lim_{n \to 1} S_A^{(n)}$, that is the replica limit for the entanglement entropy.
The scalar quantities  $S_A^{(n)}$ with $n\geqslant 1$ are usually called 
entanglement entropies, assuming that $S_A^{(1)} \equiv  S_A$.

The symplectic spectrum of $\gamma_A$ is obtained by diagonalising $(\textrm{i} J \gamma_A)^2$ or $(\textrm{i} \gamma_A J)^2$.
Indeed, by employing that $W$ is symplectic and that 
$J^{\textrm t}\mathcal{D}^r_\textrm{\tiny d}  J= \mathcal{D}^r_\textrm{\tiny d}$ for any non negative integer $r$, 
it is not difficult to realise that 
\be
\label{Jgamma relations}
(\textrm{i} J \gamma_A)^2 
=
W^{-1} \,\mathcal{D}_\textrm{\tiny d}^2\, W
\;\;\;\qquad \;\;\;
(\textrm{i} \gamma_A J)^2 
=
\widetilde{W}^{-1}\,  \mathcal{D}_\textrm{\tiny d}^2 \, \widetilde{W}
\ee
where we have introduced
\be
\label{W widetildeW relation}
\widetilde{W} \equiv J^{\textrm t}  \,W J
= 
W^{-\textrm t} .
\ee
Being $ W\,  \widetilde{W}^{\textrm t}  = \widetilde{W}^{\textrm t} \, W = \boldsymbol{1}$,
the matrix $W$ is not orthogonal.

The Williamson's theorem can be employed to decompose also the
entanglement hamiltonian matrix $H_A$ defined in (\ref{ent-ham HC}), namely
\be
\label{H_A Williamson_dec}
H_A = W_H^{\textrm t}  \; \mathcal{E}_\textrm{\tiny d}   \,W_H
\ee
where $W_H \in \textrm{Sp}(\ell)$ and $\mathcal{E}_\textrm{d} \equiv \mathcal{E} \oplus \mathcal{E}$,
being $ \mathcal{E} =\textrm{diag}(\varepsilon_1 , \dots , \varepsilon_\ell)$
the symplectic spectrum of $H_A$.
The symplectic eigenvalues $\varepsilon_k$ are often called single particle entanglement energies.

The symplectic spectrum of $H_A$ is related to the symplectic spectrum of $\gamma_A$ as follows 
\be
\label{Ediag_Ddiag}
\mathcal{E} 
=
2 \,\textrm{arccoth}(2\,\mathcal{D})
=
\log\!\bigg( \frac{\mathcal{D} +1/2}{\mathcal{D} - 1/2} \bigg) \,.
\ee
Being $\mathcal{E}$ and $\mathcal{D}$ diagonal matrices, we have
$\varepsilon_k = 2 \,\textrm{arccoth}(2\sigma_k) = \log[(\sigma_k +1/2)/(\sigma_k -1/2)]$, 
whose inverse reads
$ \sigma_k =  \coth (\varepsilon_k/2)/2 = [(e^{\varepsilon_k} + 1)/(e^{\varepsilon_k} - 1)]/2$
for $1\leqslant k \leqslant \ell$.
Let us remark that the uncertainty principle leads to $\sigma_k \geqslant 1/2$, and this implies that $\varepsilon_k >0$.

The determinant of $H_A$ is determined only by the symplectic spectrum.
Indeed, by employing the Williamson's decomposition (\ref{H_A Williamson_dec}) and the fact that
a symplectic matrix has determinant equal to one, one finds
$\textrm{det} \,H_A = \textrm{det}\, \mathcal{E}_\textrm{\tiny d}  = \prod_{k=1}^\ell \varepsilon_k^2$.

In \cite{banchi-pirandola-15} the entanglement hamiltonian matrix $H_A$ has been expressed in terms of 
the reduced covariance matrix $\gamma_A$ as follows
\be
\label{banchi-EH}
H_A 
=
 2\,\textrm{i} \, J  \,  \textrm{arccoth}(2\,\textrm{i}\gamma_A J) 
=
 2\,\textrm{i}   \,  \textrm{arccoth}(2\,\textrm{i}J\gamma_A) \, J\,.
\ee
This relation can be obtained from (\ref{H_A Williamson_dec})
by employing (\ref{Ediag_Ddiag}) and 
\be
\label{W_H is Wtilde}
W_H =  \widetilde{W}
\ee
where $ \widetilde{W}$ is given by (\ref{W widetildeW relation}).
The details of this derivation have been reported in the Appendix\;\ref{sec_app:EH derivation}.
Notice that, if $B = A^{-1}(\Lambda \oplus\Lambda)  A$ for some diagonal matrix $\Lambda $, 
then  also $B = (J^{\textrm t} A)^{-1} (\Lambda \oplus\Lambda) (J^{\textrm t} A)$.
This implies that we can replace $W$ with $J^{\textrm t} W$, $-W$ or $-J^{\textrm t} W$ in (\ref{Jgamma relations}), 
and similarly for $\widetilde{W}$.
From (\ref{williamson th gammaA}) and (\ref{H_A Williamson_dec}), one observes that 
$\gamma_A$ and $H_A$ are not affected by this ambiguity.

The expression (\ref{banchi-EH}) can be also written as 
\be
\label{eh-cov-mat-B}
H_A
=
J^{\textrm t} \gamma_A\, J  \;
h \big( \sqrt{ (\textrm{i} \gamma_A J)^2}\,\big) 
=
h \big( \sqrt{ (\textrm{i} J \gamma_A)^2}\,\big)
\, J^{\textrm t} \gamma_A\, J 
\ee
where we have introduced the function $h(y)  \equiv  y^{-1} \log[ (y +1/2)/(y - 1/2)]$.
The equivalence between (\ref{eh-cov-mat-B}) and (\ref{banchi-EH}) is shown 
in Appendix\;\ref{sec_app:EH derivation}.
Notice that the expression (\ref{banchi-EH}) or its equivalent form (\ref{eh-cov-mat-B}) cannot be applied for pure states, 
where $(2\textrm{i} J \gamma)^2 =\boldsymbol{1}$.

By employing (\ref{williamson th gammaA}), (\ref{W widetildeW relation}), (\ref{H_A Williamson_dec}) and (\ref{W_H is Wtilde}), 
one finds 
$\gamma_A \, H_A  = W^{\textrm t} \big( \mathcal{D}_\textrm{\tiny d}\, \mathcal{E}_\textrm{\tiny d}  \big) \widetilde{W}$
and its transpose $ H_A \, \gamma_A  = \widetilde{W}^{\textrm t} \big( \mathcal{D}_\textrm{\tiny d}\, \mathcal{E}_\textrm{\tiny d}  \big)  W$.
This implies
$\textrm{Tr}((H_A  \gamma_A )^n)  = 2 \sum_k (\sigma_k\, \varepsilon_k)^n $ for $n\geqslant 1$.
Thus, for a generic function $f(x)$ we have
$ \textrm{Tr}(f(H_A \, \gamma_A) ) =
\textrm{Tr}(f(\gamma_A\, H_A)  ) =
 2 \sum_k f( \sigma_k\, \varepsilon_k) $.

The above discussion tells that the entanglement hamiltonian matrix $H_A$ can be obtained from $\gamma_A$
either through the explicit expressions in (\ref{H_A Williamson_dec}) and (\ref{banchi-EH}) or 
by finding the symplectic spectrum of the reduced covariance matrix $\gamma_A$ and the symplectic matrix $W$ first
and then use (\ref{H_A Williamson_dec}), (\ref{Ediag_Ddiag}) and (\ref{W_H is Wtilde}).
In the latter procedure, the needed data can be obtained through a standard diagonalisation, as indicates in (\ref{Jgamma relations}).
In our numerical analysis, this step is not straightforward because the matrix 
$(\textrm{i} J \gamma_A)^2$ to diagonalise is not symmetric. 
In the Appendix\;\ref{app:cholesky} we describe the procedure that we have employed to construct 
the symplectic matrix $W$, that is based on the Cholesky decomposition of $\gamma_A$.

The Williamson's theorem (\ref{H_A Williamson_dec}) and the relation (\ref{W_H is Wtilde}) 
allow to write entanglement hamiltonian (\ref{ent-ham HC}) 
in the following diagonal form
\be
\label{K_A s_diag}
\widehat{K}_A 
= 
\frac{1}{2}\; \boldsymbol{\hat{s}}^{\textrm t}  \,\mathcal{E}_{\textrm{\tiny d}}\,  \boldsymbol{\hat{s}}
\;\;\qquad\;\;
\widetilde{W}  \boldsymbol{\hat{r}}
\equiv
\boldsymbol{\hat{s}}
\equiv
\bigg( \!
\begin{array}{c}
\hat{\boldsymbol{\mathfrak{q}}} \\  \hat{\boldsymbol{\mathfrak{p}}}
\end{array} \! \bigg)
\ee
where the hermitian operators $\hat{\mathfrak{q}}_k$ and $\hat{\mathfrak{p}}_k$ collected in the vector $\boldsymbol{s}$ satisfy the canonical commutation relations, 
being $\widetilde{W} $ symplectic.
It is convenient to introduce the annihilation and creation operators 
$\boldsymbol{\hat{b}}^{\textrm t} \equiv (\hat{\mathfrak{b}}_1 \, \dots  \, \hat{\mathfrak{b}}_\ell \;\, \hat{\mathfrak{b}}_1^\dagger\,  \dots \, \hat{\mathfrak{b}}_\ell^\dagger\,)$ as follows
\be
\label{b_operators def}
\boldsymbol{\hat{b}}
\equiv 
\Omega^{-1} \boldsymbol{\hat{s}}
\;\;\;\qquad\;\;\;
\hat{\mathfrak{b}}_k \equiv \frac{\hat{\mathfrak{q}}_k +\textrm{i}\,\hat{\mathfrak{p}}_k}{\sqrt{2}}
\qquad
\Omega \equiv 
\frac{1}{\sqrt{2}}
\bigg( \!\! \begin{array}{cc}
 \boldsymbol{1} &  \boldsymbol{1} \\
 -\textrm{i} \boldsymbol{1} & \textrm{i} \boldsymbol{1} \\
\end{array}  \! \bigg)
\ee
where the $2\ell \times 2\ell$ matrix $\Omega$ is unitary. 
Since $\Omega^{-1} (\textrm{i} J) \Omega^{-\textrm t} = J$, 
it is straightforward to check that $[\hat{b}_i, \hat{b}_j] =  J_{ij}$.
In terms of these operators, the entanglement hamiltonian (\ref{K_A s_diag}) becomes
\be
\label{K_A HC eps-form}
\widehat{K}_A 
= 
\frac{1}{2}\; \boldsymbol{\hat{b}}^{\textrm t} 
\bigg( \! \begin{array}{cc}
 \boldsymbol{0} &  \mathcal{E} \\
\mathcal{E} &  \boldsymbol{0} \\
\end{array}  \! \bigg) \,
\boldsymbol{\hat{b}}
=
\sum_{k=1}^\ell 
\varepsilon_k \! \left( \hat{\mathfrak{b}}_k^\dagger\, \hat{\mathfrak{b}}_k +\frac{1}{2}\, \right).
\ee

This operator has a well known form; hence it can be treated in the standard way. 
By introducing the eigenstates $|n_k\rangle$ of the occupation number operator $  \mathfrak{b}_k^\dagger \mathfrak{b}_k $,
whose eigenvalues are given by non negative integers $n_k$,
the reduced density matrix (\ref{rho_A}) can be written through the projectors
$|\boldsymbol{n}\rangle\equiv\bigotimes_{k=1}^\ell| n_k \rangle$
as follows \cite{holevo}
\be
\label{rho_A lambda_n}
\rho_A
=
\sum_{\boldsymbol{n}} 
\frac{e^{-\sum_{k=1}^\ell \varepsilon_k ( n_k+1/2 )} }{\mathcal{Z}_A}\;
|\boldsymbol{n}\rangle \langle \boldsymbol{n}|
\,\equiv
\sum_{\boldsymbol{n}}
\lambda_{\boldsymbol{n}}  | \boldsymbol{n}\rangle \langle \boldsymbol{n}|
\ee
where $n_k$ is the $k$-th element of the $\ell$-dimensional vector $\boldsymbol{n}$.
The coefficients $\lambda_{\boldsymbol{n}}$ provide the entanglement spectrum.
The normalisation condition $\Tr \rho_A =1$ leads to
\be
\label{normalisation final}
\mathcal{Z}_A
=
\sum_{\boldsymbol{m}}
 \langle \boldsymbol{m}| 
 \bigg(\!
 \sum_{\boldsymbol{n}} 
e^{-\!\sum_{k=1}^\ell \varepsilon_k ( n_k+1/2 )}
|\boldsymbol{n}\rangle \langle \boldsymbol{n}|
 \bigg)
 | \boldsymbol{m}\rangle
=
\prod_{k=1}^\ell \frac{ e^{-\varepsilon_k / 2}}{1- e^{-\varepsilon_k }}
=
\prod_{k=1}^\ell \sqrt{\sigma_k^2- 1 /4}
\ee
where the orthonormality of the states $| \boldsymbol{n}\rangle$ has been employed
to write $\mathcal{Z}_A$ as the product of $\ell$ geometric series (one for each $n_k$).
Combining (\ref{rho_A lambda_n}) and (\ref{normalisation final}), 
the generic element of the entanglement spectrum can be expressed in terms of the symplectic spectrum $\mathcal{D}$ as follows
\be
\label{ent_spec_1}
\lambda_{\boldsymbol{n}}
=
\prod_{k=1}^\ell
\left[\left(1- e^{-\varepsilon_k }\right) e^{- n_k \varepsilon_k }\right]
=
\prod_{k=1}^\ell
\frac{1}{\sigma_k+1/2}
\left( \frac{\sigma_k-1/2}{\sigma_k+ 1/2}\right)^{n_k}.
\ee
The largest eigenvalue is given by
$\lambda_{\text{\tiny max}} = \lambda_{\boldsymbol{0}}$, namely it has $n_k=0$
for all $k$.
Its explicit expression reads
$\lambda_{\text{\tiny max}} = \prod_{k=1}^\ell (1- e^{-\varepsilon_k }) = \prod_{k=1}^\ell  (\sigma_k+1/2)^{-1}$.
Thus, (\ref{ent_spec_1}) can be written as
\begin{equation}
\label{lambda_n over lambda_max hc}
\lambda_{\boldsymbol{n}}
=
\lambda_{\text{\tiny max}} \, e^{- \sum_{k=1}^\ell  n_k \varepsilon_k}
=
\lambda_{\text{\tiny max}}
\prod_{k=1}^\ell
\left( \frac{\sigma_k-1/2}{\sigma_k+1/2}\right)^{n_k}.
\ee

The gap between two eigenvalues in the entanglement spectrum 
characterised by the vectors $\boldsymbol{m}$ and $\boldsymbol{n}$ is given by 
\begin{equation}
\label{hc_gaps_generic}
g_{\boldsymbol{m},\boldsymbol{n}}
\equiv
\log\lambda_{\boldsymbol{m}} -  \log\lambda_{\boldsymbol{n}}
=
\sum_{k=1}^\ell
(n_k - m_k) \, \varepsilon_k
=
\sum_{k=1}^\ell
(m_k - n_k)\log\!\left(\frac{\sigma_k-1/2}{\sigma_k+1/2} \right).
\end{equation}
The gaps with respect to the largest eigenvalue are
$\log\lambda_{\text{\tiny max}} -  \log\lambda_{\boldsymbol{n}} = \sum_{k=1}^\ell \varepsilon_k \,n_k$
and correspond to the special case $\boldsymbol{m}=\boldsymbol{0}$ in (\ref{hc_gaps_generic}).
By arranging the symplectic spectrum in decreasing order $\sigma_1 \geqslant \sigma_2 \geqslant \dots \geqslant \sigma_\ell$,
we have $\varepsilon_1 \leqslant \varepsilon_2 \leqslant \dots \leqslant \varepsilon_\ell$ for the single particle entanglement energies.
In order to find the gaps $0< g_1 < g_2 < \dots$ with respect to the largest eigenvalue, one can first introduce
$\mathcal{B}_p^{(r)} \equiv \big\{ \sum_{j=1}^p \varepsilon_{k_j}  \, , 1\leqslant k_j \leqslant r \big\}$ for $r\geqslant 1$
and then compute the $r$-th gap as $ g_r = \textrm{min}[ ( \cup_{p=1}^r \mathcal{B}_p^{(r)}  ) \setminus   \{ g_j, 1\leqslant j \leqslant r-1\} ] $.
This procedure is not optimal, but it allows to get $g_r$ easily also for high values of $r$.
The smallest gaps are given by $g_1 = \varepsilon_1$, $g_2  = \textrm{min}\{2\varepsilon_1, \varepsilon_2   \}$ and 
$g_3 = \textrm{min} \big\{ 3\varepsilon_1, \textrm{max}\{2\varepsilon_1, \varepsilon_2   \} , \varepsilon_3 \big\}$.
The curves in Fig.\;\ref{fig:SpectrumBoson} have been found by employing the above expressions.

It is useful to partition 
the symmetric matrices $\gamma_A $ and $H_A$ into $\ell \times \ell $ blocks as follows
\be
\label{gamma-W-block-dec}
\gamma_A \equiv
\bigg( \hspace{-.02cm} 
\begin{array}{cc}
 Q &  R \\
 R^{\textrm t} &  P \\
\end{array}  \hspace{-.02cm}  \bigg)
\;\;\; \qquad \;\;\;
H_A \equiv
\bigg( \hspace{-.02cm} 
\begin{array}{cc}
 M &  E \\
 E^{\textrm t} &  N \\
\end{array}  \hspace{-.02cm}  \bigg)\,.
\ee
The blocks of $\gamma_A$ are the correlators
$Q_{ij} =  \langle \hat{q}_i \,\hat{q}_j \rangle$, 
$P_{ij} = \langle \hat{p}_i \,\hat{p}_j \rangle$ and 
$R_{ij} = \textrm{Re} \langle \hat{q}_i \,\hat{p}_j \rangle$, 
with $1\leqslant i \leqslant \ell$ and $1\leqslant j \leqslant \ell$.
Correspondingly,  the partitions in $\ell \times \ell$ blocks of
the symplectic matrices $W $ and $\widetilde{W}$ occurring in the Williamson's decompositions
of $\gamma_A $ and $H_A$ (see (\ref{williamson th gammaA}), (\ref{H_A Williamson_dec}) and (\ref{W widetildeW relation}))
read
\be
\label{W Wtilde block form}
W \equiv
\bigg( \hspace{-.02cm} 
\begin{array}{cc}
 U &  Y \\
 Z &  V \\
\end{array}  \hspace{-.02cm}  \bigg)
\;\;\; \qquad \;\;\;
\widetilde{W} \equiv
\bigg( \hspace{-.02cm} 
\begin{array}{cc}
 \widetilde{U} &  \widetilde{Y} \\
 \widetilde{Z} &  \widetilde{V} \\
\end{array}  \hspace{-.02cm}  \bigg)
\ee
and therefore the relation  (\ref{W widetildeW relation}) becomes
\be
\label{W-Wtilde block rel}
\bigg( \hspace{-.02cm} 
\begin{array}{cc}
 \widetilde{U} &  \widetilde{Y} \\
 \widetilde{Z} &  \widetilde{V} \\
\end{array}  \hspace{-.02cm}  \bigg)
=
\bigg( \hspace{-.03cm} 
\begin{array}{cc}
 V &  \!\!-Z \\
 -Y &  \!\! U \\
\end{array}  \hspace{-.0cm}  \bigg)\,.
\ee

These partitions in $\ell \times \ell$ blocks and the relation (\ref{W_H is Wtilde})
lead to write the Williamson's decompositions (\ref{williamson th gammaA}) and (\ref{H_A Williamson_dec})
of  $\gamma_A $ and $H_A$ respectively as
\be
\gamma_A 
=
\bigg( \hspace{.0cm} 
\begin{array}{cc}
  U^{\textrm t} \, \mathcal{D} \, U +  Z^{\textrm t} \, \mathcal{D} \, Z \hspace{.3cm}
  &  
  U^{\textrm t} \, \mathcal{D} \, Y + Z^{\textrm t} \, \mathcal{D} \, V
   \\
 Y^{\textrm t} \, \mathcal{D} \, U + V^{\textrm t} \, \mathcal{D} \, Z \hspace{.3cm}
 &  
 Y^{\textrm t} \, \mathcal{D} \, Y +  V^{\textrm t} \, \mathcal{D} \, V 
\end{array}  \hspace{-.02cm}  \bigg)
\ee 
and
\be
\label{HA-block-dec-extended-final}
H_A 
=
\bigg( \hspace{-.1cm} 
\begin{array}{cc}
  V^{\textrm t} \, \mathcal{E} \, V +  Y^{\textrm t} \, \mathcal{E} \, Y \hspace{.3cm}
  &  
  - \,V^{\textrm t} \, \mathcal{E} \, Z - Y^{\textrm t} \, \mathcal{E} \, U
   \\
 -\,Z^{\textrm t} \, \mathcal{E} \, V - U^{\textrm t} \, \mathcal{E} \, Y \hspace{.3cm}
 &  
 Z^{\textrm t} \, \mathcal{E} \, Z +  U^{\textrm t} \, \mathcal{E} \, U 
\end{array}  \hspace{-.05cm}  \bigg)\,.
\ee 
Since the symplectic spectrum $\mathcal{E}$ of $H_A$ is related to
the symplectic spectrum $\mathcal{D}$ of $\gamma_A$ through (\ref{Ediag_Ddiag}),
the expression (\ref{HA-block-dec-extended-final}) provides the
entanglement hamiltonian matrix $H_A$ in terms of the blocks of the matrices occurring in the 
Williamson's decomposition of $\gamma_A$.

In the Appendix\;\ref{app:EH-block-diag}  we discuss the special case where $\gamma_A$ and $H_A$
are block diagonal.
This class is very important because it includes the entanglement hamiltonians matrices
corresponding to static configurations \cite{ch-rev, Arias-16, Arias-17, Arias-18}.

\subsection{A contour for the entanglement entropies}
\label{sec:HC_contour}

The contour for the entanglement entropies is a function of the position inside the region $A$
(and of the time, when the system is out of equilibrium) that has been introduced to understand the spatial 
structure of the bipartite entanglement in pure states. 
For free lattice models, it has been studied in \cite{br-04, chen-vidal, frerot-roschilde, cdt-17-contour, trs-18-rainbow}.
Like the entanglement hamiltonian, also 
the contour for the entanglement entropies 
cannot be determined from the entanglement spectrum only.

The contour function for the entanglement entropies $s_A^{(n)} : A \to \mathbb{R}$ must satisfy 
the following two conditions
\be
\label{contour_lattice_def}
S^{(n)}_A = \sum_{i\,\in\, A} s^{(n)}_A(i)
\;\;\qquad \;\;
s^{(n)}_A(i) \geqslant 0\,.
\ee
In \cite{chen-vidal} other three requirements have been introduced and
a contour function for the entanglement entropy of free fermions has been constructed,
verifying that it fulfils all these five properties. 
This set of constraints does not fix uniquely the contour function.

In the harmonic lattices, a contour function has been proposed 
in \cite{cdt-17-contour}, that satisfies the two conditions in (\ref{contour_lattice_def}) 
and a weakened version of the other three conditions proposed in \cite{chen-vidal}. 
In this manuscript we will employ this prescription.
Another expression fulfilling (\ref{contour_lattice_def}) has been suggested in \cite{cdt-17-contour}, but the remaining three properties have not been proved for this proposal. 
A third contour function has been considered in \cite{br-04, frerot-roschilde}: it satisfies the first condition in (\ref{contour_lattice_def})
but numerical violations of the positivity condition $s^{(n)}_A(i) \geqslant 0$ have been observed for some configurations \cite{cdt-17-contour}.

Let us consider the case where $A$ is an  interval made by $\ell$ sites, for simplicity.
A contour function $s^{(n)}_A(i)$ can be constructed by associating $\ell$ real numbers $p_k(i)$ to every symplectic eigenvalue $\sigma_k$,
where $1\leqslant i, k\leqslant \ell$.
The function $p_k(i)$ is often called mode participation function \cite{br-04} and it fulfils  the following conditions
\be
\label{pki condition}
\sum_{i=1}^{\ell}p_k(i)=1
\;\;\;\;\qquad\;\;\;\;
p_k(i)\geqslant  0\,.
\ee
A mode participation function $p_k(i)$ allows 
to write the entanglement entropies (\ref{SA and SAn from nuk}) as
in (\ref{contour_lattice_def}) with
\be
\label{contour from mpf}
s_A^{(n)}(i)
=
\sum_{k=1}^{\ell}
p_k(i)\, s_n(\sigma_k)
\ee
being $s_n(y)$ the functions defined in (\ref{sx def}) and (\ref{snx def}) for $n\geqslant 1$.

We find it natural to consider also the contribution to the entanglement entropies given by an interval $(i_1, i_2) \subset A$, namely
\be
\label{integrated contour x1x2 discrete}
\mathcal{S}^{(n)}_A(i_1, i_2)
=\!
\sum_{i \, =\, i_1}^{i_2}  s^{(n)}_A(i)
\;\; \qquad \;\;
i_1, i_2 \in A\,.
\ee
The entanglement entropies $S^{(n)}_A$ correspond to the special case where $i_1$ and $i_2$ are the sites at the boundary of $A$.

We are interested in the temporal evolution of the above quantities after a global quench and
it is often useful to study their increments with respect to their initial value.
In particular, for $\mathcal{S}^{(n)}_A(i_1, i_2;t) $ given by (\ref{integrated contour x1x2 discrete}), let us introduce
\be
\label{integrated contour x1x2 discrete sub}
\Delta\mathcal{S}^{(n)}_A(i_1, i_2;t)
\equiv
\mathcal{S}^{(n)}_A(i_1, i_2;t) -\mathcal{S}^{(n)}_A(i_1, i_2;0) \,.
\ee

In harmonic lattices,  the contour function (\ref{contour from mpf}) is not unique because any mode participation function 
fulfilling (\ref{pki condition}) provides a contour function (\ref{contour from mpf})
that satisfies (\ref{contour_lattice_def}).
Furthermore, any $2\ell \times 2\ell$ orthogonal matrix  $O$ can be employed to construct a mode participation 
function $p_k(i)$.
In order to make contact with the reduced covariance matrix $\gamma_A$,
in \cite{cdt-17-contour} the Euler decomposition of the symplectic matrix $W$ in (\ref{williamson th gammaA}) 
has been  employed to construct an orthogonal matrix $K$ leading to a reasonable candidate for the contour 
function $s_A^{(n)}(i)$.
This orthogonal matrix can be written in terms of $W$ as follows
\be
K
\equiv 
(W W^{\mathrm{t}})^{-1/2}\, W
=
W(W^{\mathrm{t}}W)^{-1/2}.
\ee
Partitioning this $2\ell \times 2\ell$ orthogonal matrix into $\ell \times \ell$ blocks
\be
K =
\bigg(\!
\begin{array}{cc}
 U_K &  \!Y_K \\
 Z_K &  \!V_K \\
\end{array}  \! \bigg)
\ee
the contour function for the entanglement entropies in harmonic lattices proposed in \cite{cdt-17-contour}
is (\ref{contour from mpf}) with the mode participation function given by 
\be
\label{mpf from W}
p_k(i)
=
\frac{1}{2}\bigg (  [(U_K)_{k,i}]^2  +  [(Y_K)_{k,i}]^2  + [(Z_K)_{k,i}]^2  +  [(V_K)_{k,i}]^2   \bigg)\,.
\ee

Thus, the symplectic spectrum of $\gamma_A$ and the symplectic matrix $W$ occurring in its
Williamson's decomposition (\ref{williamson th gammaA}) provide both  
the entanglement hamiltonian matrix $H_A$ of \cite{banchi-pirandola-15} (see e.g. (\ref{HA-block-dec-extended-final}))
and the contour function for the entanglement entropies of \cite{cdt-17-contour}.

Let us remark that, although the discussion about the contour function $s_A^{(n)}(i)$ has been done for an interval, 
it is straightforward to extend the above formulas to higher spatial dimensions or to subregions $A$ made by many disconnected components.

\section{Free fermionic lattices}
\label{sec-williamson-fermion}

In this section we consider a fermionic system described by a quadratic hamiltonian.
In \S\ref{sec:fermions-eh} we study the entanglement hamiltonian of a subsystem, that is quadratic as well.
In \S\ref{sec-contour-fermions} we review the construction of the contour for the entanglement entropy in these systems proposed in \cite{chen-vidal}.
Relevant special cases are discussed in \S\ref{sec:fermions-special cases}.

\subsection{Entanglement hamiltonian}
\label{sec:fermions-eh}

In a chain of free fermions described by a quadratic hamiltonian and in a Gaussian state,
the entanglement hamiltonian $\widehat{K}_A$  takes the following quadratic form
\cite{peschel-03-modham}
\be
\label{K_A fermions generic}
\widehat{K}_A 
\,=\,
\frac{1}{2}
\sum_{i,j=1}^\ell \!
\Big(
B_{i,j}\, \hat{c}_i^\dagger \, \hat{c}_j
+ D_{i,j}\, \hat{c}_i \, \hat{c}_j^\dagger
+ F_{i,j}\, \hat{c}_i \, \hat{c}_j
+ G_{i,j}\, \hat{c}_i^\dagger \, \hat{c}_j^\dagger
\, \Big)
\ee
where for the operators $\hat{c}_i$ and $\hat{c}_i^\dagger$ the canonical anticommutation relations 
$\{\hat{c}_i, \hat{c}_j^\dagger\} = \delta_{i,j}$ and $\{\hat{c}_i, \hat{c}_j\} = \{\hat{c}_i^\dagger, \hat{c}_j^\dagger\}  =0$ hold.
The complex matrices in (\ref{K_A fermions generic}) satify  $D = - B^\ast$ and $G =-F^\ast$ to ensure that $\widehat{K}_A$ is hermitian. 
Collecting the operators into the $2\ell$ dimensional vector $\boldsymbol{\hat{a}}$ defined by
$\boldsymbol{\hat{a}}^{\textrm t} \equiv (\hat{c}_1 \, \dots  \, \hat{c}_\ell \;\, \hat{c}_1^\dagger\,  \dots \, \hat{c}_\ell^\dagger\,)$,
the canonical anticommutation relations become $\{ \hat{a}_i, \hat{a}_j\} = I_{i,j}$, 
being $I$ the $2\ell \times 2\ell$ symmetric matrix with the $\ell \times \ell$ matrix $\boldsymbol{0}$
along the diagonal and the $\ell \times \ell$ identity matrix off diagonal. 
Standard manipulations allow to write the entanglement hamiltonian (\ref{K_A fermions generic}) in the following form 
 \cite{peschel-03-modham}
\be
\label{K_A fermion c-step}
\widehat{K}_A 
\,=\,
\frac{1}{2}\; \boldsymbol{\hat{a}}^{\textrm t} H_a \, \boldsymbol{\hat{a}}
\;\;\qquad\;\;
H_a = 
\bigg( \! \begin{array}{cc}
X \!\! &  -T^\ast  \\
T \!\!  &  -X^\ast 
\end{array}  \!\! \bigg) 
\ee
up to constant terms, where the $\ell \times \ell$ matrices $T$ and $X$ are
\be
T \equiv  \frac{1}{4} \Big( B + B^\dagger - \big(D + D^\dagger \big)^\ast \Big) 
\qquad 
X \equiv\frac{1}{4} \Big(F  - F^{\textrm t} - \big(  G - G^{\textrm t}\big)^\ast \Big)
\ee
and satisfy the conditions $T^\dagger = T$ and $X^{\textrm t} = - X$.

The matrix $\Omega$ defined in (\ref{b_operators def}) allows to introduce the vector 
$\boldsymbol{\hat{r}} \equiv \Omega \, \boldsymbol{\hat{a}}$, whose elements are hermitian operators
that satisfy $\{\hat{r}_i , \hat{r}_j \} = \delta_{i,j}$, being $\Omega \,I\, \Omega^{\textrm t} = \boldsymbol{1}$.
In terms of these Majorana operators, the entanglement hamiltonian (\ref{K_A fermion c-step}) becomes
\be
\label{K_A fermion r-step}
\widehat{K}_A 
=
\frac{1}{2}\; \boldsymbol{\hat{r}}^{\textrm t} H_r \, \boldsymbol{\hat{r}}
= 
\frac{\textrm{i}}{2}\;
\boldsymbol{\hat{r}}^{\textrm t} H_A \, \boldsymbol{\hat{r}}
\;\;\;\qquad\;\;\;
H_r \equiv \Omega^{-\textrm t} H_a \,\Omega^{-1}
\qquad
H_A \equiv - \textrm{i} H_r
\ee
where $H_A$ is the following matrix
\be
\label{H_A XT}
H_A
\equiv
\bigg( \!\begin{array}{cc}
\textrm{Im}(T+X) &  \textrm{Re}(T+X)  \\
- \,\textrm{Re}(T-X)    &  \textrm{Im}(T-X) 
\end{array}  \! \bigg) \,.
\ee
Since $H_A$ is $2\ell \times 2\ell$ real and antisymmetric, 
a $2\ell \times 2\ell$  real orthogonal matrix $\widetilde{O}$ exists such that
\be
\label{diagonalisation_Hr}
\widetilde{O}\, H_A  \,\widetilde{O}^{\textrm t} 
=
\bigg( \!\! \begin{array}{cc}
 \boldsymbol{0} \!&  \mathcal{E} \\
-\mathcal{E} \!&  \boldsymbol{0} \\
\end{array}  \! \bigg) 
\ee
where $\mathcal{E} = \textrm{diag}\{ \varepsilon_1 , \dots ,  \varepsilon_\ell \}$ with $\varepsilon_k > 0$  
 \cite{matrix-analysis}.
 The $\varepsilon_k$ are sometimes called single particle entanglement energies. 
From (\ref{diagonalisation_Hr}) one observes that 
$\widetilde{O} \,(\textrm{i}H_A)^2 \, \widetilde{O}^{\textrm t} =  \mathcal{E}^2 \oplus \mathcal{E}^2$.
Thus, the real symmetric matrix $-H_A^2$ has a degenerate spectrum given by $\{ \,\varepsilon_k^2\, , 1\leqslant k \leqslant \ell\, \}$ and is diagonalised by the orthogonal matrix $\widetilde{O}$.

The second crucial $2\ell \times 2\ell$ real antisymmetric matrix in the analysis 
of the fermionic Gaussian state is the following correlation matrix
\cite{lieb-schulz-ExactlySolvModels, lieb-gaussian-fermion-states, ee-lattice-fermion, br-fermions} 
\be
\label{Williamson dec Gamma-fermion}
(  \Gamma_A)_{i,j} 
\equiv 
-\,\textrm{i}\,
 \langle [  \hat{r}_i , \hat{r}_j ]  \rangle
\ee
which gives $ \langle  \hat{r}_i \hat{r}_j  \rangle = [\,\textrm{i} \, (  \Gamma_A)_{i,j} +\delta_{i,j} \,]/2$.
The same theorem employed above for $H_A$ allows to claim that a $2\ell \times 2\ell$ real orthogonal matrix $O$ exists such that
\be
\label{diagonalisation_Gamma_r}
O\, \Gamma_A \,O^{\textrm t} 
=
\bigg( \!\! \begin{array}{cc}
 \boldsymbol{0} \!&  \mathcal{N} \\
-\mathcal{N} \!&  \boldsymbol{0} \\
\end{array}  \! \bigg) 
\ee
where $\mathcal{N} \equiv \textrm{diag}\big\{ \nu_1 , \dots ,  \nu_\ell \big\}$ with $\nu_k > 0$.

The single particle entanglement spectrum $\mathcal{E}$ and the orthogonal matrix $\widetilde{O}$  
in (\ref{diagonalisation_Hr}) 
are related to the diagonal matrix $\mathcal{E}$ and the orthogonal matrix $O$ in (\ref{diagonalisation_Gamma_r}) 
as follows \cite{peschel-03-modham} 
\be
\label{eps spectrum Omatrix fermion}
\mathcal{E} =  2 \,\textrm{arctanh} \big(\mathcal{N}\big)
\hspace{.6cm} \qquad \hspace{.6cm}
O = \widetilde{O}
\ee
where the first expression is equivalent to $\varepsilon_k = 2 \,\textrm{arctanh} (\nu_k) =\log[(1+\nu_k)/(1-\nu_k)]$
with $1\leqslant k \leqslant \ell$.
By employing (\ref{diagonalisation_Gamma_r}), the relations in (\ref{eps spectrum Omatrix fermion}) 
lead to\footnote{Notice that, for an odd function 
$f_{\textrm{\tiny odd}}(x) \equiv \sum_{p\geqslant 0} t_p \,x^{2p+1}$, by using (\ref{diagonalisation_Gamma_r}), one finds
\be
f_{\textrm{\tiny odd}}\big(  O\, \Gamma_A \,O^{\textrm t}  \big) 
=
\bigg( \!\! \begin{array}{cc}
 \boldsymbol{0} \!&  \tilde{f}_{\textrm{\tiny odd}}(\mathcal{N}) \\
-\,\tilde{f}_{\textrm{\tiny odd}}(\mathcal{N})  \!&  \boldsymbol{0} \\
\end{array}  \! \bigg) 
\ee
where $\tilde{f}_{\textrm{\tiny odd}}(x) \equiv \sum_{p\geqslant 0} (-1)^p \,t_p \,x^{2p+1} $.}
\be
\label{H_A final fermion}
H_A = 2 \arctan \!\big(\Gamma_A\big)\,.
\ee
This formula plays the same role of (\ref{banchi-EH}) in the bosonic case (see \S\ref{sec:EHwilliamson}).

The relation (\ref{diagonalisation_Hr}) leads to introduce $\boldsymbol{\hat{d}} \equiv O \, \boldsymbol{\hat{r}}$,
whose elements satisfy $\{\hat{d}_i , \hat{d}_j \} = \delta_{i,j}$, being $O $ orthogonal.
Furthermore, it is convenient to define fermionic creation and annihilation operators $\hat{\mathfrak{f}}_k^\dagger$ and $\hat{\mathfrak{f}}_k$
as $\boldsymbol{\hat{f}} \equiv \Omega^{-1}  \boldsymbol{\hat{d}} $,
where $\boldsymbol{\hat{f}}^{\textrm t} = (\, \hat{\mathfrak{f}}_1 \,\dots\, \hat{\mathfrak{f}}_\ell \;\, \hat{\mathfrak{f}}_1^\dagger \, \dots \,\hat{\mathfrak{f}}_\ell^\dagger \,)$, whose elements satisfy $\{ \hat{f}_i, \hat{f}_j\} = I_{i,j}$.
Combining the linear maps introduced above, one obtains
\be
\label{omega-O-omega map}
\boldsymbol{\hat{f}} = \Omega^{-1} \, O \, \Omega \, \boldsymbol{\hat{a}}
\ee
 $\Omega^{-1} \, O \, \Omega$ is unitary, being $O$ real orthogonal and $ \Omega$ unitary. 
 The relation (\ref{diagonalisation_Hr}) and the anticommutation relations of $\boldsymbol{\hat{f}}$ allow to write 
 the entanglement hamiltonian (\ref{K_A fermion r-step}) in terms of the operators in $\boldsymbol{\hat{d}}$ 
 and $\boldsymbol{\hat{\mathfrak{f}}}$ respectively as 
\be
\label{K_A fermions final}
\widehat{K}_A = 
\frac{\textrm{i}}{2}\;
\boldsymbol{\hat{d}}^{\textrm t} \bigg( \!\! \begin{array}{cc}
 \boldsymbol{0} \!&  \mathcal{E} \\
-\,\mathcal{E} \!&  \boldsymbol{0} \\
\end{array}  \! \bigg)  \, \boldsymbol{\hat{d}}
\,=
\sum_{k=1}^\ell \varepsilon_k \!\left( \hat{\mathfrak{f}}_k^\dagger\, \hat{\mathfrak{f}}_k - \frac{1}{2}\, \right)
\ee
that makes explicit the role of the single particle entanglement energies.

Let us introduce the partition in $\ell \times \ell$ blocks of the orthogonal matrix defined in (\ref{diagonalisation_Gamma_r}), namely
\be
\label{O-block-dec}
O =
\bigg(\!
\begin{array}{cc}
 U_O &  \!Y_O \\
 Z_O &  \!V_O \\
\end{array}  \! \bigg)
\ee
where the blocks are constrained by the orthogonality of $O$ as follows
\be
\label{ortho-cond}
 O O^{\textrm{t}}
 =
\bigg(\!
\begin{array}{cc}
 U_O U_O^{\textrm{t}} + Y_O Y_O^{\textrm{t}} &  \!U_O Z_O^{\textrm{t}} + Y_O V_O^{\textrm{t}}  \\
 Z_O U_O^{\textrm{t}} + V_O Y_O^{\textrm{t}} &  \!Z_O Z_O^{\textrm{t}} + V_O V_O^{\textrm{t}}  \\
\end{array}  \! \bigg)
=
\bigg(\!
\begin{array}{cc}
 \boldsymbol{1} &  \!\boldsymbol{0} \\
 \boldsymbol{0} &  \!\boldsymbol{1} \\
\end{array}  \! \bigg)\,.
\ee
In terms of the blocks of $O$ in (\ref{O-block-dec}), the unitary matrix in (\ref{omega-O-omega map}) becomes
\be
\label{omOom-general}
\Omega^{-1} \, O \, \Omega 
=
\frac{1}{2}
\bigg(\!
\begin{array}{cc}
 U_O+V_O +\textrm{i}( Z_O-Y_O) &     U_O-V_O +\textrm{i}( Z_O+Y_O) \\
 U_O-V_O -\textrm{i}( Z_O+Y_O) &     U_O+V_O -\textrm{i}( Z_O-Y_O)  \\
\end{array}  \! \bigg)\,.
\ee
By employing that $\boldsymbol{\hat{d}} \equiv O \, \Omega \,\boldsymbol{\hat{a}}$ 
in (\ref{K_A fermions final}) and comparing the resulting expression with (\ref{K_A fermion c-step}), 
one writes $T$ and $X$ in terms of the blocks  of $O$ in (\ref{O-block-dec}) as follows
\bea
\label{T=UVec}
T &=& \frac{1}{2} \,\Big\{
 U_O^{\textrm{t}} \,\mathcal{E}  \,V_O + V_O^{\textrm{t}} \,\mathcal{E}  \,U_O
 - Y_O^{\textrm{t}} \,\mathcal{E}  \,Z_O - Z_O^{\textrm{t}} \,\mathcal{E}  \,Y_O
 \\
 & & \hspace{1cm}
 + \textrm{i} 
 \big[\,
 U_O^{\textrm{t}} \,\mathcal{E}  \,Z_O - Z_O^{\textrm{t}} \,\mathcal{E}  \,U_O
 + Y_O^{\textrm{t}} \,\mathcal{E}  \,V_O -  V_O^{\textrm{t}} \,\mathcal{E}  \,Y_O
\, \big]
 \Big\}
 \nonumber
 \\
 \label{X=UVec}
 \rule{0pt}{.8cm}
 X &=& \frac{1}{2} \,\Big\{
 U_O^{\textrm{t}} \,\mathcal{E}  \,V_O - V_O^{\textrm{t}} \,\mathcal{E}  \,U_O
  + Y_O^{\textrm{t}} \,\mathcal{E}  \,Z_O - Z_O^{\textrm{t}} \,\mathcal{E}  \,Y_O
   \\
 & & \hspace{1cm}
 + \textrm{i} 
 \big[\,
 U_O^{\textrm{t}} \,\mathcal{E}  \,Z_O - Z_O^{\textrm{t}} \,\mathcal{E}  \,U_O
 - Y_O^{\textrm{t}} \,\mathcal{E}  \,V_O + V_O^{\textrm{t}} \,\mathcal{E}  \,Y_O
\, \big]
 \Big\}\,.
 \nonumber
\eea

The expression (\ref{K_A fermions final}) for the entanglement hamiltonian naturally leads to introduce the 
fermionic occupation number operators $\hat{\mathfrak{f}}_k^\dagger\, \hat{\mathfrak{f}}_k $,
whose eigenvalues and eigenvectors allow to write the reduced density matrix
as $\rho_A = \sum_{\boldsymbol{n}} \lambda_{\boldsymbol{n}}  | \boldsymbol{n}\rangle \langle \boldsymbol{n}|$.
Thus, one can perform an analysis similar to the one reported for the harmonic lattices in \S\ref{sec:EHwilliamson}, 
with the crucial difference that the elements of the vector $\boldsymbol{n}$ are $n_k \in \{0,1\}$ in these fermionic models. 
In particular, the reduced density matrix can be written as
\be
\label{rho_A lambda_n fermion}
\rho_A
=
\sum_{\boldsymbol{n}} 
\frac{e^{-\sum_{k=1}^\ell \varepsilon_k ( n_k-1/2 )} }{\mathcal{Z}_A}\;
|\boldsymbol{n}\rangle \langle \boldsymbol{n}|
\,\equiv
\sum_{\boldsymbol{n}}
\lambda_{\boldsymbol{n}}  | \boldsymbol{n}\rangle \langle \boldsymbol{n}|
\ee
whose normalisation condition $\Tr \rho_A =1$ provides the following normalisation constant
\be
\label{normalisation final fermion}
\mathcal{Z}_A
=
\sum_{\boldsymbol{m}}
 \langle \boldsymbol{m}| 
 \bigg(\!
 \sum_{\boldsymbol{n}} 
e^{-\!\sum_{k=1}^\ell \varepsilon_k ( n_k-1/2 )}
|\boldsymbol{n}\rangle \langle \boldsymbol{n}|
 \bigg)
 | \boldsymbol{m}\rangle
=
\prod_{k=1}^\ell e^{\varepsilon_k / 2}\big(1+ e^{-\varepsilon_k }\big)\,.
\ee
Plugging this expression in (\ref{normalisation final fermion}), one obtains the entanglement spectrum 
\be
\label{spectrum_fermion}
\lambda_{\boldsymbol{n}} 
= 
\frac{e^{-\sum_{k=1}^\ell \varepsilon_k \,n_k} }{\prod_{k=1}^\ell (1+ e^{-\varepsilon_k } )}
=
\lambda_{\text{\tiny max}}\, e^{-\sum_{k=1}^\ell \varepsilon_k \,n_k}
\ee
where the largest eigenvalue corresponds to $n_k =0$ for all $k$, being $\varepsilon_k > 0$.
This gives $ \lambda_{\text{\tiny max}} = \prod_{k=1}^\ell (1+ e^{-\varepsilon_k})^{-1}$, 
that has been used to get the last expression in (\ref{spectrum_fermion}).
The gap between two eigenvalues can be written as follows
\be
g_{\boldsymbol{m},\boldsymbol{n}}
\equiv
\log\lambda_{\boldsymbol{m}} -  \log\lambda_{\boldsymbol{n}}
=
\sum_{k=1}^\ell
(n_k - m_k) \, \varepsilon_k
=
\sum_{k=1}^\ell
(n_k - m_k) \, \log\!\left( \frac{1+\nu_k}{1-\nu_k} \right).
\ee
We find it convenient to consider the gaps with respect to the largest eigenvalue. 
These gaps and their temporal evolution after a global quench are discussed in \S\ref{sec:fermion-quench}.

\subsection{A contour for the entanglement entropies}
\label{sec-contour-fermions}

An exhaustive analysis of a contour function for the entanglement entropy in chains of free fermions 
has been carried out in \cite{chen-vidal}.
We employ this prescription in our analysis and 
refer the reader interested in further details to this reference.

From (\ref{diagonalisation_Gamma_r}) and the matrix $\Omega$ introduced in (\ref{b_operators def}), 
it is straightforward to find
\be
\label{iGamma/2 dec}
\frac{\boldsymbol{1} + \textrm{i} \, \Gamma_A}{2}
\,=\,
O^{\textrm{t}}\, \Omega 
\left( 
\frac{\boldsymbol{1} + \mathcal{N}}{2} \oplus \frac{\boldsymbol{1} - \mathcal{N}}{2}
\,\right) 
\Omega^\dagger O\,.
\ee
Similarly, considering (\ref{diagonalisation_Hr}) and (\ref{eps spectrum Omatrix fermion}), 
one also obtains
\be
\frac{\boldsymbol{1} + \textrm{i} \, H_A}{2}
\,=\,
O^{\textrm{t}}\, \Omega 
\left( 
\frac{\boldsymbol{1} + \mathcal{E}}{2} \oplus \frac{\boldsymbol{1} - \mathcal{E}}{2}
\,\right) 
\Omega^\dagger O\,.
\ee
Thus, by employing (\ref{iGamma/2 dec}), for a generic function $f$, we have 
\cite{cramer-plenio-rev, chen-vidal} 
\be
\label{ent-fermions-trace formula}
\Tr\bigg[ f\bigg(\frac{\boldsymbol{1} + \textrm{i} \, \Gamma_A}{2} \! \bigg) \bigg]
=
\sum_{k=1}^\ell
\bigg[\,
f\bigg(\frac{1+\nu_k}{2}\bigg) + f\bigg(\frac{1-\nu_k}{2}\bigg)
\,\bigg]
\ee
The entanglement entropy can be obtained by specifying this formula to $f(x) = - \,x \log(x)$, while
the $n$-th moment $\Tr \rho_A^n$ of the reduced density matrix corresponds to $f(x) = x^n$.
We remark that $\nu_k <1$ \cite{lieb-gaussian-fermion-states}.

A natural contour for the entanglement entropies is obtained as explained in \S\ref{sec:HC_contour} 
(see (\ref{contour from mpf}))
with the mode participation function given by \cite{chen-vidal} 
\be
\label{mpf from O chen-vidal}
p_k(i)
=
\frac{1}{2}\bigg (  [(U_O)_{k,i}]^2  +  [(Y_O)_{k,i}]^2 + [(Z_O)_{k,i}]^2  +  [(V_O)_{k,i}]^2   \bigg)\,.
\ee
In \cite{chen-vidal} it has been shown that the contour function constructed from (\ref{mpf from O chen-vidal}) satisfies 
other three properties beside (\ref{contour_lattice_def}) that we do not discuss in this manuscript. 
All these constraints can be part of a more detailed definition (still unknown) that could lead to identify 
the contour function for the entanglement entropies in a unique way.

\subsection{Special cases}
\label{sec:fermions-special cases}

It is worth focussing the expressions discussed in \S\ref{sec:fermions-eh} and \S\ref{sec-contour-fermions} in some special cases. 

First, let us consider the class of entanglement hamiltonians having
\be
\label{special-case-diag-unitary}
V_O = U_O
\;\;\qquad\;\;
Y_O = - Z_O
\ee
that includes the entanglement hamiltonian after the global quench  discussed in \S\ref{sec:fermion-quench}.

The unitary matrix (\ref{omOom-general}) that provides the map (\ref{omega-O-omega map})
becomes the diagonal matrix
$\Omega^{-1} \, O \, \Omega = (U_O +\textrm{i}\, Z_O) \oplus (U_O -\textrm{i}\, Z_O)$,
with $U_O +\textrm{i}\, Z_O$ unitary.
The latter condition can be obtained also by specialising the orthogonality condition (\ref{ortho-cond}) to this case. 
Furthermore, (\ref{T=UVec}) and (\ref{X=UVec}) simplify respectively to
\be
\label{Tmatrix-peschel}
T = 
U_O^{\textrm{t}} \,\mathcal{E}  \,U_O +  Z_O^{\textrm{t}} \,\mathcal{E}  \,Z_O
+ \textrm{i}
\big(
U_O^{\textrm{t}} \,\mathcal{E}  \,Z_O  - Z_O^{\textrm{t}} \,\mathcal{E}  \,U_O
\big)
\;\; \qquad \;\;
X= \boldsymbol{0}
\ee
where we remind that $T$ is hermitian.  
Hence, (\ref{K_A fermion c-step}) becomes a fermionic hamiltonian with hopping terms given by \cite{peschel-03-modham}
\be
\label{K_A peschel}
\widehat{K}_A 
=\!
\sum_{i,j=1}^\ell 
T_{i,j} \, \hat{c}_i^\dagger \, \hat{c}_j
\ee
 and (\ref{H_A XT}) simplifies to
\be
H_A
=
\bigg( \! \begin{array}{cc}
\textrm{Im}\,T  \!&  \textrm{Re}\,T  \\
- \,\textrm{Re}\,T   \! &  \textrm{Im}\,T
\end{array}  \! \bigg) \,.
\ee
The hermitian matrix $T$ in (\ref{K_A peschel}) 
can be diagonalised by a unitary matrix $\widetilde{U}$ 
and it has real eigenvalues $\eta_k$, i.e. 
$\widetilde{U} \,T \, \widetilde{U}^\dagger = \textrm{diag}\{ \eta_i , \dots \eta_\ell\}$.
This decomposition allows to write 
\be
\label{T-decomposition-peschel}
T_{i,j} = \sum_{k=1}^\ell  \eta_k \,   \widetilde{U}^\ast_{k,i}\, \widetilde{U}_{k,j}\,.
\ee
Plugging this result into (\ref{K_A peschel}), the entanglement hamiltonian
becomes $\widehat{K}_A  = \sum_{k=1}^\ell  \eta_k \, \hat{\mathfrak{f}}^\dagger_k \,\hat{\mathfrak{f}}_k$,
where we have defined $\boldsymbol{\hat{\mathfrak{f}}} = \widetilde{U} \, \boldsymbol{\hat{c}} $.
Comparing this result with the map in (\ref{omega-O-omega map}), that is diagonal in this special case and 
it is written in the text below (\ref{special-case-diag-unitary}), we conclude that $\widetilde{U} = U_O +\textrm{i}\, Z_O$.
These observations have been done in \cite{peschel-03-modham}, where it has also found that
\be
\label{peschel02 H_A_gen}
T^{\textrm t} = \log(C_A^{-1} - \boldsymbol{1} ) 
\ee
being $C_A$ the correlation matrix whose generic element is $(C_A)_{i,j} = \langle \hat{c}_i^\dagger  \hat{c}_j \rangle $, with $i,j\in A$.
From (\ref{peschel02 H_A_gen}), we have that the eigenvalues $\eta_k$ of $T$ are related to the eigenvalues $\zeta_k$ of $C_A$
as follows
\be
\label{etak-zetak rel}
\eta_k = \log(1/ \zeta_k - 1)\,.
\ee
Comparing with \S\ref{sec:fermions-eh}, one finds $\varepsilon_k = |\eta_k|$ and $\nu_k = |2\zeta_k -1|$.

The R\'enyi entropies are obtained from the eigenvalues of $C_A$ as 
\cite{ep-rev, ch-rev,fazio-review, cramer-plenio-rev, weedbrook}
\be
\label{renyi-fermions}
S_A^{(n)} =  \sum_{k=1}^\ell  s_n(\zeta_k)
\qquad
s_n(y) \equiv \frac{1}{1-n} \,\log \! \big[ y^n +(1-y)^n \big]\,.
\ee
The limit $n\to 1^+$ of these expressions provides the entanglement entropy as
\be 
\label{EE fermions}
S_A =  \sum_{k=1}^\ell  s(\zeta_k)
\qquad
s(y) \equiv - \,y \log(y) - (1-y) \log(1-y)\,.
\ee

As for the contour for the entanglement entropies, in the special case (\ref{special-case-diag-unitary}) 
the mode participation function (\ref{mpf from O chen-vidal}) simplifies to
\be
\label{mpf fermion A-case}
p_k(i)
=
  [(U_O)_{k,i}]^2   + [(Z_O)_{k,i}]^2    
  =
   \big| \widetilde{U}_{k,i} \big|^2\,.
\ee
As consistency check, let us observe that the $i$-th element of the diagonal of the matrix relation  
$\widetilde{U}^\dagger \widetilde{U} = \boldsymbol{1}$ gives the condition $\sum_{k=1}^\ell p_k(i) = 1$ 
for $1\leqslant i \leqslant \ell$.

Another relevant special case corresponds to $Y_O = Z_O = \boldsymbol{0}$.
For these entanglement hamiltonian matrices the unitary matrix (\ref{omOom-general}) reduces to
\be
\Omega^{-1} \, O \, \Omega 
=
\frac{1}{2}
\bigg(\!
\begin{array}{cc}
 U_O+V_O  &     U_O-V_O  \\
 U_O-V_O &     U_O+V_O  \\
\end{array}  \! \bigg)
\ee
where $U_O$ and $V_O$ are orthogonal.
In this case (\ref{T=UVec}) and (\ref{X=UVec}) become respectively
\be
T = \frac{1}{2} \,\Big\{
 U_O^{\textrm{t}} \,\mathcal{E}  \,V_O + V_O^{\textrm{t}} \,\mathcal{E}  \,U_O
\Big\}
\;\;\qquad\;\;
X = \frac{1}{2} \,\Big\{
 U_O^{\textrm{t}} \,\mathcal{E}  \,V_O - V_O^{\textrm{t}} \,\mathcal{E}  \,U_O
\Big\}\,.
\ee

Let us remarks that in the formulas described above we have not specified the explicit form of the 
two-point correlators, hence they can be used to study more complicated configurations like e.g. 
the ones where $A$ is the union of many disjoint blocks of sites.

Finally, an interesting class of entanglement hamiltonian matrices 
included in the one defined by (\ref{special-case-diag-unitary})
corresponds to $V_O = U_O$ and $Y_O = Z_O = \boldsymbol{0}$.
In this case the unitary matrix (\ref{omOom-general}) becomes the diagonal matrix
$\Omega^{-1} \, O \, \Omega = U_O \oplus U_O $,
where $U_O$ is orthogonal.
Furthermore, the expression of $T$ in (\ref{Tmatrix-peschel}) 
simplifies to $T = U_O^{\textrm{t}} \,\mathcal{E}  \,U_O$.
For instance, at equilibrium, 
the entanglement hamiltonians of an interval in the ground state 
considered in \cite{eisler-peschel-17, ep-18} belongs to this class.

\section{Insights from CFT}
\label{sec:cft-naive}

In this section we employ the analytic results obtained for the entanglement hamiltonian 
of a semi-infinite line after a global quench in CFT \cite{ct-16}
to get some insights about the qualitative behaviour of the entanglement spectrum
and of the contour for the entanglement entropies. 

The CFT analysis performed in \cite{cc-05-global quench} (see also the recent review \cite{cc-16-quench rev}) leads to the following linear growth of the entanglement entropies
before the saturation
\be
\label{renyi linear growth}
S_A^{(n)} \simeq \frac{\pi c}{3\tau_0}\left(1+\frac{1}{n}\right) t
\;\;\;\qquad\;\;\;
t/\ell < 1/2
\ee
(comparing the notation with \cite{cc-16-quench rev}, 
we have $\tau_0 = 4 \tau_{0,\textrm{\tiny there}}$),
where $c$ is the central charge of the model.
This gives $S_A \simeq 2\pi c\, t /(3\tau_0)$ for the entanglement entropy
and, by fitting this linear growth, we can get $\tau_0$ numerically.
A slight dependence on $n$ has been observed in $\tau_0$ (see e.g. \cite{coser-14-quench-neg}).
We find worth remarking that a factor of $2$ in (\ref{renyi linear growth}) 
is due to the fact that the interval has two endpoints.
Taking the limit $n \to \infty$ in (\ref{renyi linear growth})  one obtains
$-\log\lambda_{\textrm{\tiny max}} \simeq \pi c\, t /(3\tau_0)$.
The parameter $\tau_0$ encodes some features of the initial state 
(we refer to \cite{cardy-15} for a complete discussion).

The temporal evolution of the entanglement hamiltonian of a semi-infinite line
after a global quantum quench in CFT has been studied in \cite{ct-16}
by employing methods and results of CFT with boundaries \cite{bcft}.
This analysis provides also the entanglement spectrum and,
in particular, the above linear growth for $-\log\lambda_{\textrm{\tiny max}} $ is recovered up to a factor of 
$2$, which is due to the fact that the semi-infinite line has only one endpoint. 
This result suggests to explore the qualitative behaviour of the entanglement spectrum 
before the saturation at $t/\ell \simeq 1/2$ 
by taking the analytic CFT expressions for the semi-infinite line and introducing properly 
a factor of $2$ to take into account the occurrence of two endpoints.
As for the gaps in the entanglement spectrum, from \cite{ct-16} we obtain
\be
\label{gap_r cft naive}
g_{a,0}
\simeq 
\frac{\pi \tau_0 \, \Delta_a}{2\,t}
\;\;\;\qquad\;\;\;
t/\ell < 1/2
\ee
which is $1/2$ of the corresponding quantity for the semi-infinite line.
In (\ref{gap_r cft naive}), the coefficients $\Delta_a$ are the non vanishing conformal dimensions 
(including also the ones of the descendants) of a boundary CFT with the proper boundary condition,
as discussed in \cite{ct-16}.
We find it worth considering also the ratios $g_{b,0}/g_{a,0} = \Delta_b / \Delta_a$
because they are independent of $\tau_0$.
From (\ref{renyi linear growth}) and (\ref{gap_r cft naive}) it is straightforward to construct another expression where $\tau_0$ does not occur, namely
\be
\label{g_a S_A naive cft}
g_{a,0}\, S_A^{(n)} \simeq 
\frac{\pi^2 c}{6}  \left(\!1+\frac{1}{n}\right)  \Delta_a
\;\;\;\qquad\;\;\;
t/\ell < 1/2\,.
\ee
The temporal evolutions of (\ref{gap_r cft naive}), of (\ref{g_a S_A naive cft})
and of the ratios of the entanglement gaps with respect to the first one
after the global quenches that we consider in this manuscript 
are shown in Fig.\;\ref{fig:SpectrumBoson} and Fig.\;\ref{fig:SpectrumFermion}. 
From these data we notice that the best agreement with CFT is observed 
for the ratios of the entanglement gaps.

The above analysis can be easily adapted to the temporal evolution after a local quantum quenches \cite{cc-07-local quench} by employing the corresponding results of \cite{ct-16}.
The crucial difference is that logarithmic growths occur in these cases.  
For instance, for local quenches (\ref{gap_r cft naive}) has $\log t$ instead of $t$ in the denominator.
Instead, (\ref{g_a S_A naive cft}) holds also for local quenches.

In the remaining part of this section we exploit the CFT results for the 
entanglement hamiltonian of the semi-infinite line after a global quench \cite{ct-16}
to get insights about the temporal evolution
of the contour function for the entanglement entropies of an interval.

For some particular one-dimensional spatial bipartitions  at equilibrium 
(e.g. an interval in the infinite line at zero and finite temperature, 
or in a finite periodic system for the ground state) 
a natural candidate for the contour function for the entanglement entropies in CFT
can be obtained from the weight function occurring in the entanglement hamiltonian 
\cite{cdt-17-contour}.
By applying this idea to the half line after a global quench,
one finds the following CFT ansatz for the contour function \cite{ct-16} 
\be
\label{cont_func_cft_halfline}
s^{(n)}_{\textrm{h.l.}}(x,t) 
=
\frac{c}{12} \!\left(\! 1+\frac{1}{n} \right) \mathcal{F}_{\textrm{h.l.}}(x,t) 
\ee
where $n \geqslant 1$ and
\be
\label{Fhl def}
\mathcal{F}_{\textrm{h.l.}}(x,t) 
\equiv
\frac{2\pi\,[\cosh(2\pi  t/\tau_0)]^2 \coth(\pi x/\tau_0) }{\tau_0\,[ \cosh(4\pi  t/\tau_0) + \cosh(2\pi  x/\tau_0)]}\,.
\ee

It is straightforward to observe that
$\tau_0\, s^{(n)}_{\textrm{h.l.}}(x,t) $ depends on $x/\tau_0$, $t/\tau_0$ and the central charge $c$.
Furthermore, $\mathcal{F}_{\textrm{h.l.}}(x,t)  = 1/x +O(1)$ as $x \to 0^+$, independently of time. 
This implies that the expected logarithmic divergence of $S_A^{(n)}$ as the UV cutoff $\epsilon \to 0$ 
is independent of time as well. 
Indeed, this divergence can be obtained by removing an infinitesimal disk around the endpoint of the half line
and integrating the contour function (\ref{cont_func_cft_halfline}) from $\epsilon$ to some point inside the half line.
Instead,  by integrating (\ref{cont_func_cft_halfline}) in the entire half line $(\epsilon,+\infty)$ with $\epsilon/\tau_0 \ll 1$, one obtains \cite{ct-16}
\be
\label{integral_contour_half_line}
\int_{\epsilon}^{\infty} \!\! s^{(n)}_{\textrm{h.l.}}(x,t) \, dx 
= 
\frac{c}{12} \left(1+\frac{1}{n} \right)
\log \!\left( \frac{\tau_0}{\pi \epsilon} \, \cosh(2\pi t/\tau_0) \right)
+ O(\epsilon^2)
\ee
that gives the linear growth found in \cite{cc-05-global quench}, namely $\tfrac{\pi c}{3\tau_0} t$ when $t\gg \tau_0$.
The r.h.s. of (\ref{integral_contour_half_line}) provides only the leading term of the entanglement entropy as $\epsilon \to 0$.
A subleading $O(1)$ term comes from the contribution of the conformal boundary states
introduced through the regularisation procedure \cite{ct-16, tachikawa-14}
(see \cite{cdt-17-contour} for a discussion on this term in relation to the contour function at equilibrium).

 \begin{figure}[t!]
\vspace{.2cm}
\hspace{-1.1cm}
\includegraphics[width=1.1\textwidth]{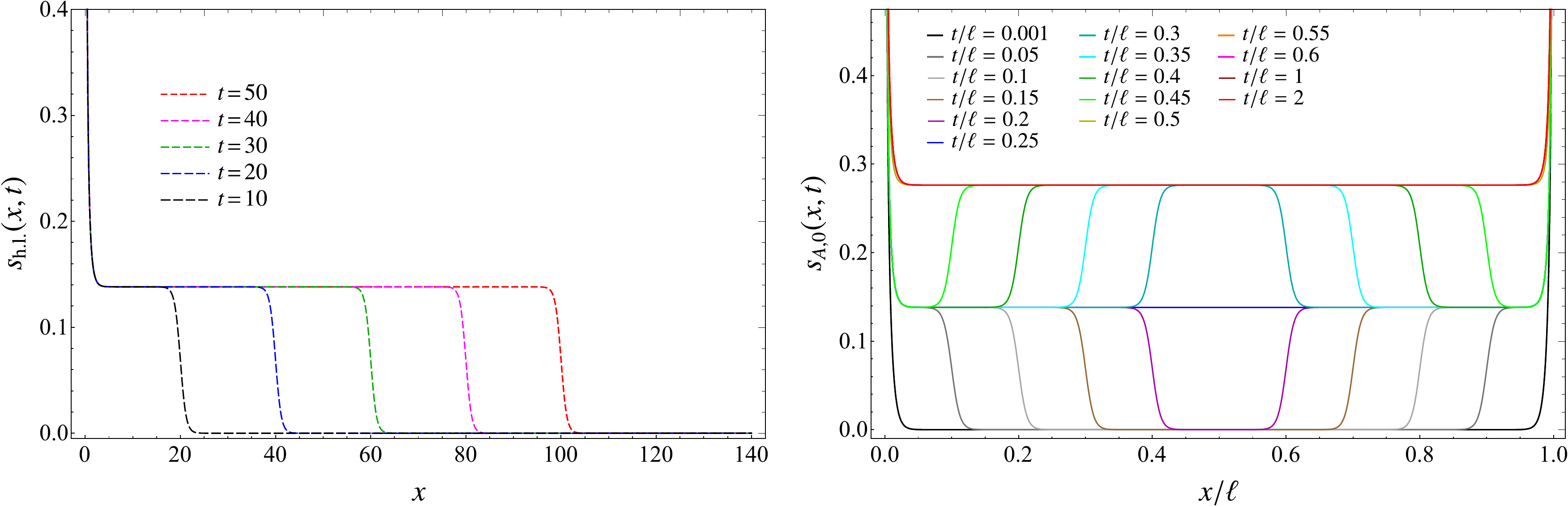}
\vspace{-.4cm}
\caption{ 
The contour function for the entanglement entropy of half line (left panel) from (\ref{cont_func_cft_halfline}) 
and of an interval (right panel) from the naive formula (\ref{contour_cft_naive}).
In these plots $c=1$ and $\tau_0 = 3.7905$.
}
\label{fig:CFTnaive}
\end{figure}

Considering $t/\tau_0$ fixed in (\ref{cont_func_cft_halfline}), 
we find that $s^{(n)}_{\textrm{h.l.}}(x,t) \sim e^{-2\pi x/\tau_0}$ as $x\to +\infty$ 
and that the height of the horizontal plateau is given by $\tfrac{\pi \,c}{12 \tau_0}(1+\tfrac{1}{n})$
(this can be done by taking $t\to \infty$ first and then $x \to +\infty$).

A typical temporal evolution of the contour function for the entanglement entropy $s_{\textrm{h.l.}}(x,t) \equiv s^{(1)}_{\textrm{h.l.}}(x,t)$
is shown in the left panel of Fig.\,\ref{fig:CFTnaive}. 
The linear divergence as $x \to 0^+$ and the exponential decay as $x \to +\infty$ are independent of time.
These two regimes are connected for the intermediate values of $x$ by a smooth step at $x \simeq 2t$ whose height is 
$\tfrac{\pi \,c}{6 \,\tau_0}$.
The occurrence of this step moving in time has been first observed in the numerical analysis performed in \cite{chen-vidal}.

In this manuscript we are interested in the contour function for the entanglement entropies of an interval $A$ of length $\ell$
in the infinite line after a global quantum quench. 
A CFT analysis for this case is not available in the literature. 
Hence, let us consider the naive contour function obtained by
superposing the two contour functions (\ref{cont_func_cft_halfline}) associated to the half lines $(0,+\infty)$ and $(-\infty, \ell)$,
namely
\be
\label{contour_cft_naive}
s^{(n)}_{A,0}(x,t) 
\equiv 
s^{(n)}_{\textrm{h.l.}}(x,t) + s^{(n)}_{\textrm{h.l.}}(\ell - x,t) 
=
\frac{c}{12} \!\left(\! 1+\frac{1}{n} \right)\!
\big[ \mathcal{F}_{\textrm{h.l.}}(x,t) +  \mathcal{F}_{\textrm{h.l.}}(\ell- x,t) \big]
\ee
with $\mathcal{F}_{\textrm{h.l.}}(x,t)$ given by  (\ref{Fhl def}).
The observations made above about $\mathcal{F}_{\textrm{h.l.}}(x,t)$ lead to notice that
$\tau_0 \, s^{(n)}_{A,0}(x,t) $ obtained from (\ref{contour_cft_naive}) 
is a function of $t/\ell$, $x/\ell$ and $\tau_0/\ell$;
and that $s^{(n)}_{A,0}(x,t) $ displays a linear divergence for $x\to 0^+$ and for $x \to \ell^-$, 
with the same time independent coefficient. 
For $t=0$, the naive analytic result (\ref{contour_cft_naive}) gives
\be
\label{contour CFT t=0}
s^{(n)}_{A,0}(x,0) 
=
\frac{\pi c}{6\tau_0} \!\left(\! 1+\frac{1}{n} \right)
\left[\,
\frac{1}{\sinh(2\pi  x/\tau_0)}
+ 
\frac{1}{\sinh(2\pi  (\ell-x)/\tau_0)} 
\,\right]
\ee
and in the asymptotic regime of long time we find
\be
\label{contour CFT large t}
\lim_{t\to \infty} s^{(n)}_{A,0}(x,t)  = 
\frac{\pi c}{12\,\tau_0} \!\left(\! 1+\frac{1}{n} \right)
\Big[
\coth(\pi x/\tau_0) +\coth(\pi (\ell-x)/\tau_0) 
\Big]\,.
\ee

The naive contour function for the entanglement entropy is given by
(\ref{contour_cft_naive}) with $n=1$ and its typical temporal evolution 
is shown in the right panel of Fig.\,\ref{fig:CFTnaive}.
This evolution can be explained qualitatively in terms of two fronts with height $\tfrac{\pi c}{6\tau_0}$ 
that start at the two different endpoints of the interval and then propagate 
in opposite directions with velocity equal to $2$.
The two fronts cross and superpose at $t/\ell \simeq 1/4$ 
and each front takes $t \simeq \ell/2$ to travel across the entire interval. 
This qualitative features have been first observed in \cite{chen-vidal}
by analysing the numerical data obtained in a chain of free fermions.
Further numerical data supporting this picture 
are presented in \S\ref{sec:HCnumerics} and \S\ref{sec:fermion-quench}.

Given two points $0< x_1 < x_2 < \ell$ inside the interval $A$, let us
integrate  the naive contour function (\ref{contour_cft_naive}) in the interval $(x_1, x_2)$.
The result reads
\bea
\label{integrated contour x1x2}
& &\hspace{-2.4cm}
\mathcal{S}^{(n)}_{A,0}(x_1, x_2;t)
 \equiv 
\int_{x_1}^{x_2} \! s^{(n)}_{A,0}(x,t)\, dx
\\
&& \hspace{-.5cm}
=
\frac{c}{24} \!\left(\! 1+\frac{1}{n} \right) 
\log \left(
\frac{\big[ \cosh(4\pi t/\tau_0) + \cosh(2\pi (\ell -x) /\tau_0)\big] \big( \!\sinh(\pi x/\tau_0) \big)^2}{
\big[ \cosh(4\pi t/\tau_0) + \cosh(2\pi x /\tau_0)\big] \big(\! \sinh(\pi (\ell-x)/\tau_0) \big)^2}
\right)\!
\Bigg|^{x_2}_{x_1}
\nonumber
\eea
which is a function of $t/\ell$, $x_1/\ell$, $x_2/\ell$ and $\tau_0 / \ell$.
Subtracting to (\ref{integrated contour x1x2}) its value at $t=0$, one finds
\be
\label{IntegContS2sub}
\Delta\mathcal{S}^{(n)}_{A,0}(x_1, x_2;t) 
\equiv
\mathcal{S}^{(n)}_{A,0}(x_1, x_2;t) -\mathcal{S}^{(n)}_{A,0}(x_1, x_2;0) 
=
\frac{c}{24} \!\left(\! 1+\frac{1}{n} \right)
\log\!\left(\frac{\eta(x_2;t)}{\eta(x_1;t)}\right)
\ee
where we have introduced the following function
\be
\eta(x;t)
\equiv
\frac{
\cosh(\pi (2t -\ell +x )/\tau_0) \, \cosh(\pi (2t +\ell - x )/\tau_0) \,[ \cosh(\pi x/\tau_0) ]^2
}{
\cosh(\pi (2t-x) /\tau_0) \, \cosh(\pi ( 2t + x)/\tau_0) \,[ \cosh(\pi (\ell-x)/\tau_0) ]^2
}\,.
\ee

In Fig.\;\ref{fig:IC_bosons} and Fig.\;\ref{fig:IC_fermions} this formula 
(where the value of $\tau_0$ is obtained by fitting the linear growth of the entanglement entropy)
gives the grey dashed-dotted curves,
that can be compared against numerical data corresponding  to 
particular quenches in a harmonic chain (see \S\ref{sec:HCnumerics}) 
and in a chain of free fermions (see \S\ref{sec:fermion-quench}) 
respectively.

The expression in (\ref{IntegContS2sub}) provides the entanglement entropies when $(x_1,x_2) = (\epsilon, \ell - \epsilon)$.
In this case, we have 
\be
\Delta \mathcal{S}^{(n)}_{A,0}(0, \ell ;t)
=
\lim_{\epsilon \,\to\, 0} \Delta \mathcal{S}^{(n)}_{A,0}(\epsilon, \ell-\epsilon ;t)
\,=\,
\frac{c}{12} \!\left(\! 1+\frac{1}{n} \right)
\log\!\left(\frac{2\, [\cosh(\pi\ell/\tau_0) \cosh(2\pi t/\tau_0)]^2 }{ \cosh(2\pi\ell/\tau_0) + \cosh(4\pi t/\tau_0) }\right).
\ee
Taking $t\gg \tau_0$ and $\ell \gg \tau_0$ first,
and then considering the regimes $t\ll \ell /2$ and $t\gg \ell /2$, this expression simplifies respectively to
\be
\Delta \mathcal{S}^{(n)}_{A,0}(0, \ell ;t)
\simeq
\frac{\pi c}{3 \tau_0} \!\left(\! 1+\frac{1}{n} \right)
t
\;\; \qquad \;\;
\Delta \mathcal{S}^{(n)}_{A,0}(0, \ell ;t)
\simeq
\frac{\pi c}{3 \tau_0} \!\left(\! 1+\frac{1}{n} \right)\frac{\ell}{2}
\ee
in agreement with \cite{cc-05-global quench}.

Let us remark that the results reported above have been obtained by employing in a
naive way the CFT analytic expressions of \cite{ct-16} corresponding to the half line
to study the case of a finite interval, 
assuming that the dynamics before the thermalisation at $t/\ell \simeq 1/2$ 
is governed by a neighbourhood of the endpoints, 
where the interval is indistinguishable from the half line.
It would be interesting to extend the analysis of \cite{ct-16} for the quantum quenches 
to a finite interval.

\section{Interval in a harmonic chain}
\label{sec:HCnumerics}

In this section we study the temporal evolution of the 
entanglement hamiltonian matrix $H_A$ of an interval $A$ in an infinite harmonic chain
after a global quench of the frequency parameter.

The hamiltonian (\ref{HC ham}) with periodic boundary conditions $\hat{q}_L =\hat{q}_0$ and $\hat{p}_L =\hat{p}_0$
can be diagonalised in the Fourier space by introducing the
annihilation and creation operators $a_k$ and $a_k^\dagger$ in the standard way. 
This procedure can be understood also in terms of canonical transformations
through the Williamson's theorem applied to the matrix defining the quadratic hamiltonian  (\ref{HC ham}).
The diagonalised form of (\ref{HC ham}) is given by
\be
\label{HC hamiltonian diag}
\widehat{H}(\omega) = \sum_{k=0}^{L-1} \omega_k \! \left( \hat{a}^\dagger_k \hat{a}_k +\frac{1}{2} \,\right)
\ee
where the dispersion relation reads
\be
\label{dispersion relation periodic}
\omega_k \equiv \sqrt{\omega^2 + \frac{4\kappa}{m} \, \big[ \sin(\pi k/L) \big]^2}
\;\; \qquad\;\;
k=0, \dots , L-1\,.
\ee

The quench protocol is the following. 
The system is prepared in the ground state $| \psi_0 \rangle$ of the hamiltonian (\ref{HC ham}) with periodic boundary conditions
and non vanishing mass $\omega=\omega_0\neq 0$, whose diagonalised form is 
\be
\label{HC hamiltonian diag 0}
\widehat{H}(\omega_0) = \sum_{k=0}^{L-1} \omega_{0,k} 
\left( \hat{a}^\dagger_{0,k} \,\hat{a}_{0,k} +\frac{1}{2} \,\right) 
\ee
being the dispersion relation $\omega_{0,k}$ given by (\ref{dispersion relation periodic}) with $\omega=\omega_0$.
At $t=0$ the frequency parameter is suddenly quenched to a different value $\omega \neq \omega_0$;
hence the unitary time evolution of $| \psi_0 \rangle$ for $t>0$ is determined by  the hamiltonian (\ref{HC ham}) 
with $\omega \neq \omega_0$, namely
\be
\label{time_evolved_state}
| \psi(t) \rangle =  e^{-\textrm{i} \widehat{H}(\omega) t} \, | \psi_0 \rangle
\qquad
t>0\,.
\ee

The temporal evolution of several quantities after this quench have been studied in various papers 
(see the reviews \cite{cc-16-quench rev, fagotti-essler-review} for an exhaustive list of references).

We are interested in the temporal evolution of the entanglement hamiltonian matrix $H_A$ discussed in \S\ref{sec:EHwilliamson} after the global quench of the frequency parameter described above. 
The entanglement hamiltonian matrix $H_A(t)$ can be evaluated once the reduced covariance matrix $\gamma_A(t)$ is known
(see (\ref{banchi-EH}), (\ref{eh-cov-mat-B}) and (\ref{HA-block-dec-extended-final})).
Thus, the first step is to construct the covariance matrix $\gamma(t)$ of the entire system, whose blocks 
are given by the following two-point correlators
\be
\label{time dep corrs}
\begin{array}{l}
Q_{i,j}(t) \equiv \langle \psi_0 |\, \hat{q}_i(t)  \,\hat{q}_j(t) \, | \psi_0 \rangle 
\\
\rule{0pt}{.6cm}
P_{i,j}(t) \equiv \langle \psi_0 | \,\hat{p}_i(t)  \,\hat{p}_j(t) \, | \psi_0 \rangle 
\\
\rule{0pt}{.6cm}
R_{i,j}(t) \equiv 
\textrm{Re} \big[ \langle \psi_0 | \,\hat{q}_i(t)  \,\hat{p}_j(t)  \,| \psi_0 \rangle \big]
\end{array}
\ee
where the time evolved operators in the Heisenberg picture $ \hat{q}_j(t)$ and $ \hat{q}_j(t) $ read
\be
 \hat{q}_j(t) = e^{\textrm{i} \widehat{H} t}   \hat{q}_j(0) e^{-\textrm{i} \widehat{H} t} 
\qquad
 \hat{q}_j(t) = e^{\textrm{i} \widehat{H} t}   \hat{q}_j(0) e^{-\textrm{i} \widehat{H} t} \,.
\ee

For periodic boundary conditions, the explicit expressions of the correlators in (\ref{time dep corrs}) 
in the thermodynamic limit read \cite{cc-07-quench-extended}
\be
\label{QPRmat t-dep}
\begin{array}{l}
\displaystyle
Q_{i,j}(t) 
=
\frac{1}{4\pi} \int_{0}^{2\pi} Q_{\tilde{k}}(t) \cos\big[(i-j)\,\tilde{k}\,\big] \, d\tilde{k}
\\
\rule{0pt}{.9cm}
\displaystyle
P_{i,j}(t) 
=
\frac{1}{4\pi} \int_{0}^{2\pi} P_{\tilde{k}}(t) \cos\big[(i-j)\,\tilde{k}\,\big] \, d\tilde{k}
 \\
 \displaystyle
 \rule{0pt}{.9cm}
 \displaystyle
R_{i,j}(t) 
=
-\,\frac{1}{4\pi} \int_{0}^{2\pi} R_{\tilde{k}}(t) \cos\big[(i-j)\,\tilde{k}\,\big] \, d\tilde{k}
\end{array}
\ee
where  the functions $Q_k(t)$, $P_k(t) $ and $R_k(t)$ containing the dependence on $\omega$, $\omega_0$ and $t$
are defined as follows
\be
\label{QPRmat t-dep-k}
\begin{array}{l}
\displaystyle
Q_{k}(t)
\equiv  
\frac{1}{m \omega_k}
\left( \,\frac{\omega_k}{\omega_{0,k}} \cos^2(\omega_k t) 
+ \frac{\omega_{0,k}}{\omega_k} \sin^2(\omega_k t) \right)
\\
\displaystyle
\rule{0pt}{.9cm}
P_{k}(t)
\equiv  
m \omega_k
\left( \,\frac{\omega_k}{\omega_{0,k}}  \sin^2(\omega_k t)
+ \frac{\omega_{0,k}}{\omega_k}  \cos^2(\omega_k t) \right)
 \\
 \displaystyle
 \rule{0pt}{.9cm}
R_{k}(t)
\equiv
\left(\,\frac{\omega_k}{\omega_{0,k}} - \frac{\omega_{0,k}}{\omega_k}  \right)
\sin(\omega_k t) \cos(\omega_k t) 
\end{array}
\ee
where $\omega_k \equiv \sqrt{\omega^2 + \tfrac{4\kappa}{m} [ \sin(k/2) ]^2}$
and $\omega_{0,k} $ is given by the same expression with $\omega$ replaced by $\omega_0$.
At $t=0$ these functions become respectively
\be
Q_{k}(0) =  \frac{1}{m \omega_{0,k}} 
\;\;\qquad\;\;
P_{k}(0) = m \omega_{0,k} 
\;\;\qquad\;\;
R_{k}(0) = 0 \,.
\ee
Plugging these expressions into (\ref{QPRmat t-dep}),
one recovers the correlators associated to the ground state of $\widehat{H}_0$, as expected.

It is important to remark that for $t>0$ all the modes labelled by $k$ in (\ref{QPRmat t-dep}) and (\ref{QPRmat t-dep-k})
provide a finite contribution when $\omega =0$.
The zero mode, i.e. the one corresponding to $k=0$, deserves a particular attention because,
for static configurations, the occurrence of this mode leads to a divergence in the correlator $\langle \hat{q}_i \hat{q}_j  \rangle$
as $\omega \to 0^+$.
This is related to the translation invariance of the periodic chain. 
Instead, the correlators (\ref{QPRmat t-dep}) are finite in the massless regime;
hence, we are allowed to consider a global quench where the evolution hamiltonian has $\omega=0$.

The time evolved state (\ref{time_evolved_state}) is also Gaussian, hence it can be described through the formalism discussed in \S\ref{sec:EHwilliamson}.
In particular, the correlators (\ref{QPRmat t-dep}) provide the blocks of the covariance matrix $\gamma(t)$ after the global quench. 
The correlators (\ref{QPRmat t-dep}) or their modifications 
have been employed to study different aspects of entanglement in harmonic lattices 
(see e.g. \cite{sotiriadis-cardy, coser-14-quench-neg, mezei-muller-16-quench-scalar}).

\begin{figure}[t!]
\vspace{.0cm}
\hspace{.0cm}
\includegraphics[width=.968\textwidth]{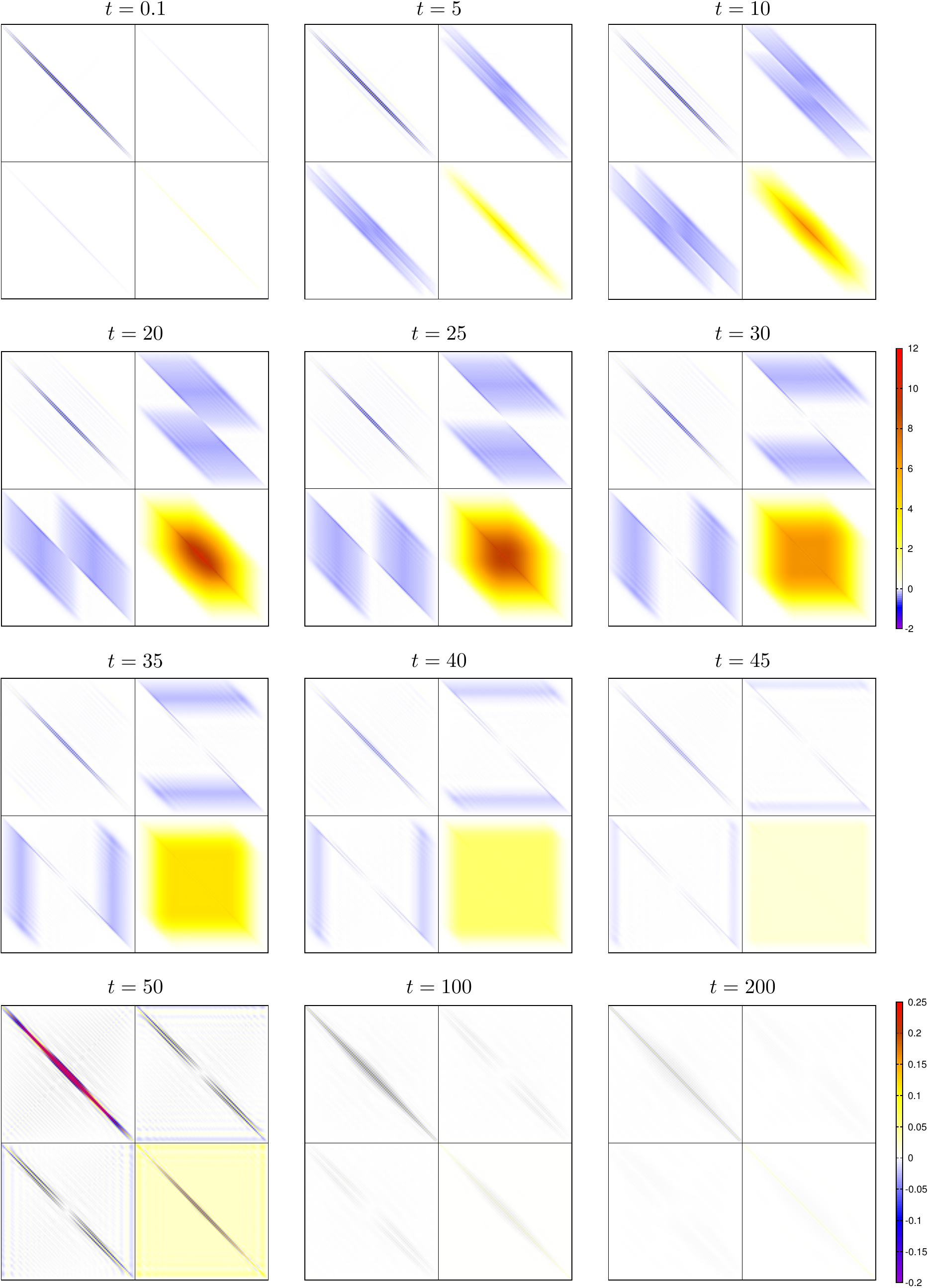}
\vspace{.0cm}
\caption{
Entanglement hamiltonian matrix $H_A(t)$ in (\ref{gamma_A H_A t-dep}) 
for an interval having $\ell = 100$ sites in the infinite harmonic chain 
after the global quench given by $\omega_0=1$ and $\omega=0$.
}
\label{fig:HquenchDensityHC}
\end{figure}

\subsection{Numerical results for the entanglement hamiltonian matrix}
\label{sec:numerics HC}

We are interested in the temporal evolution of the entanglement hamiltonian matrix $H_A(t)$ corresponding to an interval $A$ 
made by $\ell $ sites after the global quench of the frequency parameter. 
The reduced covariance matrix $\gamma_A(t)$ and the entanglement hamiltonian matrix $H_A(t)$ can be partitioned into
$\ell \times \ell$ blocks (see (\ref{gamma-W-block-dec})) that depend on $t>0$, namely
\be
\label{gamma_A H_A t-dep}
\gamma_A(t) 
=\,
\bigg( \!
\begin{array}{cc}
Q(t)  & \!\! R(t) \\
R(t)^{\textrm{t}}  & \!\!  P(t)  \\
\end{array}  \! \bigg)
\;\;\qquad\;\;
H_A(t) 
=
\bigg( \!
\begin{array}{cc}
M(t)  &  \!  \!  E(t) \\
E(t)^{\textrm{t}}  &  \!  \!  N(t)  \\
\end{array}  \!  \bigg)\,.
\ee
The relation between $H_A(t)$ and the reduced covariance matrix $\gamma_A(t)$ can be expressed as in (\ref{banchi-EH}), 
(\ref{eh-cov-mat-B}) or (\ref{HA-block-dec-extended-final}).

In our numerical analysis, the entanglement hamiltonian matrix $H_A(t)$ 
has been constructed by employing the Williamson's decomposition (\ref{H_A Williamson_dec}) for $H_A$ and (\ref{W_H is Wtilde}), 
where the symplectic matrix $W$ entering in the Williamson's decomposition (\ref{williamson th gammaA}) for $\gamma_A$
has been obtained as explained in the Appendix\;\ref{app:cholesky}.
The symplectic eigenvalues of $H_A(t)$ have been found from the symplectic spectrum of $\gamma_A$ through the relation (\ref{Ediag_Ddiag}).
The correlators employed to define $\gamma_A(t)$ have been introduced in (\ref{QPRmat t-dep}).
In this manuscript we mainly consider the global quench of the frequency parameter 
described above with $\omega_0 =1$ and $\omega=0$.
Some data corresponding to an initial state having $\omega_0 =5$ are also provided.

The relation (\ref{Ediag_Ddiag}), that provides the symplectic spectrum of $H_A$ in terms of the one of $\gamma_A$,
requires $\sigma_k >1/2$. Since many elements of the symplectic spectrum of $\gamma_A$ are very close to $1/2$,
the software approximates them to $1/2$, spoiling the applicability of (\ref{Ediag_Ddiag}) to get the symplectic spectrum of $H_A$.
This forces to obtain $\sigma_k$'s with very high numerical precision. 
The number of digits depends on various parameters of the configuration
like $\ell$ and $t$. 
In our numerical analysis for the harmonic chain we worked with precisions between $200$ and $1000$ digits.

In Fig.\;\ref{fig:HquenchDensityHC} we show the temporal evolution of the entanglement hamiltonian matrix
$H_A(t)$ after the global quench written in the form (\ref{gamma_A H_A t-dep}), i.e. in terms of its $\ell \times \ell$ blocks.
The main feature to highlight with respect to the entanglement hamiltonian matrices corresponding to static 
configurations \cite{ch-rev} is that the off diagonal block $E$
is not vanishing and therefore it contributes in a non trivial way 
to determine the out of equilibrium dynamics.
The magnitudes of the elements of $H_A(t)$ become very small for long time.
Indeed, a smaller scale has  been chosen in the three bottom panels of Fig.\;\ref{fig:HquenchDensityHC}
to show the structure of the non vanishing elements.
At early times only the diagonal blocks are non vanishing and mostly around their main diagonals.
As time evolves, also the off diagonal elements of the blocks become relevant.  
The most evident feature that can be observed from the time evolution in Fig.\;\ref{fig:HquenchDensityHC}
is the occurrence of bands in the different blocks of $H_A(t)$ whose widths increase with time. 
We consider these widths in more detail below, during the discussion of 
Fig.\;\ref{fig:MNRantidiags}, Fig.\;\ref{fig:MNRantidiags_LC} and Fig.\;\ref{fig:HC_corr}.

The reflection symmetry of the configuration with respect to the center of the interval leads 
to a corresponding reflection symmetry with respect to the center 
in any given diagonal of the blocks $M(t)$ and $N(t)$ at any fixed time.
More explicitly, we checked numerically that $M_{i,i} = M_{\ell-i+1, \ell-i+1}$ and $N_{i,i} = N_{\ell-i+1, \ell-i+1}$ with $1\leqslant i \leqslant \ell$ for the main diagonals 
and that $M_{i,i+p} = M_{\ell- p-i+1, \ell-i+1} $ and $N_{i,i+p} = N_{\ell- p-i+1, \ell-i+1} $ along the $p$-th diagonal, with $p>0$ and $1\leqslant i \leqslant \ell-p$.
As for the off diagonal block $E(t)$, we observe a symmetry with respect to the center of the block,
namely $E_{i,j} = E_{\ell-i+1, \ell-j+1}$ for $1\leqslant i,j \leqslant \ell$. 

The  temporal evolution of the blocks composing $H_A(t)$ is qualitatively different. 
As for the symmetric $\ell \times \ell$ matrix $M(t)$ in (\ref{gamma_A H_A t-dep}), the largest contributions come from the main diagonal, that decrease as time evolves and remain positive during all the evolution, 
reaching a stationary curve for long times  (see the top left panel of Fig.\;\ref{fig:MNRdiags}).
The first diagonals are mostly negative and in some point become positive at some time 
(see the top right panel of Fig.\;\ref{fig:MNRdiags} and the corresponding inset).
In the top left panel of Fig.\;\ref{fig:MNRantidiags} we show the antidiagonal of $M(t)$, that displays oscillations around zero.

 \begin{figure}[t!]
\vspace{.5cm}
\hspace{-1.1cm}
\includegraphics[width=1.1\textwidth]{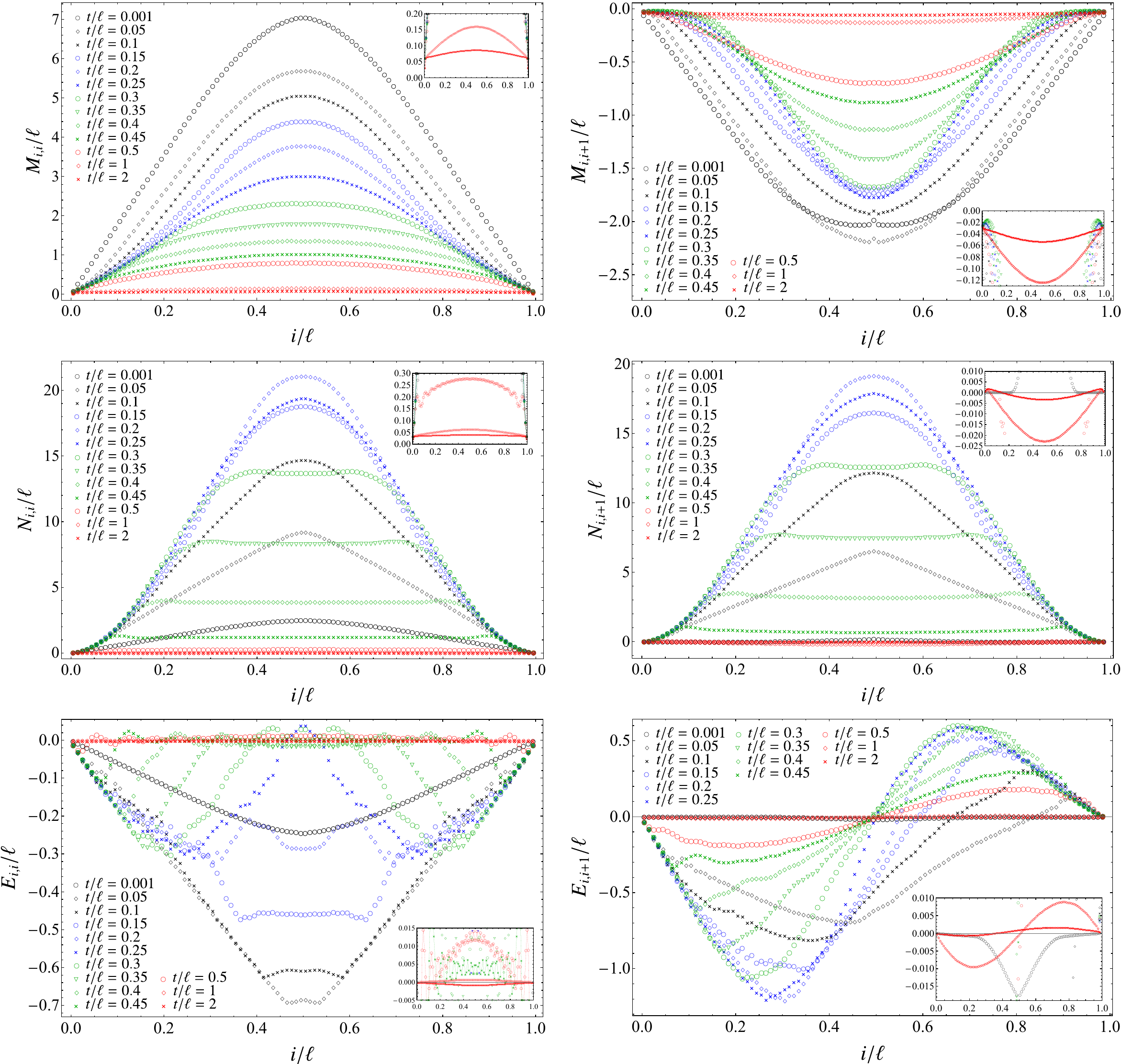}
\vspace{-.5cm}
\caption{
Harmonic chain:
Temporal evolution of the main diagonals (left panels) 
and of the first diagonals (right panels) 
composing the $\ell \times \ell$ blocks of the entanglement hamiltonian matrix $H_A(t)$ 
in (\ref{gamma_A H_A t-dep}) for an interval with $\ell =100$ sites.
The insets zoom in on small values, in order to show the curves for large times.
This figure, that is complementary to Fig.\;\ref{fig:HquenchDensityHC} and Fig.\;\ref{fig:MNRantidiags},
has been discussed in \S\ref{sec:numerics HC}.
}
\vspace{2.5cm}
\label{fig:MNRdiags}
\end{figure}

 \begin{figure}[t!]
\vspace{.2cm}
\hspace{-1.1cm}
\includegraphics[width=1.1\textwidth]{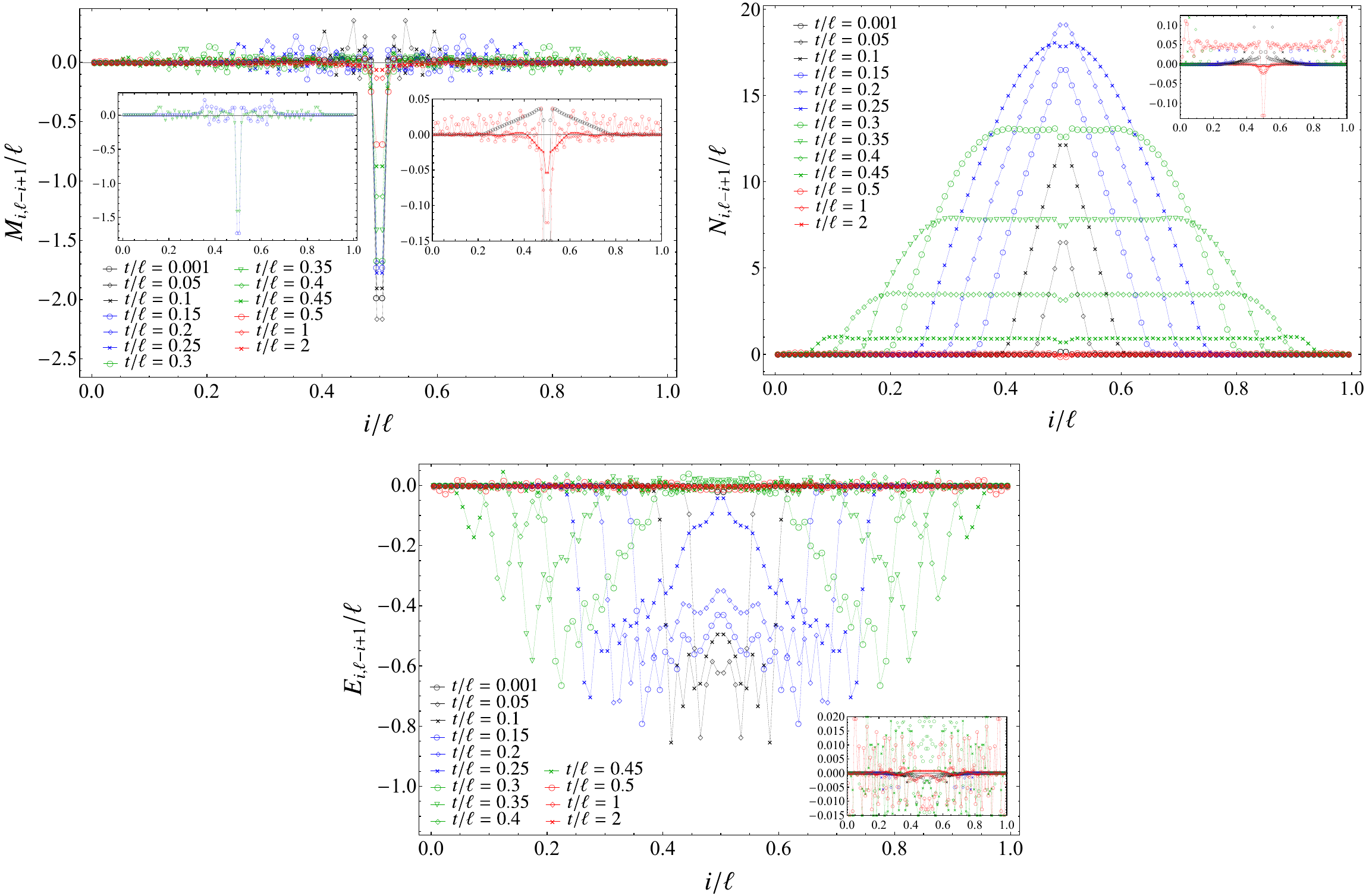}
\vspace{-.5cm}
\caption{
Harmonic chain:
Temporal evolution of the antidiagonals of the $\ell \times \ell$ blocks 
composing the entanglement hamiltonian matrix $H_A(t)$
in (\ref{gamma_A H_A t-dep})
for an interval with $\ell =100$ sites. 
The insets containing red curves zoom in on small values, in order to show the curves for large times.
In the second zoom of the top left panel we show only two curves to highlight 
the fact that the antidiagonals are very small for $|i/\ell-1/2| \gtrsim t/\ell$.
A similar behaviour occurs in the other panels, as shown in Fig.\;\ref{fig:MNRantidiags_LC}.
}
\label{fig:MNRantidiags}
\end{figure}

Interesting features can be observed also for the temporal evolution of the symmetric 
block $N(t)$ in (\ref{gamma_A H_A t-dep}).
The diagonal of $N(t)$ is shown in the middle left panel of  Fig.\;\ref{fig:MNRdiags}:
first it grows until a maximal curve and then it relaxes to a positive curve for long time.
The elements of $N(t)$ are mostly positive but some of them become negative for large times 
(see e.g. the inset of the top right panel in Fig.\;\ref{fig:MNRantidiags}
and the one of the middle right panel in Fig.\;\ref{fig:MNRdiags}).
The block $N(t)$ is the only one where the evolution of the main diagonal and of the first diagonal are qualitatively 
similar: a smooth positive wedge grows from the center of the interval until $t/\ell \simeq 0.2$, 
then its tip decreases and gets smooth forming a plateau whose height decreases in time.

The evolution of the block $E(t)$ is qualitatively different from the one of $M(t)$ and $N(t)$
because this matrix is not symmetric. Furthermore, it vanishes for $t=0$ 
and it seems that its contribution is zero also for long times.
The elements of this block are mostly negative but without a definite sign for all the times 
(see e.g. its first diagonal in the bottom right panel of Fig.\;\ref{fig:MNRdiags}).
From Fig.\;\ref{fig:HquenchDensityHC}, we observe the formation of two mostly negative bands 
close to the diagonal for small times having the shape of two parallelograms 
(see also the top right panel in Fig.\;\ref{fig:antidiagonals_d0})
whose height along the antidiagonal increases in time until $t/\ell \simeq 0.5$, when they vanish.
As for the main diagonal of $E(t)$, a smooth negative wedge develops until it reached a minimum value; 
then it goes back to zero in a peculiar way;
indeed a small plateau is formed whose width increases and then decreases,
leading to two fronts that move in the opposite directions towards the endpoints of the interval
(see the bottom left panel of Fig.\;\ref{fig:MNRdiags}).

In Fig.\;\ref{fig:MNRantidiags} we show the temporal evolution of the antidiagonals of the blocks 
composing the entanglement hamiltonian matrix $H_A(t)$ in (\ref{gamma_A H_A t-dep}).
Although the temporal evolutions of the antidiagonals in the three blocks look quite different,
the main common feature is the fact that at a fixed time they become very small (and not vanishing) 
around the same points.
The distance $2d_0$ between these points provides the total width of the bands occurring at any given time
in each block (see Fig.\;\ref{fig:HquenchDensityHC}).
The width $d_0$ increases linearly in time with velocity equal to one,
as shown in the top left panel of Fig.\;\ref{fig:antidiagonals_d0}.
This feature has been highlighted in Fig.\;\ref{fig:MNRantidiags_LC}, where each panel
shows all the three antidiagonals together at a given time. 

 \begin{figure}[t!]
\vspace{.5cm}
\hspace{-1.1cm}
\includegraphics[width=1.1\textwidth]{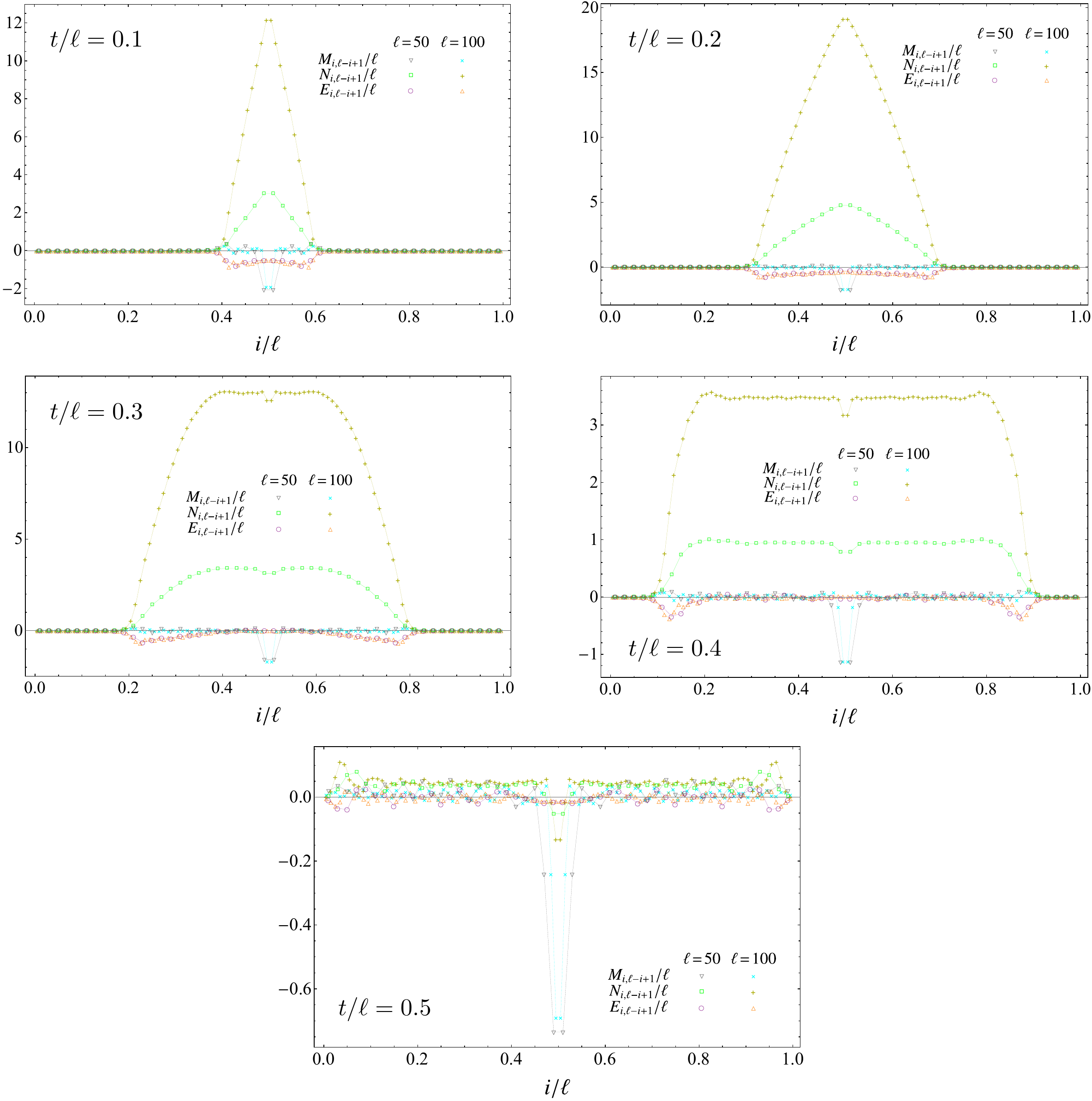}
\vspace{-.5cm}
\caption{
Harmonic chain:
Antidiagonals of the blocks in the entanglement hamiltonian matrix (\ref{gamma_A H_A t-dep})
for various times and two values for $\ell$.
The points where all these antidiagonals become very small evolve linearly 
in opposite direction travelling from the center towards the endpoints of the interval
with velocity equal to one 
(see top panels in Fig.\;\ref{fig:antidiagonals_d0}, where $2d_0$ denotes the width of the curves shown here).
}
\vspace{1cm}
\label{fig:MNRantidiags_LC}
\end{figure}

 \begin{figure}[t!]
\vspace{.5cm}
\hspace{-1.1cm}
\includegraphics[width=1.1\textwidth]{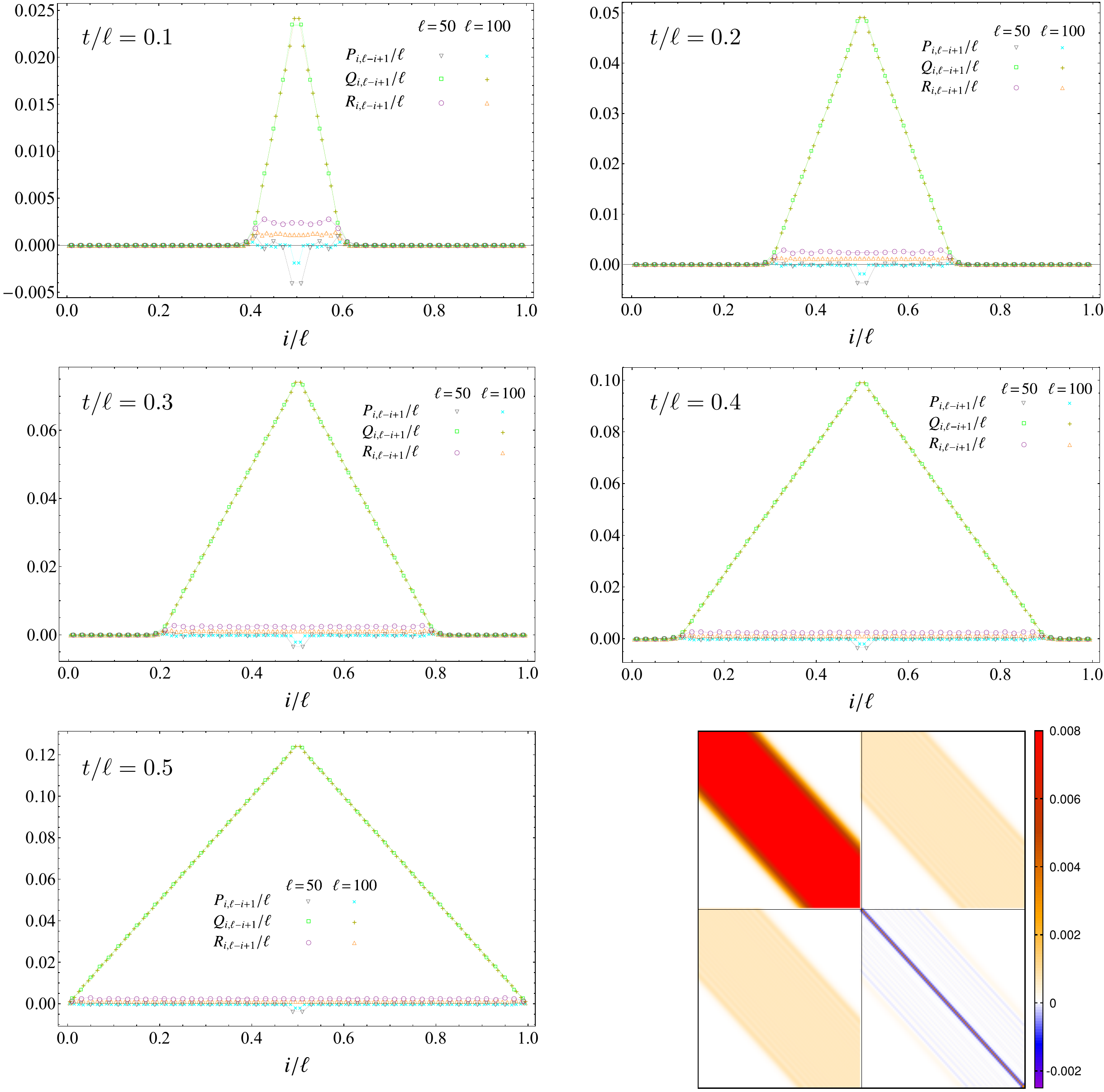}
\vspace{-.3cm}
\caption{
Harmonic chain:
Antidiagonals of the blocks in the correlation matrix (\ref{gamma_A H_A t-dep})
for an interval of length $\ell$.
The points where all these antidiagonals become very small evolve linearly 
in opposite direction travelling from the center towards the endpoints of the interval
with velocity equal to one. This evolution is observed also for the entanglement hamiltonian matrix 
(see Fig.\;\ref{fig:MNRantidiags_LC}).
The bottom right panel shows $C_A$ for $t=20$ and $\ell = 100$.
This should be compared with the left panel of the second line in Fig.\;\ref{fig:HquenchDensityHC}.
}
\vspace{1cm}
\label{fig:HC_corr}
\end{figure}

As for the block $M(t)$, its main diagonal (top left panel of Fig.\;\ref{fig:MNRdiags}) is not captured in 
the top left panel of Fig.\;\ref{fig:MNRantidiags} because $\ell$ is even. 
In $N(t)$ both the temporal evolutions of the diagonal (middle left panel in Fig.\;\ref{fig:MNRdiags}) 
and of the antidiagonal (top right panel in Fig.\;\ref{fig:MNRantidiags}) display a plateau but 
we remark that the behaviour before the formation of this plateau is very different. 

It is instructive to compare the temporal evolution of the blocks in entanglement hamiltonian matrix $H_A(t)$
with the one of the blocks in the reduced covariance matrix $\gamma_A(t)$ (see (\ref{gamma_A H_A t-dep})).
The main difference is the fact that the blocks of $H_A(t)$ are not homogeneous along their diagonals
(see Fig.\;\ref{fig:HquenchDensityHC} and Fig.\;\ref{fig:MNRdiags}),
while the blocks of $\gamma_A(t)$ are constant along any given diagonal
(see the bottom right panel of Fig.\;\ref{fig:HC_corr}, that corresponds to a given time),
as it can be easily inferred from (\ref{QPRmat t-dep}).
In Fig.\;\ref{fig:HC_corr} we consider the temporal evolution of the antidiagonals of the blocks in $\gamma_A(t)$.
Comparing this figure with Fig.\;\ref{fig:MNRantidiags_LC}, 
one concludes that also the width of the bands occurring in the three blocks of $\gamma_A(t)$
increase linearly in time with velocity equal to one. 
Nonetheless, let us stress that the shape of the antidiagonals occurring in $H_A(t)$ and in $\gamma_A(t)$
is very different.

\begin{figure}[t!]
\vspace{.5cm}
    \hspace{-1.1cm}
\includegraphics[width=1.1\textwidth]{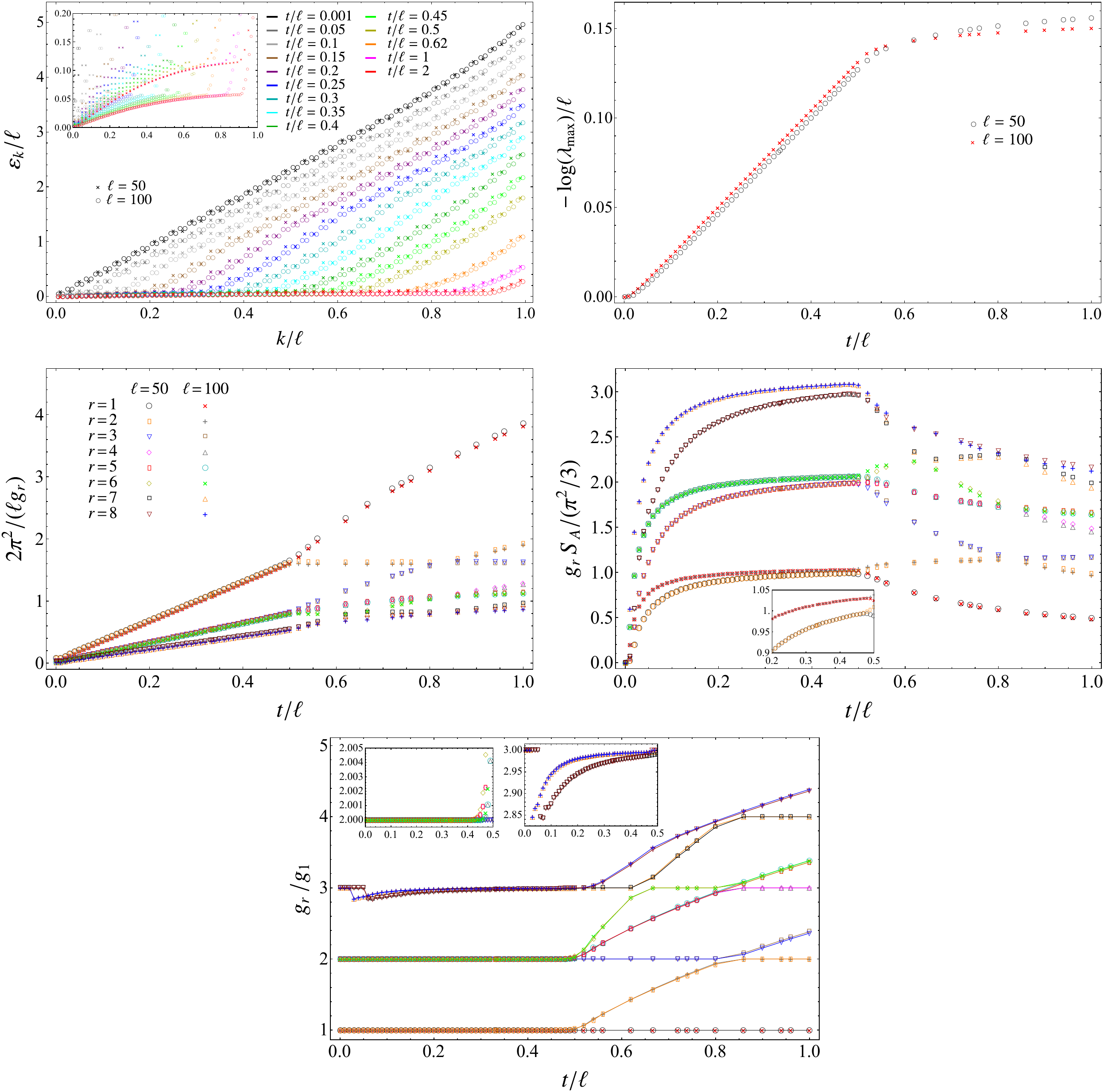}
\vspace{-.5cm}
\caption{
Harmonic chain:
Entanglement spectrum for an interval of length $\ell$ after the quench of the frequency parameter 
given by $\omega_0=1$ and $\omega=0$.
Top left: Single particle entanglement spectrum (see (\ref{Ediag_Ddiag})).
Top right: Temporal evolution of the largest eigenvalue of the entanglement spectrum (see the text below (\ref{ent_spec_1})).
Middle left: Temporal evolution of the first gaps in the entanglement spectrum (see the text below (\ref{hc_gaps_generic})). The legenda of this panel holds also in the remaining ones.
Middle right: Temporal evolution of $g_r S_A$ 
(the inset zooms in on the lowest plateau, showing that the data having $\ell =50$
and $\ell=100$ do not overlap).
Bottom: Temporal evolution of the ratios $g_r/g_1$ between the gaps in the entanglement spectrum
(the insets zoom in on the two higher plateaux).
}
\label{fig:SpectrumBoson}
\end{figure}

\begin{figure}[t!]
\vspace{.2cm}
    \hspace{-1.1cm}
\includegraphics[width=1.1\textwidth]{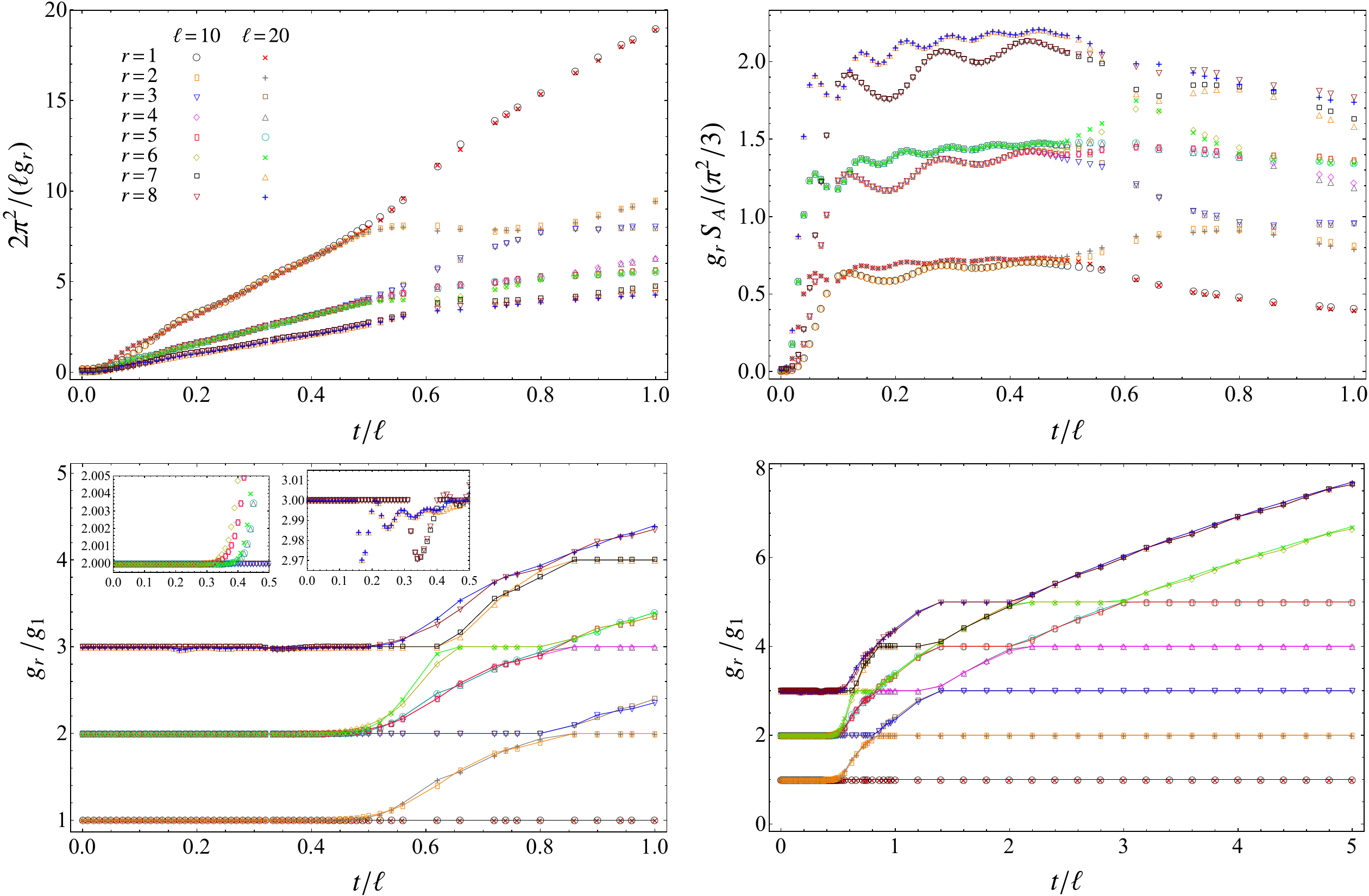}
\vspace{-.5cm}
\caption{
Harmonic chain: Temporal evolution of the gaps in the entanglement spectrum 
for an interval of length $\ell$ after the quench of the frequency parameter 
given by $\omega_0=5$ and $\omega=0$ (see also the corresponding panels in 
Fig.\;\ref{fig:SpectrumBoson}).
Top left: Temporal evolution of the first gaps in the entanglement spectrum (see the text below (\ref{hc_gaps_generic})). The legenda of this panel holds also for the other ones.
Top right: Temporal evolution of $g_r S_A$.
Bottom: Temporal evolution of the ratios $g_r/g_1$ between the gaps in the entanglement spectrum.
The panel on the left focuses on the regime $t/\ell <1$ (the insets zoom in on the two higher plateaux),
while the panel on the right highlights the behaviour for long times. 
The curves in the left panel are very similar to the ones in the bottom panel of Fig.\;\ref{fig:SpectrumBoson}. 
}
\label{fig:SpectrumBoson_omega5}
\end{figure}

As first quantitative check,
we have compared the linear growth of the entanglement entropy obtained from our numerical data 
for the harmonic chain
with the CFT formula $S_A \simeq 2\pi c\, t /(3\tau_0)$ \cite{cc-05-global quench, cc-16-quench rev} (see also \S\ref{sec:cft-naive}) with $c=1$, 
finding $\tau_0 \simeq 3.79$ (which agrees e.g. with the numerical value obtained in \cite{cdt-17-contour})

\subsection{Entanglement spectrum}
\label{sec-ES-HC}

A numerical analysis for some quantities related to the entanglement spectrum has been reported in 
Fig.\;\ref{fig:SpectrumBoson}.
The top left panel shows the symplectic spectrum of $H_A$ for various times and two different values of $\ell$. 
For the low-lying part of this spectrum, i.e. small $k$'s, we have $\varepsilon_k \simeq \varepsilon_{k+1}$ 
and this degeneracy disappears for high values of $k$.
At small times, $\varepsilon_k$ is linear in terms of $k$
while for long time the curve of $\varepsilon_k$ bends towards zero, although it never vanishes. 
It would be interesting to consider $\varepsilon_k/\ell$ in the limit of large values of $\ell$.
In the top right panel of Fig.\;\ref{fig:SpectrumBoson} we consider the temporal evolution of the largest 
eigenvalue of the entanglement spectrum (see the text below (\ref{ent_spec_1})).
Since $S_A^{(n)} \to -\log \lambda_{\textrm{\tiny max}}$ as $n \to \infty$,
this curve can be compared with the CFT results for the R\'enyi entropies \cite{cc-05-global quench, cc-16-quench rev} and a good agreement is found. 
Fitting the linear growth of $ -\log \lambda_{\textrm{\tiny max}}$ we get a slope of $\tfrac{1}{2.1}$ times the slope of the entanglement entropy, while the one predicted by (\ref{renyi linear growth}) is $\tfrac{1}{2}$.

In the remaining panels of Fig.\;\ref{fig:SpectrumBoson} we explore the temporal evolutions of the gaps in the entanglement spectrum.
The data about these evolutions display two distinct temporal regimes separated by $t/\ell \simeq 1/2$.
For $t/\ell < 1/2$, the CFT result (\ref{gap_r cft naive}),
obtained from the analysis of \cite{ct-16} for the semi-infinite line, 
predicts linear growths in time for the inverse of the gaps with slopes proportional to the conformal spectrum 
(including also the dimensions of the descendants) allowed by the proper conformal boundary conditions.  
The numerical data in the middle left panel of Fig.\;\ref{fig:SpectrumBoson} display these linear growths for $t/\ell < 1/2$. 
By fitting the slopes of these linear growths through  (\ref{gap_r cft naive})
with $\tau_0 \simeq 3.79$ obtained above from the linear growth of the entanglement entropy, 
we find $\Delta_1=1.022  $, $\Delta_2=2.045  $ and $\Delta_3=3.058$.
In order to reduce the influence of the initial state encoded in $\tau_0$, 
in the middle right panel and in the bottom panel 
of Fig.\;\ref{fig:SpectrumBoson} we consider the temporal evolutions of 
$g_r S_A$ and the ratios $g_r/g_1$ respectively, which should be independent of $\tau_0$ 
according to (\ref{renyi linear growth}) and (\ref{gap_r cft naive}).
As for $g_r S_A$, it is evident that curves corresponding to different values of $\ell$ do not collapse;
hence more values of $\ell$ (possibly also larger than the ones considered here) are needed in order to make comparisons with CFT results. 

The temporal evolutions of the ratios $g_r/g_1$ of the entanglement gaps in the bottom panel of 
Fig.\;\ref{fig:SpectrumBoson} display interesting features. 
For $t/\ell < 1/2$ the curves having different $\ell$'s collapse forming plateaux 
whose heights are given by strictly positive integers.
This result agrees with the fact that the underlying CFT contains the primary $\partial_z \phi$ 
and its descendants.

In Fig.\;\ref{fig:SpectrumBoson_omega5} we show numerical data for the temporal evolution of 
the gaps in the entanglement spectrum when the initial state has $\omega_0 =5$,
in order to explore the robustness of the observations made above under changes of the initial state. 
For this $\omega_0$ we consider small values of $\ell$ such that the product $\omega_0 \ell$ gets the same values
corresponding to the data shown in Fig.\;\ref{fig:SpectrumBoson}.
The qualitative behaviour of $1/g_r$ is the same one observed in Fig.\;\ref{fig:SpectrumBoson}
(the slopes of the linear growths are different, as expected, being (\ref{gap_r cft naive}) 
dependent on the initial state through $\tau_0$).
In the temporal evolution of $g_r S_A$ we observe oscillations that are due to the small values of $\ell$
(indeed, they do not occur in the middle right panel of Fig.\;\ref{fig:SpectrumBoson}).
Interestingly, the temporal evolutions of the ratios $g_r/g_1$ coincide with the ones reported in 
Fig.\;\ref{fig:SpectrumBoson}, meaning that this quantity displays some independence on the initial state. 
It would be instructive to consider other values of $\omega_0$ and higher gaps in order to understand better 
how much the temporal evolutions of $g_r/g_1$ are robust under modifications of the initial state.  
In the bottom right panel of Fig.\;\ref{fig:SpectrumBoson_omega5} we have considered also long times
and from these data we can identify two regimes: $t/\ell \leqslant 1/2$ and the long time regime. 
In both these temporal regimes we observe plateaux having the same heights. 
Thus, the ratios in the CFT spectrum can be read also from the long time regime.

\subsection{A contour function from the quasi-particle picture}
\label{sec:QP}

In \cite{cc-05-global quench} a quasi-particle picture has been introduced to explain the temporal evolution of the entanglement entropy 
after a global quantum quench. 
The underlying idea is based on the fact that the initial state has very high energy with respect to the ground state of the hamiltonian governing the temporal evolution; hence it can be seen as a source of quasi-particle excitations. 
In particular, in one spatial dimension, it is assumed that at $t=0$ each point of the space emits two quasi-particles with opposite momenta $p$ and $-p$ according to certain probability distribution that depends on both the initial state and the evolution hamiltonian.
Only the particles emitted at the same point are entangled and all the points of the space emit the quasi-particles in the same way. 
For $t>0$, the positions of the quasi-particles emitted at the same point $x$ are $x+v_p t$ and $x-v_p t$, being $v_p>0$ and $v_{-p} = -v_p$
for $p>0$.

Considering a spatial bipartition $A\cup B$  and two points $x_1 \in A$ and $x_2\in B$;
at time $t$ they are entangled only if they are reached simultaneously by
two quasi-particle emitted from the same point $x$ at $t=0$.
The bipartite entanglement between $A$ and $B$ is obtained by summing the contributions of
all the points $x$ fulfilling this condition.
In particular, for the entanglement entropy we have \cite{cc-05-global quench}
\be
\label{S_A qp-general}
S_A(t)
\approx 
\int_{x_1\in A} \!d x_1
\int_{x_2\in B} \!d x_2
\int_{-\infty}^{\infty} \! dx
\int dp  \;
 \delta(x_1-x-v_p t) \, \delta(x_2-x+ v_p t)\, \tilde{s}(p)
\ee
where $\tilde{s}(p)$ is obtained by multiplying the momentum distribution function 
and the contribution of the pair of quasi-particles with momenta $p$ and $-p$ to the entanglement entropy.
The integration domain of the allowed momenta is model dependent.
Performing the spatial integrals in (\ref{S_A qp-general}), one finds
\be
\label{QPentropy}
S_A(t)
\approx
\,2 \,t \int_{2|v_p|t<\ell} dp
  \,\tilde{s}(p)   \,v_p   \,
  +
  \ell \int_{2|v_p|t>\ell} dp\, \tilde{s}(p)\,.
\ee
When a maximum velocity $v_{\textrm{\tiny max}}$ exists,
$S_A$ grows linearly in time for $ v_{\mathrm{max}} t \leqslant \ell /2$, while $S_A \propto \ell$ for $ v_{\mathrm{max}} t  \gg \ell /2$.
In a CFT we have $|v_p| = v_{\mathrm{max}} =1$ for all $p$ \cite{cc-05-global quench, cc-16-quench rev}.
Notice that (\ref{QPentropy}) does not take into account the initial value of the entanglement entropy.

 \begin{figure}[t!]
\vspace{.2cm}
\hspace{-1.1cm}
\includegraphics[width=1.1\textwidth]{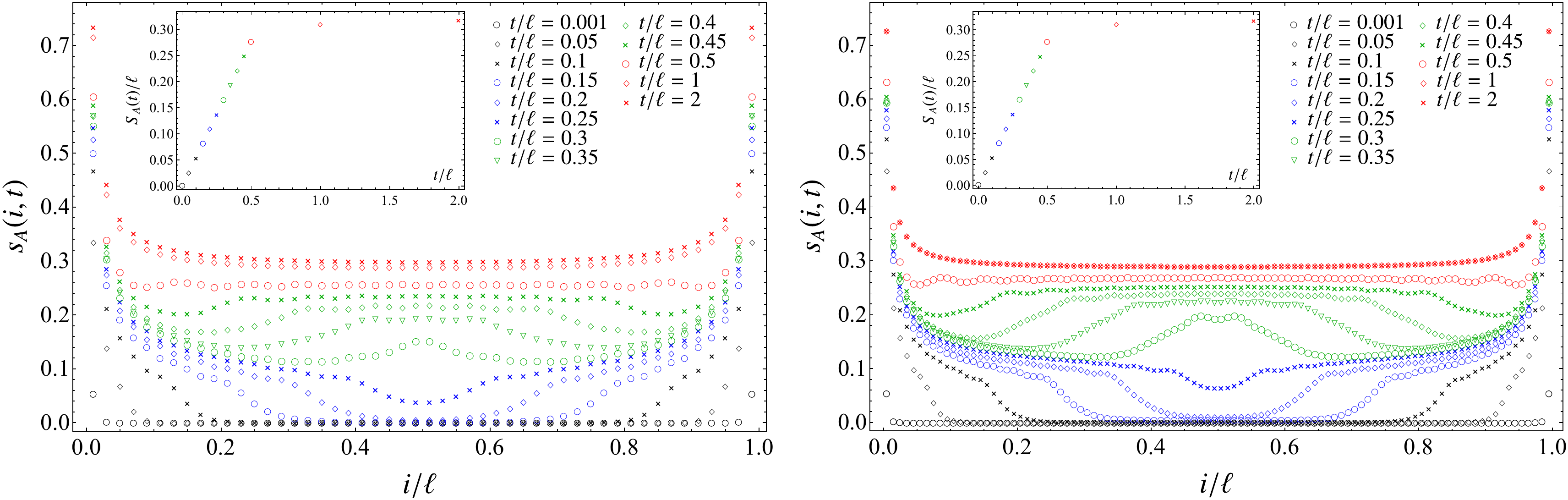}
\vspace{-.5cm}
\caption{
Temporal evolution of the contour for the entanglement entropy 
of an interval made by $\ell$ sites in the periodic harmonic chain,
evaluated from (\ref{contour from mpf}) and (\ref{mpf from W}).
Left: $\ell=50$. Right: $\ell=100$.
The insets show the temporal evolution of the entanglement entropy obtained 
through the contour function (see (\ref{contour_lattice_def}) for $n=1$). 
}
\label{fig:ContourBosonData}
\end{figure}

The expression (\ref{QPentropy}) of the entanglement entropy obtained from
this quasi-particle picture provides some functions $s_A(x,t) $ such that
$ S_A(t)=\int _A s_A(x,t) dx$. 
Let us introduce three positive functions $f_0(x)$, $f_1(x,p,t)$ and $f_2(x,p,t)$ fulfilling the following conditions
\be
\label{conf_012}
\int _A  f_0(x) \,dx = S_A\big|_{t=0}
 \qquad
\int_A  f_1(x,p,t)\,dx  =  2  v_p \,t 
 \qquad
 \int_A  f_2(x,p,t) \,dx =  \ell\,.
\ee
Taking into account the initial value $S_A|_{t=0}$ in (\ref{QPentropy})  and employing the constraints (\ref{conf_012}),
it is straightforward to write the entanglement entropy as $ S_A(t)=\int _A s_A(x,t) \,dx$ with
\be
\label{QPcontour}
s_A(x,t)
\,=
\int_{2|v_p|t<\ell} dp  \,f_1(x,p,t)\,\tilde{s}(p)\,
+
\int_{2|v_p|t>\ell} dp \, f_2(x,p,t)  \,\tilde{s}(p) + f_0(x)\,.
\ee
Since our configuration is symmetric with respect to the center of the interval, it is natural to require that
$f_0(x)=f_0(\ell - x) $ and $f_i(x,p,t)=f_i(\ell - x,p,t)$ for $ i =1,2$.
As mentioned above, (\ref{S_A qp-general}) is obtained by assuming that the emission of the quasi-particles is spatially homogeneous,
and this assumption provides a significant restriction to the form of the functions in (\ref{QPcontour}).

Let us consider the infinitesimal contribution to the entanglement entropy $dS_A(t) = s_A(x,t) dx$ provided by an infinitesimal interval $ ( x-dx/2, x+dx/2)$ centered in a point $x\in A$.
At $t>0$, this quantity is proportional to the number of quasi-particles (i) that are in this infinitesimal interval (ii) whose entangled quasi-particle is in $B$.
The quasi-particles fulfilling these conditions have been emitted at $t=0$ from the position $x-|v_p| t$ or from $x+|v_p| t$.
For $v_p\,t\le \ell /2$, the quasi-particles emitted at $x-|v_p| t$ contribute when $2|v_p|\, t> x$, 
while the ones emitted at $x+|v_p| t$ matter when $2|v_p|\, t>\ell-x$.
Instead, for $ v_p\,t>\ell/2 $ the quasi-particles coming from both $x-|v_p| t$ and $x+|v_p| t$ contribute. 
These considerations lead to the following expressions
\be
\label{f_12 from QP}
f_1(x,p,t)= \frac{1}{2}\Big[ \varphi_1(x) \,\Theta(2|v_p|\, t-x)+ \varphi_1(\ell-x)\,\Theta(2|v_p|\, t-\ell+x) \Big] 
\qquad
 f_2(x,p,t)=\varphi_2(x)
\ee
where $\varphi_j $ are non negative, $\varphi_2 (\ell -x)= \varphi_2 (x)$
and $\Theta$ is the Heaviside step function.
The spatial homogeneity in the quasi-particle production leads to 
drastic simplifications given by  $\varphi_1(x)=\varphi_2(x)=1$ identically.

 \begin{figure}[t!]
\vspace{.4cm}
\hspace{-1.1cm}
\includegraphics[width=1.1\textwidth]{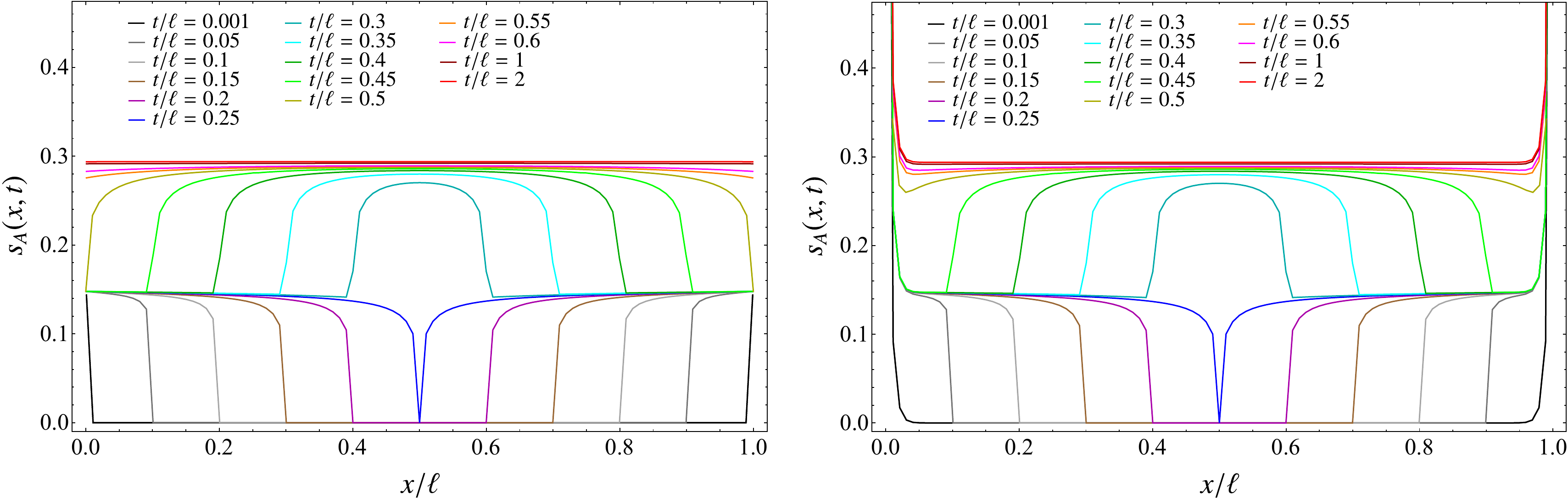}
\vspace{-.5cm}
\caption{
Temporal evolution of the contour for the entanglement entropy 
of an interval of length $\ell$ in the periodic harmonic chain
according to (\ref{QPcontourguess}), obtained 
through the quasi-particle picture (see \S\ref{sec:QP}).
Left: $f_0(x) =0$ identically. 
Right: $f_0(x) =s^{(1)}_{A,0}(x,0)$ defined in (\ref{contour CFT t=0})
has been used in (\ref{QPcontourguess}) in order to capture 
the linear divergence close to the endpoints. 
}
\label{fig:ContourBosonQP}
\end{figure}

The functions in (\ref{f_12 from QP}) fulfil (\ref{conf_012}) and the symmetry conditions introduced in the text below (\ref{QPcontour}).
Thus, the contour function (\ref{QPcontour}) becomes
\be
\label{QPcontourguess}
s_A(x,t)
=
\frac{1}{2}\,\bigg[\,\int_{x<2|v_p|t<\ell} \!\! \tilde{s}(p)\, dp\,
+
\int_{\ell\,-\, x<2|v_p|t<\ell} \!\! \tilde{s}(p)\, dp\,
\bigg]
+\!
\int_{2|v_p|t>\ell} \!\! \tilde{s}(p)\, dp + f_0(x)
\ee
Integrating this expression between two generic points $0< x_1<x_2< \ell$, we obtain
\bea
\label{integ-contour-qp-S}
& &\hspace{-2.3cm}
\mathcal{S}_A(x_1,x_2;t)=
\frac{x_2-x_1}{2}
\bigg[\,2\!\int_{2|v_p|t>\ell} \!\! \tilde{s}(p)\, dp \,+\! 
\int_{\ell-x_1<2|v_p|t<\ell}  \!\! \tilde{s}(p)\, dp \,+\! 
\int_{x_2<2|v_p|t<\ell}  \!\! \tilde{s}(p)\, dp \, \bigg]
 \\
 \rule{0pt}{.8cm}
& &
\hspace{-2.2cm}
+ \frac{1}{2}
\bigg[ \int_{\ell-x_2<2|v_p|t<\ell-x_1} \hspace{-.8cm} 
\tilde{s}(p)\,(2 |v_p| t+x_2-\ell)\, dp
   \, +\! \int_{x_1<2|v_p|t<x_2}  \hspace{-.8cm} 
   (2 |v_p| t-x_1)\, \tilde{s}(p)\, dp
   \bigg]    
   +\!
   \int_{x_1}^{x_2} \!\! f_0(x)\, dx\,.
\nonumber  
\eea

The function $\tilde{s}(p)$ can be computed by employing the fact that the density of thermodynamic entropy in the stationary state 
coincides with the one of the entanglement entropy in (\ref{QPentropy}) 
\cite{ac-18-qp-quench}.

For the global quench in the harmonic chain that we are considering,  
the stationary values of local observables can be described by a Generalised Gibbs Ensemble
\cite{rigol-07-GGE, caux-essler-prosen-15-GGE} 
(we refer the interested reader to the review \cite{vidmar-rigol-16-GGEreview} for an 
extensive list of references).
The velocity $v_p$ of the quasi-particles can be computed as $v_p=\partial_p \omega_p$, being $\omega_p$ the dispersion relation of the model. 
For the harmonic chain, in \cite{ac-18-qp-quench, cc-07-quench-extended} it has been found that
\be
\tilde{s}(p)
=
\frac{1}{2\pi}  \big[ \left(n_p+1 \right)\log \left(n_p+1 \right)-n_p\log n_p \big]
\;\;\qquad\;\;
n_p
=
\frac{1}{4}\left(\frac{\omega_p}{\omega_{0,p}}+\frac{\omega_{0,p}}{\omega_{p}} \right)-\frac{1}{2}
\label{hfunction_HC}
\ee
where $\omega_p \equiv \sqrt{\omega^2 + \tfrac{4\kappa}{m} [ \sin(p/2) ]^2}$ 
(see (\ref{dispersion relation periodic})) and $\omega_{0,p}$ is obtained by replacing $\omega$ with $\omega_0$ in $\omega_p $.
The dispersion relation (\ref{dispersion relation periodic}) provides the velocity of each momentum mode as follows
\be
\label{vk}
v_p \equiv \frac{\partial \omega_p}{\partial p} = 
\, \frac{(\kappa/m)\sin(p)}{\sqrt{\omega^2+ (4\kappa/m)\sin^2(p/2)}}
\ee
Let us remark that the above results based on the quasi-particle picture hold 
for any value of the frequencies $\omega_0$ and $\omega$,
while in our numerical analysis $\omega_0=1$ and $\omega=0$.
It would be interesting to consider also other values for these parameters.

In Fig.\;\ref{fig:ContourBosonData} we show the temporal evolution of the contour for the entanglement entropy 
evaluated through the prescription of \cite{cdt-17-contour}, outlined in \S\ref{sec:HC_contour}.
This qualitative behaviour, which has been observed also in \cite{chen-vidal} for fermionic chains after a global quench,
has been obtained through the naive analysis performed in \S\ref{sec:cft-naive} by employing the CFT formulas of \cite{ct-16}
(see the right panel of Fig.\;\ref{fig:CFTnaive}).
The curves in Fig.\;\ref{fig:ContourBosonData} can be interpreted in terms of two fronts starting from 
the endpoints and travelling in the opposite directions towards the center of the interval. 
Each front has a plateau, that is not exactly horizontal in the numerical data. 
Since the two fronts have velocity equal to $2$ (see also \S\ref{sec:cft-naive}), 
they cross each other around $t/\ell \simeq 1/4$ and superpose  until they reach the opposite endpoint around $t/\ell \simeq 1/2$.
Notice that the data shown in the two panels of Fig.\;\ref{fig:ContourBosonData}  do not overlap, 
meaning that larger values of $\ell$ are needed to obtain a prediction for this curve 
in the limit of large $\ell$.
This prediction can be done from the data obtained for the global quench 
in the chain of free fermions considered in \S\ref{sec:fermion-quench} (see Fig.\;\ref{fig:ContourFermion}).
We remark that the data in Fig.\;\ref{fig:ContourBosonData} display a divergence close to the endpoints of the interval that is independent of time. 

 \begin{figure}[t!]
\vspace{.2cm}
\hspace{2.5cm}
\includegraphics[width=.693\textwidth]{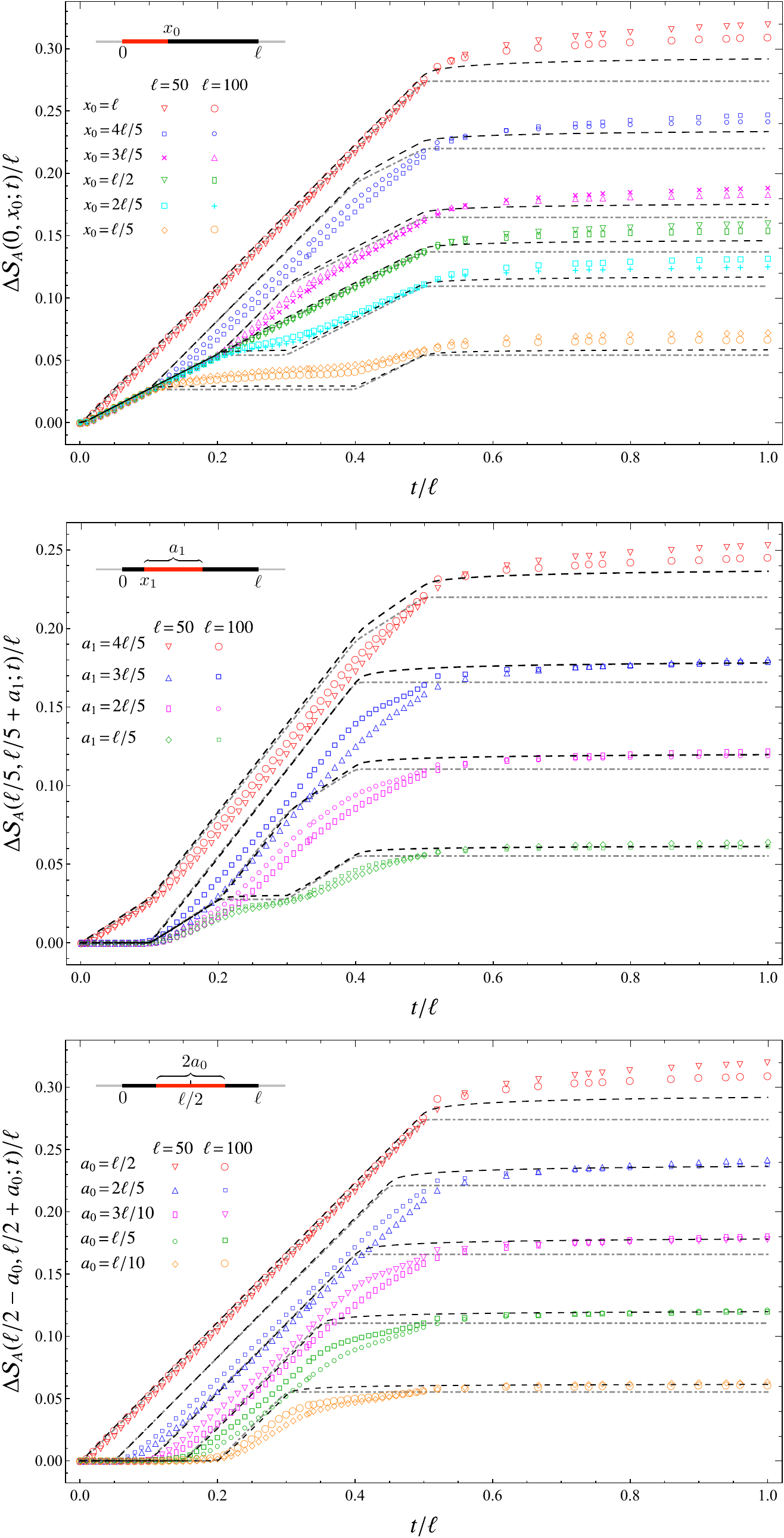}
\vspace{-.0cm}
\caption{
Harmonic chain:
Temporal evolution of (\ref{integrated contour x1x2 discrete sub}) for $n=1$.
}
\label{fig:IC_bosons}
\end{figure}

The contour function for the entanglement entropy 
obtained through the quasi-particle picture
(namely by employing  (\ref{QPcontourguess}), (\ref{hfunction_HC}) and (\ref{vk}))
 is shown in Fig.\;\ref{fig:ContourBosonQP},
 where in the left panel $f_0(x)=0$ identically, 
 while $f_0(x) =s^{(1)}_{A,0}(x,0)$ in (\ref{contour CFT t=0}) has been chosen in the right panel.
The latter choice leads to reproduce also the linear divergencies of the contour function for the
entanglement entropy near the endpoints of the interval.
The quasi-particle picture formula (\ref{QPcontourguess})
captures in a better way some features of the numerical data 
with respect to the corresponding CFT expression (\ref{contour_cft_naive}).
For instance, the curve in Fig.\;\ref{fig:ContourBosonQP} having 
$t/\ell = 1/4$ is not flat in the middle of the interval
and that the local maximum of the curve having $1/4 < t/\ell < 1$ increases with time,
while in the right panel of Fig.\;\ref{fig:CFTnaive} it is constant and equal to half of 
its asymptotic value for $t \to \infty$.

In Fig.\;\ref{fig:IC_bosons} we consider the quantity introduced in 
(\ref{integrated contour x1x2 discrete sub}) with $n=1$ 
for three different choices of $(i_1,i_2)$:
the case with $i_1 =0$ (top panel), 
the case with fixed $0 < i_1 < \ell/2$ and $i_2$ variable (middle panel)
and the case with $(i_1,i_2)$ in the middle of the interval $A$ (bottom panel).
The data points (obtained through (\ref{contour from mpf}), (\ref{integrated contour x1x2 discrete}), (\ref{integrated contour x1x2 discrete sub}) and (\ref{mpf from W}))
are compared against the corresponding formula coming from 
the quasi-particle picture expression (\ref{integ-contour-qp-S}) (black dashed curves),
where the terms containing $f_0$ cancel,
and also against the naive CFT expression (\ref{IntegContS2sub}) (grey dashed-dotted curves).
Notice that the CFT curves and the ones obtained from the quasi-particle picture change slope at the same values of $t/\ell$.
At the beginning of the evolution, $\Delta \mathcal{S}_A$ grows linearly with a slope that
depends on whether $(i_1, i_2)$ contains one or two endpoints of $A$
(notice that the red curves in the top and bottom panels correspond to the entanglement entropy),
while it vanishes whenever the endpoints of $A$ do not belong to $(i_1, i_2)$.
For $t/\ell> 1/2$ all the pairs of quasi-particles have contributed and
the curves reach a constant value proportional to the size of the interval $(i_1,i_2)$.

For $t/\ell< 1/2$, all the curves in Fig.\;\ref{fig:IC_bosons} exhibit a piecewise linear behaviour
and, for a given configuration, the slope changes at most four times
in correspondence to values of $t/\ell$ that depend on $(i_1,i_2)$
and that are captured both by the naive CFT expression (\ref{IntegContS2sub}) and 
by the quasi-particle picture expression (\ref{integ-contour-qp-S}).
In the top panel of Fig.\;\ref{fig:IC_bosons} the changes of slope occur
at $t/\ell=x_0/(2\ell)$, $t/\ell=(\ell-x_0)/(2\ell)$ and $t/\ell=1/2$;
in the middle panel at $t/\ell=x_1/2$, $t/\ell=(a_1+x_1)/(2\ell)$, $t/\ell=(\ell-a_1-x_1)/(2\ell)$ and $t/\ell=(\ell-x_1)/(2\ell)$ (in the data shown correspond to $x_1=\ell/5$);
and in the bottom panel at $t/\ell=(\ell/2-a_0)/(2\ell)$ and $t/\ell=(\ell/2+a_0)/(2\ell)$.
These values come from the quasi-particle picture.
First we observe that the segments composing any piecewise linear curve are horizontal 
or have a positive slope that can take two values such that one is twice the other one.
The quasi-particles can entangle $(i_1,i_2)$ with $B$ by crossing one or both the endpoints of $A$.
Denoting by $\mathfrak{n}_R$ ($\mathfrak{n}_L$) 
the number of particles entangling $(i_1,i_2)$ with $B$ across the right (left) endpoint of $A$,
we have that the segment is horizontal whenever
both $\mathfrak{n}_R$ and $\mathfrak{n}_L$ are constant in time. 
Instead, when only either $\mathfrak{n}_R$ or $\mathfrak{n}_L$ is increasing in time, 
the corresponding segment has a positive slope, that becomes twice this value 
whenever both $\mathfrak{n}_R$ and $\mathfrak{n}_L$ are increasing in time. 
By adapting these considerations to the configurations for $(i_1,i_2)$
considered in the three panels of Fig.\;\ref{fig:IC_bosons},
one obtains the above values of $t/\ell$ corresponding to the changes in the slope of the piecewise linear curves.

\section{Interval in a chain of free fermions}
\label{sec:fermion-quench}

In this section we study the temporal evolution of the entanglement hamiltonian matrix 
after the global quench in a chain of free fermions introduced in \cite{ep-07-local-quench, ep-rev}.

Let us consider the following inhomogeneous hamiltonian written in terms of
the fermionic creation and annihilation operators $c_n$ and ${c}_n^\dagger$,
(that satisfy the standard anticommutation relations $\{c_n^\dag,c_m^\dag\}=\{c_n,c_m\}=0$ and $\{c_n,c^\dag_m\}=\delta_{m,n}$)
\cite{ep-07-local-quench}
\be
\widehat{H}_0= -\,\frac{1}{2}\sum_{n=-\infty}^{+\infty}
\! t_n \big(\hat{c}^\dag_n \,\hat{c}_{n+1}+\hat{c}^{\dag}_{n+1}\,\hat{c}_n\big)
\ee
where $t_{2n}=1$ and $t_{2n+1}=0$, namely only pairs of sites are coupled (dimerized chain).
The system is half filled and prepared in the ground state $| \psi_0 \rangle$ of $H_0$.

At $t=0$ the inhomogeneity is removed and the unitary time evolution of $| \psi_0 \rangle$ is governed by the translation invariant hopping hamiltonian given by
\be
\label{H_ff homo}
\widehat{H}= -\,\frac{1}{2}\sum_{n=-\infty}^{+\infty}
\! \! \big(\hat{c}^\dag_n \,\hat{c}_{n+1}+\hat{c}^{\dag}_{n+1}\,\hat{c}_n\big)
\ee
which is also known as the tight binding model at half filling. 

In \cite{ep-07-local-quench}  the analytic expression of the correlation matrix after the global quench 
has been computed. 
Its generic element $C_{i,j}(t) \equiv \langle c_i^\dagger(t) \, c_j(t) \rangle$ reads
\be 
\label{correl}
C_{i,j}(t)
=
C^{(\infty)}_{i,j}
+ \textrm{i}\,\frac{i-j}{4t}e^{-\textrm{i}\frac{\pi}{2}(i+j)} J_{i-j}(2t)
\ee
where $J_{\nu}(x)$ is the Bessel function of the first kind and 
\be
\label{Cinf matrix def}
C^{(\infty)}_{i,j} \equiv \frac{1}{2}\left[\delta_{i,j}+\frac{1}{2}\big(\delta_{i,j+1}+\delta_{i,j-1}\big)\right].
\ee
By employing that $J_{-\nu}(x) = (-1)^\nu J_{\nu}(x)$, we have 
that $\textrm{Re} [ C_{i,j}(t) ]$ is symmetric and that $\textrm{Im} [ C_{i,j}(t) ]$ is antisymmetric. 
Furthermore, 
we remark that $C_{i,j}(t) \to C^{(\infty)}_{i,j} $ as $t \to +\infty$; hence the correlation matrix is real in the asymptotic regime of long time.

 \begin{figure}[t!]
\vspace{.5cm}
\hspace{-.5cm}
\includegraphics[width=1.1\textwidth]{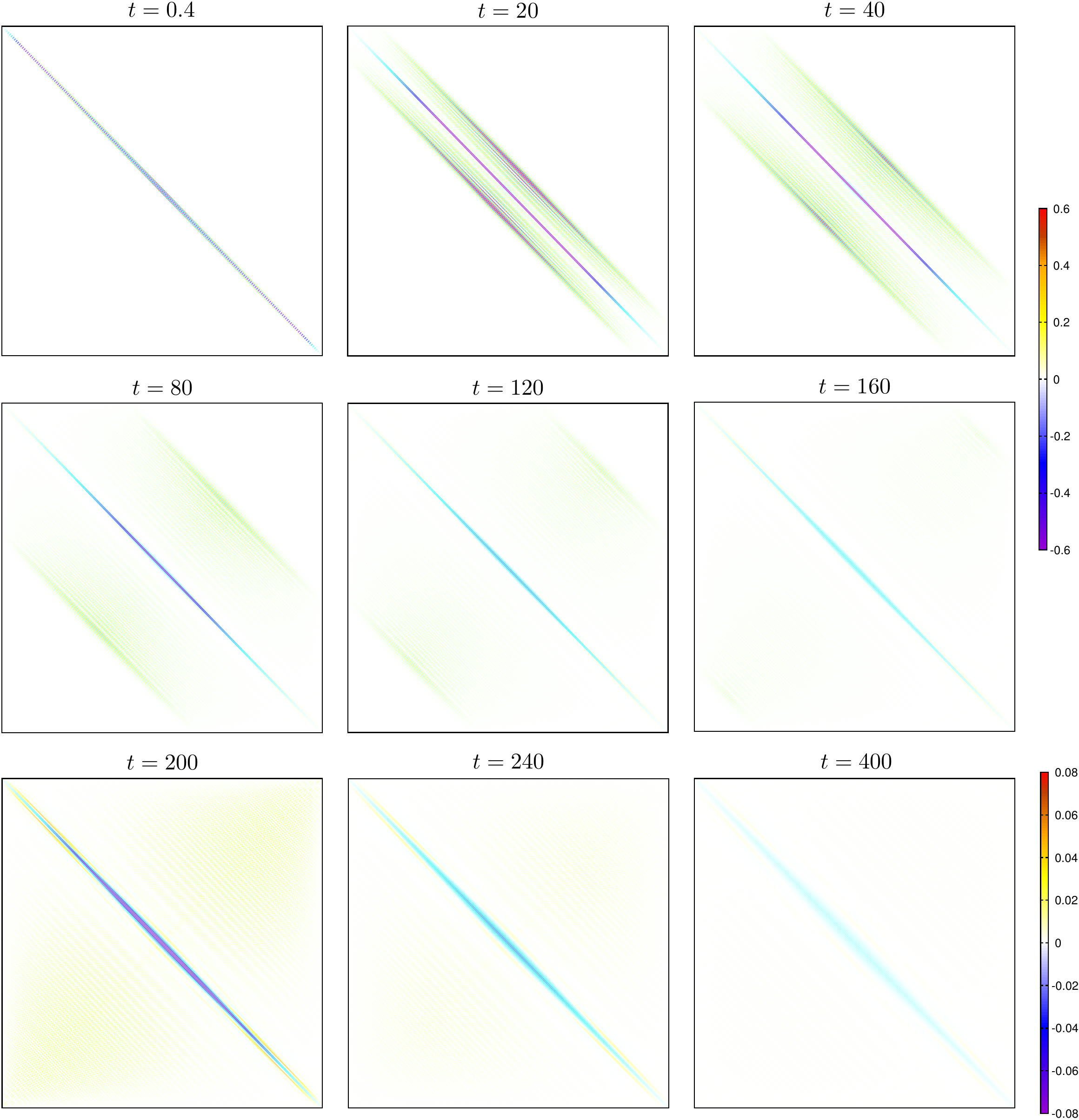}
\vspace{-.3cm}
\caption{
Temporal evolution of the 
real part of the entanglement hamiltonian matrix (\ref{peschel02 H_A global-q})
of an interval with $\ell=400$
 in the infinite chain of free fermions
after the global quench (see \S\ref{sec:fermion-quench}). 
}
\vspace{2cm}
\label{fig:H_fermion_density_realpart}
\end{figure}

 \begin{figure}[t!]
\vspace{.5cm}
\hspace{-.5cm}
\includegraphics[width=1.1\textwidth]{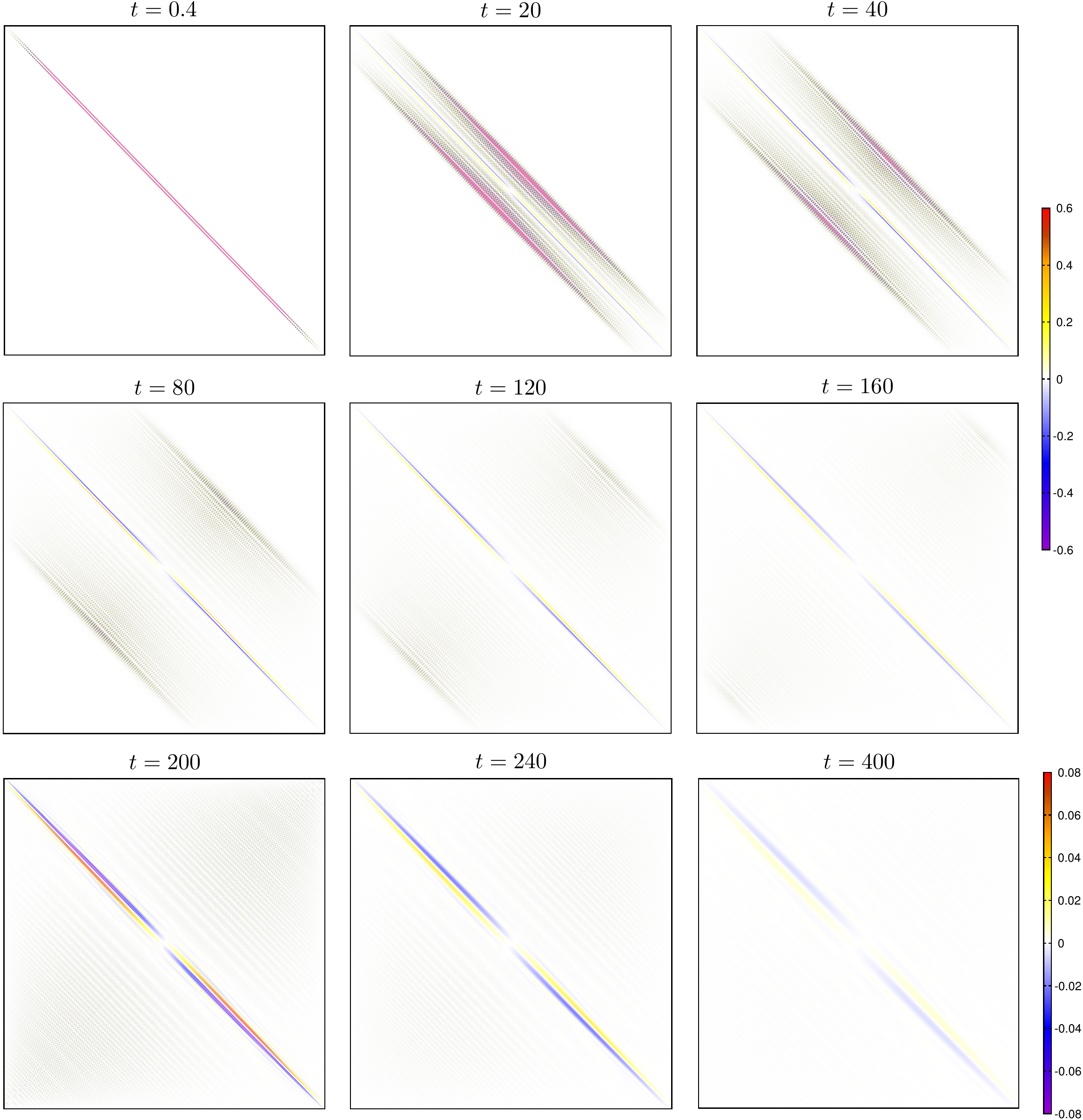}
\vspace{-.3cm}
\caption{
Temporal evolution of the 
imaginary part of the entanglement hamiltonian matrix (\ref{peschel02 H_A global-q})
of an interval with $\ell=400$
 in the infinite chain of free fermions
after the global quench (see \S\ref{sec:fermion-quench}). 
}
\vspace{2cm}
\label{fig:H_fermion_density_imaginarypart}
\end{figure}

\subsection{Entanglement hamiltonian matrix}
\label{sec-EHmatrix-FF}

Considering an interval $A$ containing $\ell$ sites labelled by $1\leqslant i \leqslant \ell$,
the generic element of the $\ell \times \ell$ correlation matrix $C_A(t)$ 
is obtained by just restricting to the rows and columns corresponding to the sites in $A$,
namely $C_A(t)_{i,j} = C(t)_{i,j}$ in (\ref{correl}) with $1\leqslant i,j \leqslant \ell$.
The entanglement hamiltonian of $A$ after this global quench is  
the operator (\ref{K_A peschel}) with the matrix $T$ given by (\ref{peschel02 H_A_gen}), i.e.
\cite{peschel-03-modham}
\be
\label{peschel02 H_A global-q}
T^{\textrm t} = \log\!\big(C_A(t)^{-1} - \boldsymbol{1} \big) \,.
\ee
This implies that the relation between the eigenvalues $\eta_k$ of $T$ and the eigenvalues $\zeta_k$ of $C_A$ reads $\eta_k = \log(1/\zeta_k -1)$, as already discussed in \S\ref{sec:fermions-special cases}. 
For a generic value of $t$, the entanglement hamiltonian matrix $T$ in (\ref{peschel02 H_A global-q}) is a complex $\ell \times \ell$ matrix.
Since $T$ is hermitian at any $t$, its real part is symmetric and its imaginary part is antisymmetric. 
Thus, we can focus on the $p$-th diagonal of these two matrices with $p \geqslant 0$.

 \begin{figure}[t!]
\vspace{.2cm}
\hspace{-.8cm}
\includegraphics[width=1.05\textwidth]{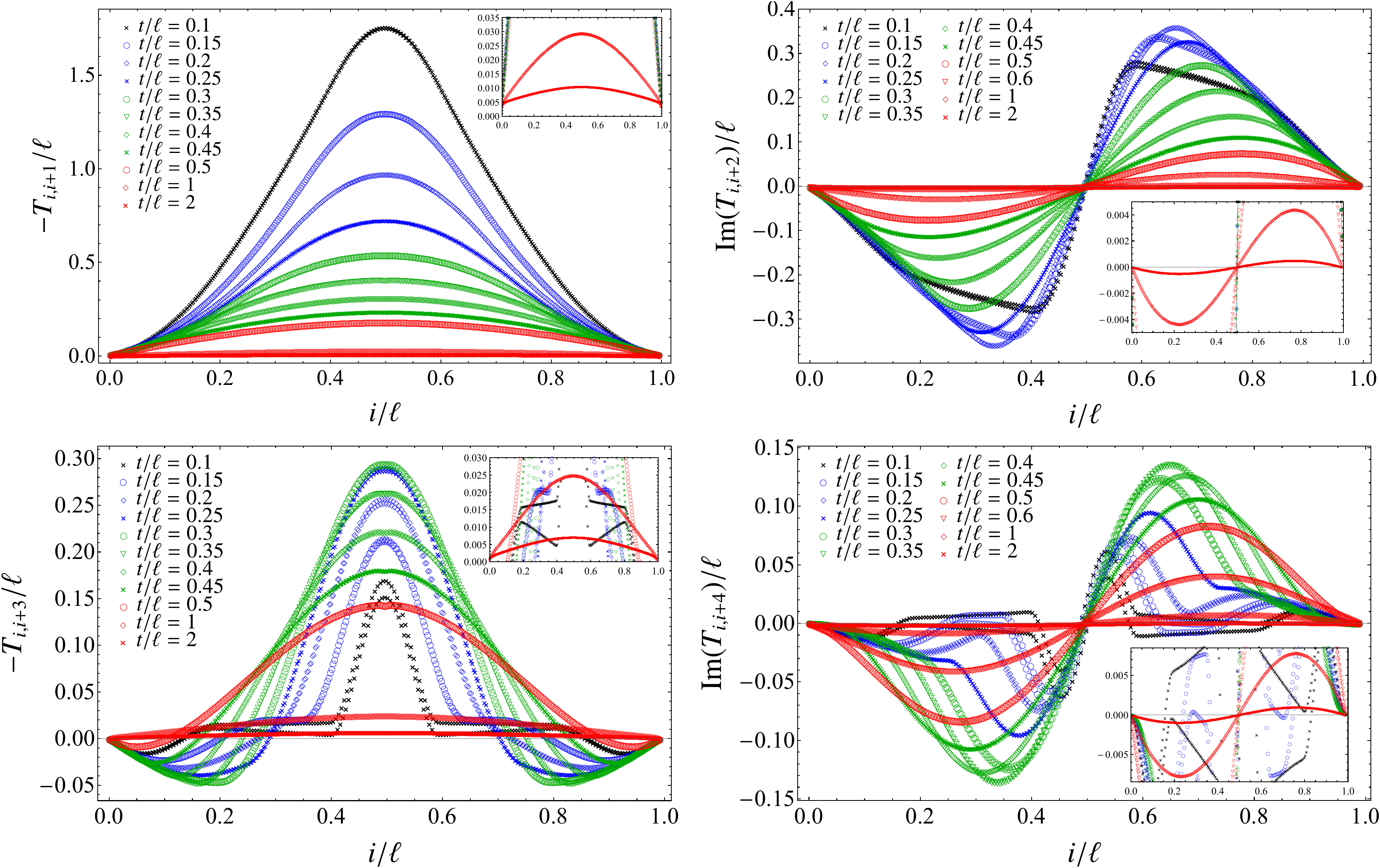}
\vspace{-.3cm}
\caption{
Chain of free fermions:
Temporal evolution of the $p$-th diagonal with $1\leqslant p \leqslant 4$
of the entanglement hamiltonian matrix (\ref{peschel02 H_A global-q})
of an interval with $\ell =400$ after the global quench.
The insets zoom in on small values, in order to show the curves for large times.
}
\label{fig:Hdiags_fermion}
\end{figure}

\begin{figure}[t!]
\vspace{.2cm}
\hspace{-1.cm}
\includegraphics[width=1.1\textwidth]{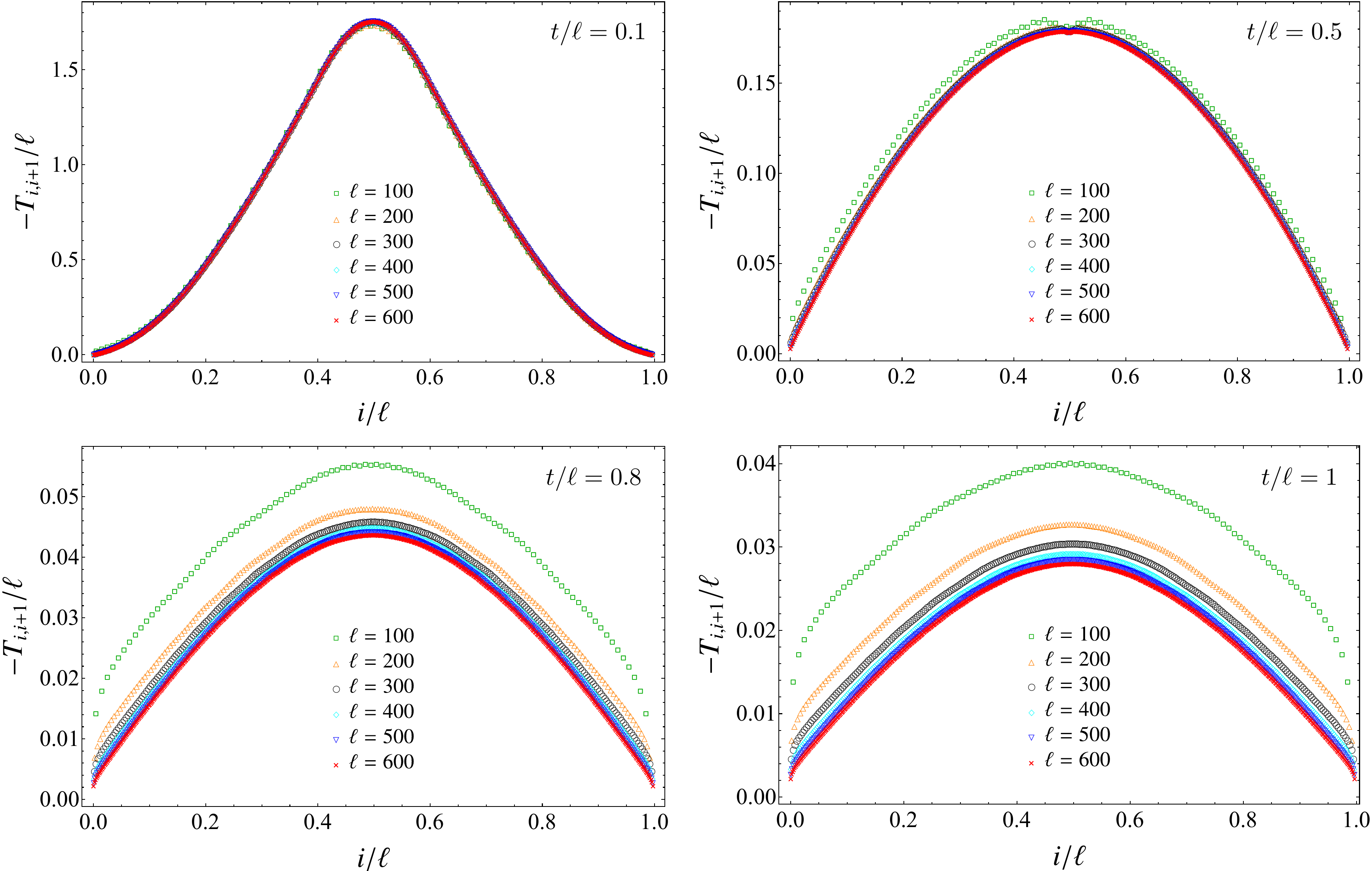}
\vspace{-.3cm}
\caption{
Chain of free fermions:
First diagonal of the entanglement hamiltonian matrix (\ref{peschel02 H_A global-q})
for four values $t$ after the global quench and for different values of $\ell$.
}
\label{fig:SpectrumDiagScaling_fermion}
\end{figure}

 \begin{figure}[h!]
\vspace{.5cm}
\hspace{-1.1cm}
\includegraphics[width=1.1\textwidth]{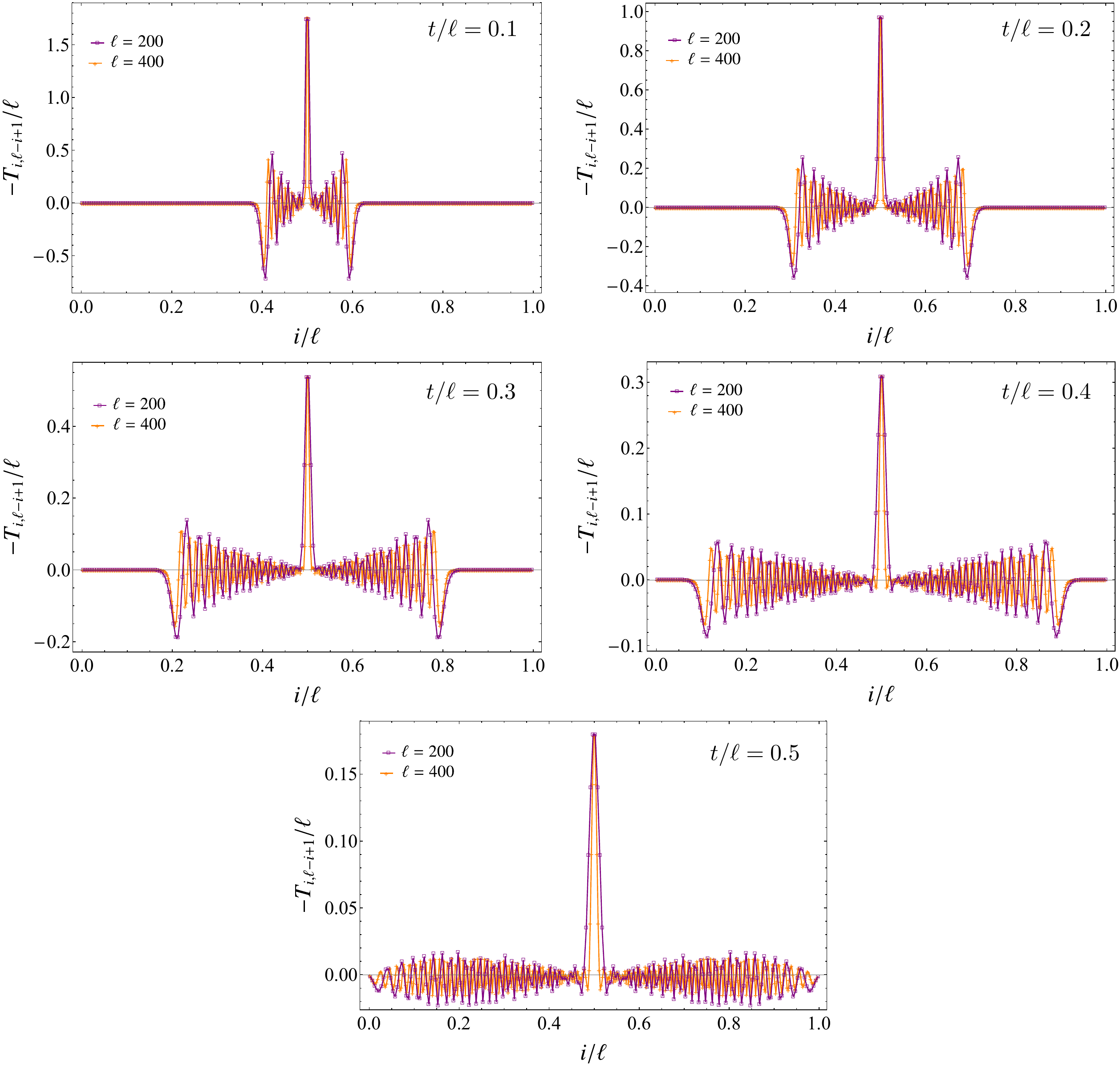}
\vspace{-.5cm}
\caption{
Chain of free fermions:
Antidiagonal of the real part of the entanglement hamiltonian matrix 
(\ref{peschel02 H_A global-q}) (see also Fig.\,\ref{fig:H_fermion_density_realpart}) 
for various times and two different lengths. 
The points corresponding to the two minima
travel in opposite directions from the center of the interval 
towards the endpoints with velocity equal to one 
(see also the bottom panels in Fig.\;\ref{fig:antidiagonals_d0}, where $2d_0$ denotes 
the distance between these two minima).
}
\vspace{2cm}
\label{fig:fermions_antidiag_LC_Re}
\end{figure}

 \begin{figure}[t!]
\vspace{.5cm}
\hspace{-1.1cm}
\includegraphics[width=1.1\textwidth]{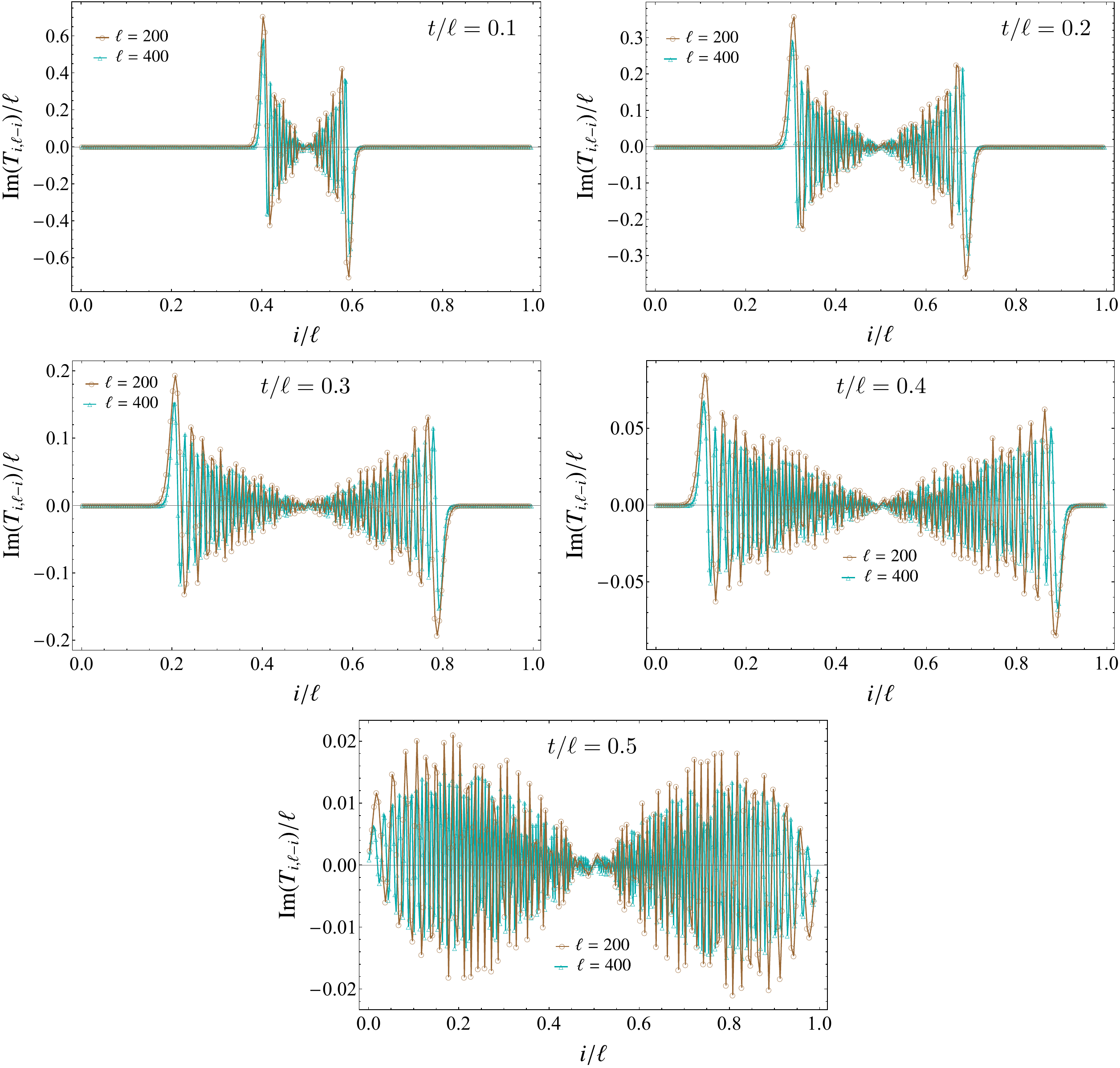}
\vspace{-.5cm}
\caption{
Chain of free fermions:
Antidiagonal of the imaginary part of the entanglement hamiltonian matrix 
(\ref{peschel02 H_A global-q}) (see also Fig.\,\ref{fig:H_fermion_density_imaginarypart}) 
for various times and two different lengths. 
The points corresponding to the maximum and to the  minimum 
travel in opposite directions from the center of the interval 
towards the endpoints with velocity equal to one 
(see also the bottom panels in Fig.\;\ref{fig:antidiagonals_d0}, where $2d_0$ denotes 
the distance between the two extrema).
}
\vspace{2cm}
\label{fig:fermions_antidiag_LC_Im}
\end{figure}

 \begin{figure}[t!]
\vspace{.2cm}
\hspace{-.8cm}
\includegraphics[width=1.1\textwidth]{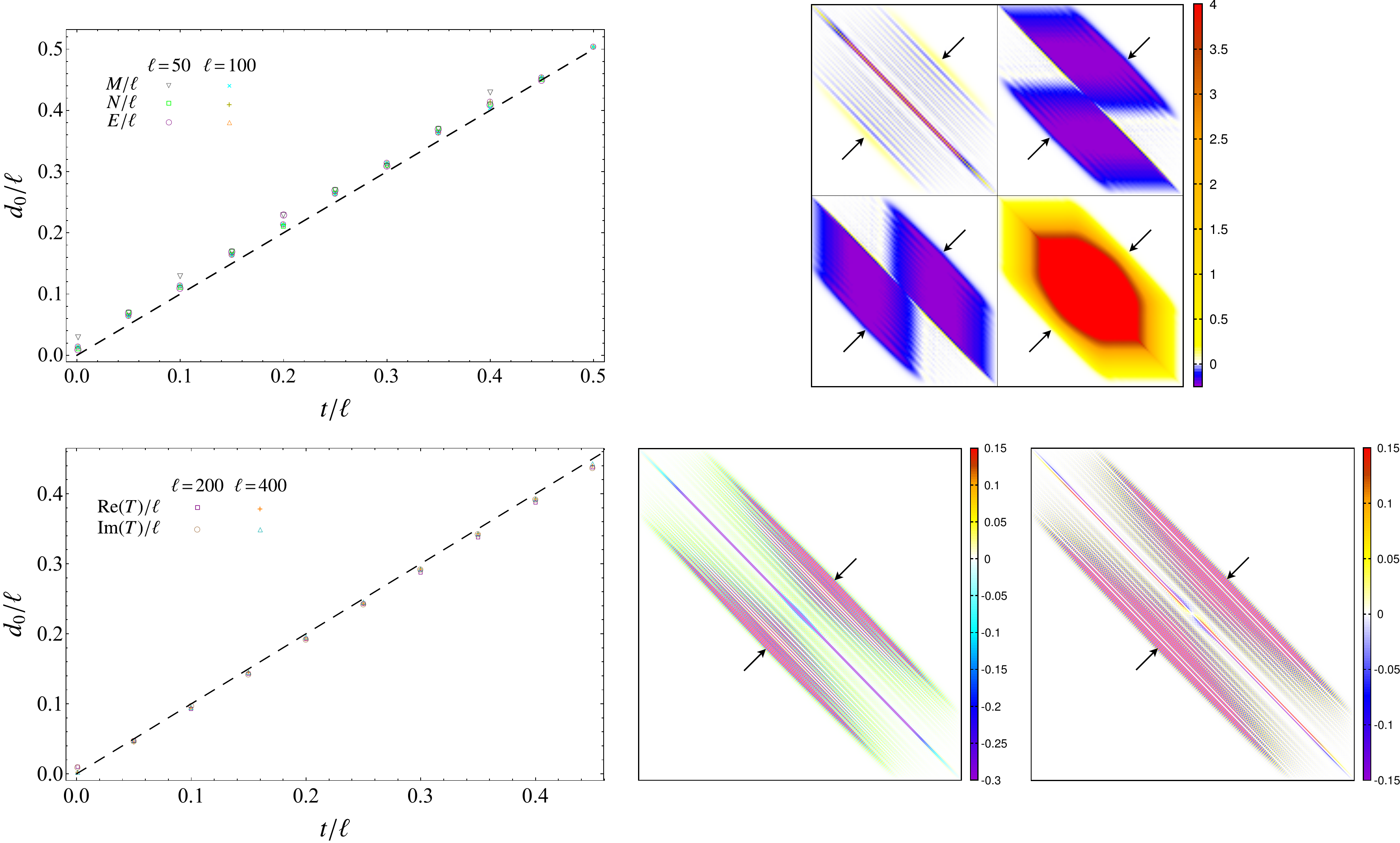}
\vspace{-.5cm}
\caption{
Linear growth of the width of the bands occurring in the 
blocks of the entanglement hamiltonian matrices after the global quench. 
The width $2d_0$ of the various bands 
is highlighted by the black arrows in the panels on the right,
where the entanglement hamiltonian matrices are shown
at a typical value $t$ such that $t/\ell \in (0,1/2)$
(see Fig.\,\ref{fig:HquenchDensityHC}, Fig.\,\,\ref{fig:H_fermion_density_realpart} 
and Fig.\,\ref{fig:H_fermion_density_imaginarypart}).
In the left panels the slope of the dashed lines is one. 
Top: Harmonic chain (see also Fig.\,\ref{fig:MNRantidiags_LC}).
Bottom: Chain of free fermions 
(see also Fig.\,\ref{fig:fermions_antidiag_LC_Re} 
and Fig.\,\ref{fig:fermions_antidiag_LC_Im}).
}
\label{fig:antidiagonals_d0}
\end{figure}

The eigenvalues $\zeta_k \in (0,1)$ and many of them lie exponentially close to 0 and 1.
This forces us to work with very high precision in order to get finite results for 
(\ref{peschel02 H_A global-q}).
In particular, the largest interval we considered has $\ell=600$ sites and
we employed between 900 and 2500 digits, depending on the value of $t$. 
The needed working precision decreases as time increases.

Also for this global quench, the configuration given by an interval in the infinite line 
is symmetric with respect to the center of the interval and this
leads to a symmetry in the elements of the $p$-th diagonal. 
In particular, the odd diagonals of $T$ are symmetric, while the even ones are antisymmetric, 
namely $T_{i,i+p}=T_{\ell-i-p+1,\ell-i-p+1}$ for odd $p$ and $T_{i,i+p}=-T_{\ell-i-p+1,\ell-i-p+1}$ for even $p$. The main diagonal vanishes identically.

In Fig.\;\ref{fig:H_fermion_density_realpart} and Fig.\;\ref{fig:H_fermion_density_imaginarypart}
we show respectively the real part of $T$ and the imaginary part of $T$ of an interval with $\ell=400$
for nine values of $t$. 
For $t\to \infty$ the imaginary part of $T$ vanishes
(a detailed analysis of this asymptotic regime is performed in \S\ref{sec:long time}).
We also checked numerically that the imaginary part of $T$ vanishes for $t\to 0$, as expected. 
In the complex matrix $T$ the odd diagonals are real, while the even diagonals are purely imaginary,
like in the correlation matrix (\ref{correl}).

As time evolves, the amplitude of the elements of $T$ decreases; indeed,
in Fig.\;\ref{fig:H_fermion_density_realpart} and Fig.\;\ref{fig:H_fermion_density_imaginarypart}
a zoom is needed for large values of $t$ in order to appreciate the fact that $T$ is non vanishing. 
It is straightforward to observe that at the beginning the main contribution is localised on the 
diagonals close to the main one and that, during the temporal evolution, also the diagonals corresponding to longer range
interactions become more important. 

In Fig.\;\ref{fig:Hdiags_fermion} we show the temporal evolution of the first four diagonals in the entanglement hamiltonian matrix. 
Notice that $T_{i,i+1}$ is always positive, while in the other panels also negative values occur. 
We find it worth remarking that all the curves displayed in this figure vanish at the endpoints of the interval.
The spatial inhomogeneity of these diagonals is a characteristic feature of the entanglement hamiltonian matrix,
if compared with the correlation matrix restricted to the interval (see Fig.\;\ref{fig:fermions_correlations}).

It is worth studying the limit $\ell \to \infty$ of the entanglement hamiltonian matrix, as done
\cite{eisler-peschel-17, ep-18, Arias-16} for a static configurations (see \cite{etp-19} for the continuum limit).
In Fig.\;\ref{fig:SpectrumDiagScaling_fermion} we consider increasing values of $\ell$ for some fixed values of $t/\ell$.
While the data have reached the asymptotic curve when $t/\ell = 0.1$ and $t/\ell = 0.5$,
it seems that larger values of $\ell$ are needed to observe this collapse for higher values of $t/\ell$.

In Fig.\;\ref{fig:H_fermion_density_realpart} and Fig.\;\ref{fig:H_fermion_density_imaginarypart}
it is straightforward to observe that the diagonals providing a long range interaction become more relevant as time
evolves by forming a band around the main diagonal whose width $2d_0$ increases with time.
A similar feature has been highlighted for the global quench in the harmonic chain discussed in \S\ref{sec:numerics HC}.
This observation naturally leads to consider the antidiagonals in $\textrm{Re}(T)$ and $\textrm{Im}(T)$,
that are shown in Fig.\;\ref{fig:fermions_antidiag_LC_Re} and  Fig.\;\ref{fig:fermions_antidiag_LC_Im},
making evident the occurrence of a band and the increasing of its width during the temporal evolution.
The width $2d_0$ of this band can be defined by taking the distance between the two minima for $\textrm{Re}(T)$
and the distance between the minimum and the maximum for $\textrm{Im}(T)$.
In the bottom panels of Fig.\;\ref{fig:antidiagonals_d0} we show that $d_0(t)$ 
defined in this way increases linearly with velocity equal to one  as time evolves. 

A similar band, whose width increases linearly with velocity equal to one as well,
occurs also in the temporal evolution of $\textrm{Re}(C_A)$ and $\textrm{Im}(C_A)$
whose antidiagonals display a behaviour like the one shown in 
Fig.\;\ref{fig:fermions_antidiag_LC_Re} and  Fig.\;\ref{fig:fermions_antidiag_LC_Im} 
for the entanglement hamiltonian matrix
(see Fig.\;\ref{fig:fermions_correlations} for $t/\ell = 0.2$).
Thus, the main difference between the temporal evolutions of $T$
and of $C_A$ is given by the inhomogeneity of the former one along the diagonals,
that can be observed by comparing the top panels in Fig.\;\ref{fig:fermions_correlations}
with the corresponding panels in 
Fig.\;\ref{fig:H_fermion_density_realpart} and Fig.\;\ref{fig:H_fermion_density_imaginarypart}.

 \begin{figure}[t!]
\vspace{.2cm}
\hspace{-1.1cm}
\includegraphics[width=1.1\textwidth]{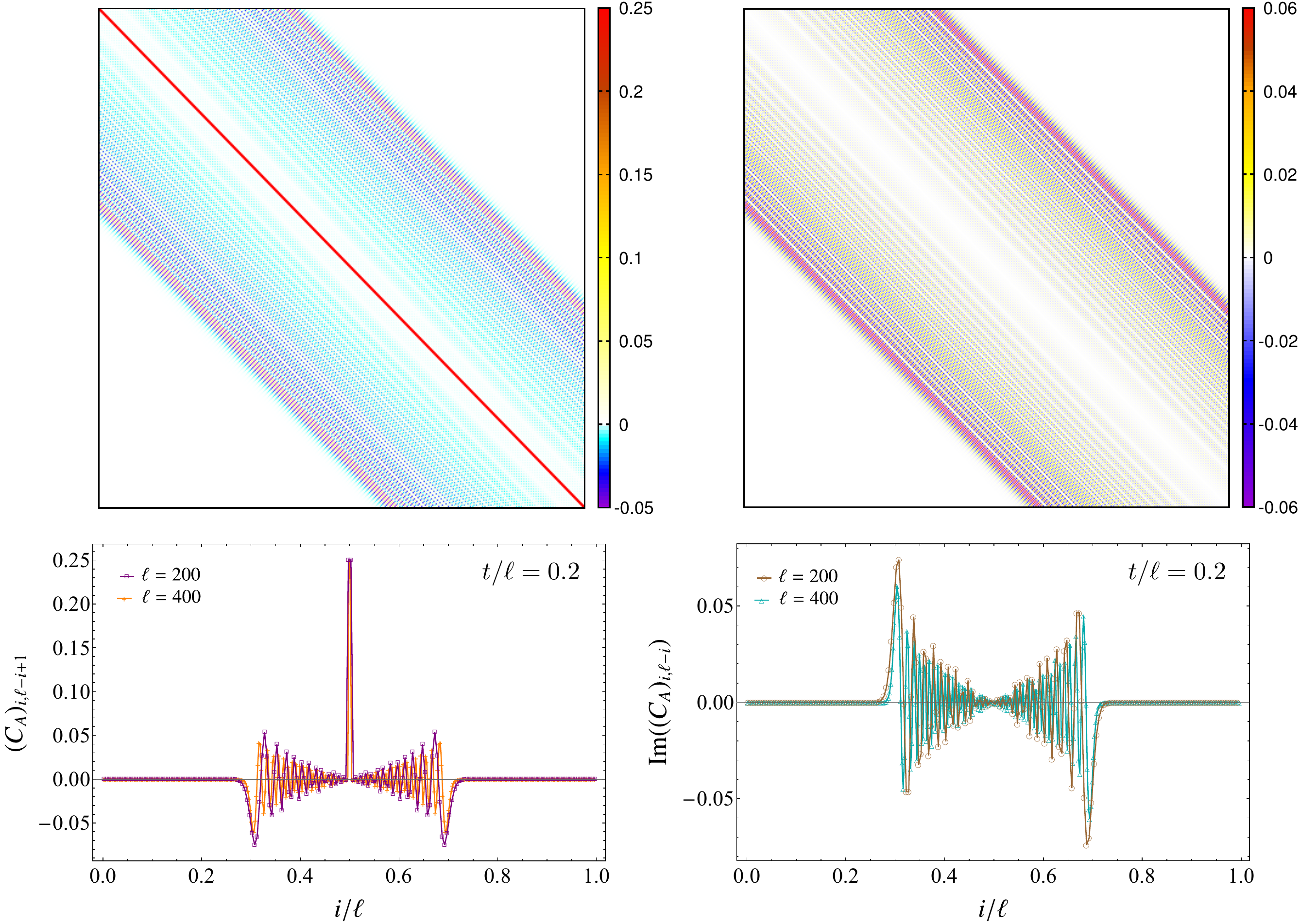}
\vspace{-.5cm}
\caption{
Chain of free fermions:
Correlation matrix $C_A$ obtained by restricting (\ref{correl}) to an interval of length $\ell$
at a fixed time after the quench given by $t/\ell=0.2$.
Top: $\textrm{Re}(C_A)$ (left) and  $\textrm{Im}(C_A)$ (right) for $\ell=400$.
Bottom: Antidiagonals of $\textrm{Re}(C_A)$ (left) and  $\textrm{Im}(C_A)$ (right) for two values of $\ell$.
}
\label{fig:fermions_correlations}
\end{figure}

\begin{figure}[t!]
\vspace{.2cm}
\hspace{-1.cm}
\includegraphics[width=1.1\textwidth]{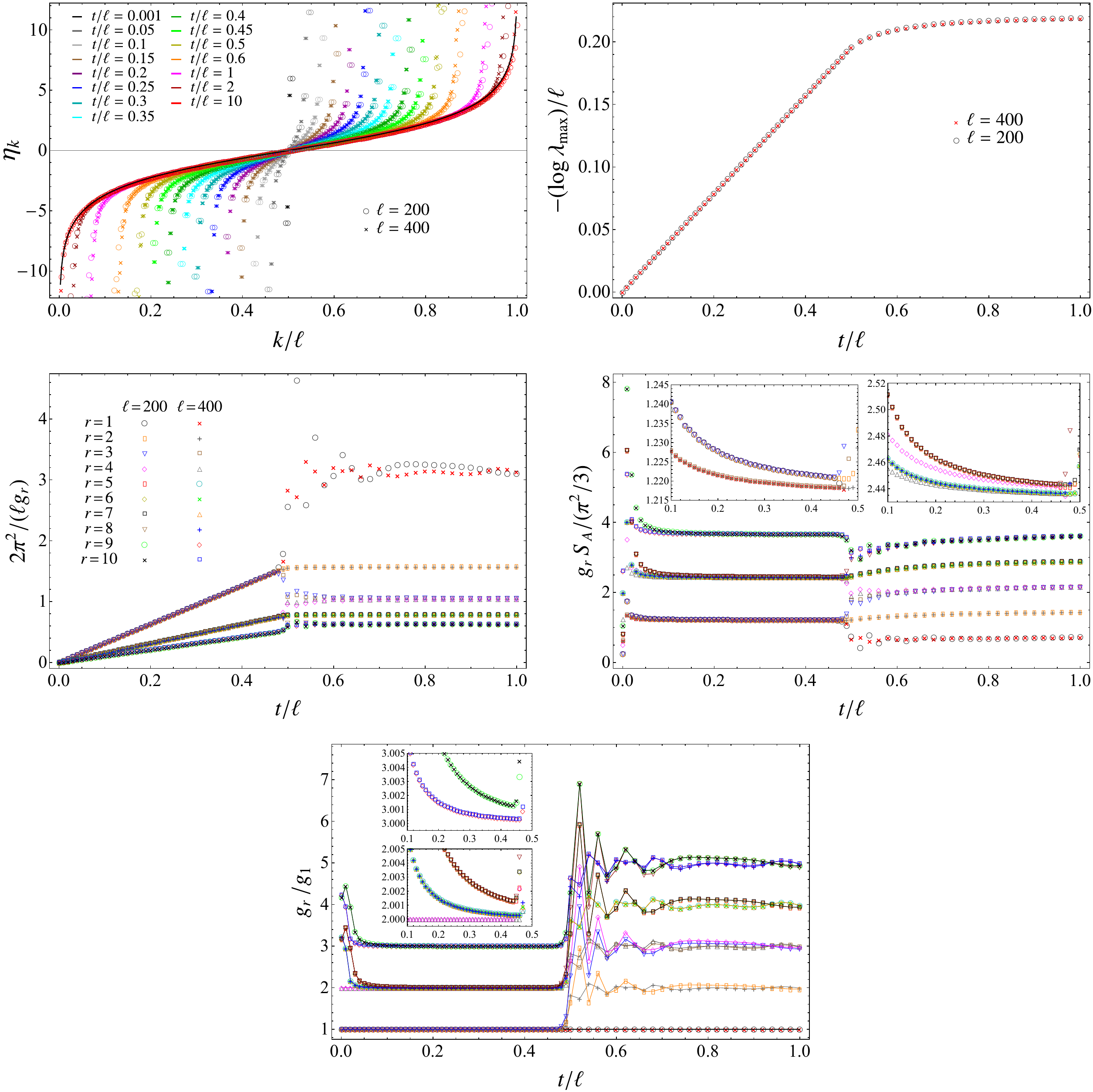}
\vspace{-.5cm}
\caption{
Chain of free fermions:
Entanglement spectrum for an interval of length $\ell$ 
after the quench discussed in \S\ref{sec:fermion-quench}.
Top left: Single particle entanglement spectrum 
for different values of $t$ (see also Fig.\;B.1 in \cite{ep-07-local-quench}).
Top right: Temporal evolution of the largest eigenvalue of the entanglement spectrum.
Middle left: Temporal evolution of the first gaps in the entanglement spectrum.
The legenda of this panel holds also in the remaining ones. 
Middle right: Temporal evolution of $g_r S_A$ 
(the insets zoom in on the two lower plateaux, 
showing that at a smaller scale the data having $\ell =200$ and $\ell=400$ do not overlap).
Bottom: Temporal evolution of the ratios $g_r/g_1$ between the gaps in the entanglement spectrum
(the insets zoom in on higher plateaux). 
The limits of the curves for $t \to \infty$ 
and large values of $\ell$ are the strictly positive integers. 
}
\label{fig:SpectrumFermion}
\end{figure}

\subsection{Entanglement spectrum}
\label{sec-ES-FF}

Also for this global quench we find it worth considering the temporal evolution 
of some quantities related to the entanglement spectrum.
In the top left panel of Fig.\;\ref{fig:SpectrumFermion}, 
the eigenvalues $\eta_k = \log(1/\zeta_k - 1)$ 
of the hermitian matrix $T$ obtained from the eigenvalues $\zeta_k$ of the reduced correlation matrix $C_A(t)$ are shown for different values of time. 
The solid black curve corresponds to the asymptotic curve for $t\to \infty$ obtained in \cite{ep-07-local-quench}.
The eigenvalues $\eta_k$ occur in the expression (\ref{T-decomposition-peschel}) of the entanglement hamiltonian matrix $T$.
Since $\eta_k=-\eta_{\ell-k}$ and only even values of $\ell$ are considered, 
we find it convenient to label these eigenvalues as
$\eta_k$ with $k=\pm \frac{1}{2}, \, \pm \frac{3}{2},\dots \pm \frac{\ell-1}{2}$. 
This gives $\eta_k=-\eta_{-k}$.
Since $\widehat{K}_A  = \sum_k  \eta_k \, \hat{\mathfrak{f}}^\dagger_k \,\hat{\mathfrak{f}}_k$,
by adapting to this model the steps that provide (\ref{spectrum_fermion}) for the harmonic chain, 
one finds that the largest eigenvalue in the entanglement spectrum is given by the configuration where
all the modes with negative $\eta_k$ are occupied.
This leads to write the largest eigenvalue as 
$\lambda_{\text{\tiny max}} = \big[\prod_k (1+ e^{-\eta_k})\big]^{-1} \prod_{k<0}\, e^{-\eta_k} $,
whose temporal evolution for $\ell=200$ and $\ell=400$ is shown in the top right panel of 
Fig.\;\ref{fig:SpectrumFermion}.

The gaps in the entanglement spectrum can be computed from the single particle entanglement energies $\eta_k$ 
by considering the particle-hole excitations with respect to the Fermi level.
For the first gaps $ 0<g_1<g_2<\dots $, this analysis gives $g_1=2\eta_{\frac{1}{2}}$,
$g_2=\eta_{\frac{1}{2}}+\eta_{\frac{3}{2}}$ and $g_3=\textrm{min}\big\{\eta_{\frac{1}{2}}+\eta_{\frac{5}{2}} \, , 2\eta_{\frac{3}{2}}\big\}$. The expressions for the higher gaps become difficult to write as $r$ increases, 
but this can be done numerically in a systematic way. 
For these entanglement gaps we carry out an analysis similar to the one performed 
for the corresponding quantities in the harmonic chain (see Fig.\;\ref{fig:SpectrumBoson}),
whose results are collected in the middle and bottom panels of Fig.\;\ref{fig:SpectrumFermion}.
Also for this global quench the data about the temporal evolution of the entanglement gaps 
display two distinct regimes separated by $t/\ell \simeq 1/2$.
In the middle left panel of Fig.\;\ref{fig:SpectrumFermion} we show the temporal evolution of the inverse of 
the entanglement gaps, in order to highlight the linear growths predicted by the CFT formula 
(\ref{gap_r cft naive}) before  $t/\ell \simeq 1/2$.
The numerical value for $\tau_0$ is obtained by fitting 
the linear growth of the entanglement entropy through the CFT formula 
$S_A \simeq 2\pi c\, t /(3\tau_0)$ of \cite{cc-05-global quench, cc-16-quench rev} (see \S\ref{sec:cft-naive}) with $c=1$ for the free massless Dirac fermion, finding $\tau_0 \simeq 3.26$.
This allows to fit the slope of the linear growths in the middle left panel of Fig.\;\ref{fig:SpectrumFermion} 
through (\ref{gap_r cft naive}), finding $\Delta_1=1.224 $, $\Delta_2=2.448  $ and $\Delta_3=3.662$.
We find it curious that the linear growths in the middle left panels of 
Fig.\;\ref{fig:SpectrumBoson} and Fig.\;\ref{fig:SpectrumFermion}
basically coincide, despite the diversity of the underlying models
(notice that the values of $\tau_0$ are different in the two quenches).

As done for the harmonic chain, in order to reduce the influence of the parameter $\tau_0$,
we consider the temporal evolutions of $g_r S_A$ and of the ratios $g_r/g_1$.
As for $g_r S_A$ (middle right panel of Fig.\;\ref{fig:SpectrumFermion}),
the curves corresponding to different values of $\ell$ collapse much better than 
the ones obtained for the quench in the harmonic chain (middle right panel of Fig.\;\ref{fig:SpectrumBoson})
because the values of $\ell$ considered for this fermionic chain are large enough. 
Nonetheless, the zooms in the insets show that these curves are distinguishable.

Also for this quench the temporal evolutions of the ratios $g_r/g_1$ in the entanglement spectrum 
(bottom panel of Fig.\;\ref{fig:SpectrumFermion}) display the most interesting features. 
When $t/\ell < 1/2$, curves having different $\ell$'s nicely collapse identifying neatly
plateaux that correspond to strictly positive integers. 
The same positive integers are obtained also in the asymptotic regime of long time and large $\ell$.
We checked numerically this result by plugging into the code employed to study $g_r/g_1$
the asymptotic values for $\eta_k$ found in \cite{ep-07-local-quench},
that will be reported later in \S\ref{sec:long time} (see (\ref{Tlarge-t-sinsin})).
The plateaux before $t/\ell \simeq 1/2$ and for long time 
should correspond to the ratios between the conformal dimensions in the spectrum
of the underlying CFT, and the data agree with the expected results for 
a massless Dirac fermion with free boundary conditions. 
We remark that in our analysis we considered only the particle-hole excitations.
It would be interesting to improve this numerical analysis by including higher gaps or by considering 
other configurations.

In some spin chains at the critical point, the entanglement spectrum for an interval at equilibrium
has been studied numerically in \cite{lauchli-spectrum} finding the conformal spectrum of a boundary CFT 
with free boundary conditions. 
In the Ising model, the entanglement spectrum after a global quench has been considered in 
\cite{tagliacozzo-torlai} and also in this model the gaps close as time evolves 
when the evolution is determined by a critical hamiltonian.

 \begin{figure}[t!]
\vspace{.2cm}
\hspace{.5cm}
\includegraphics[width=.9\textwidth]{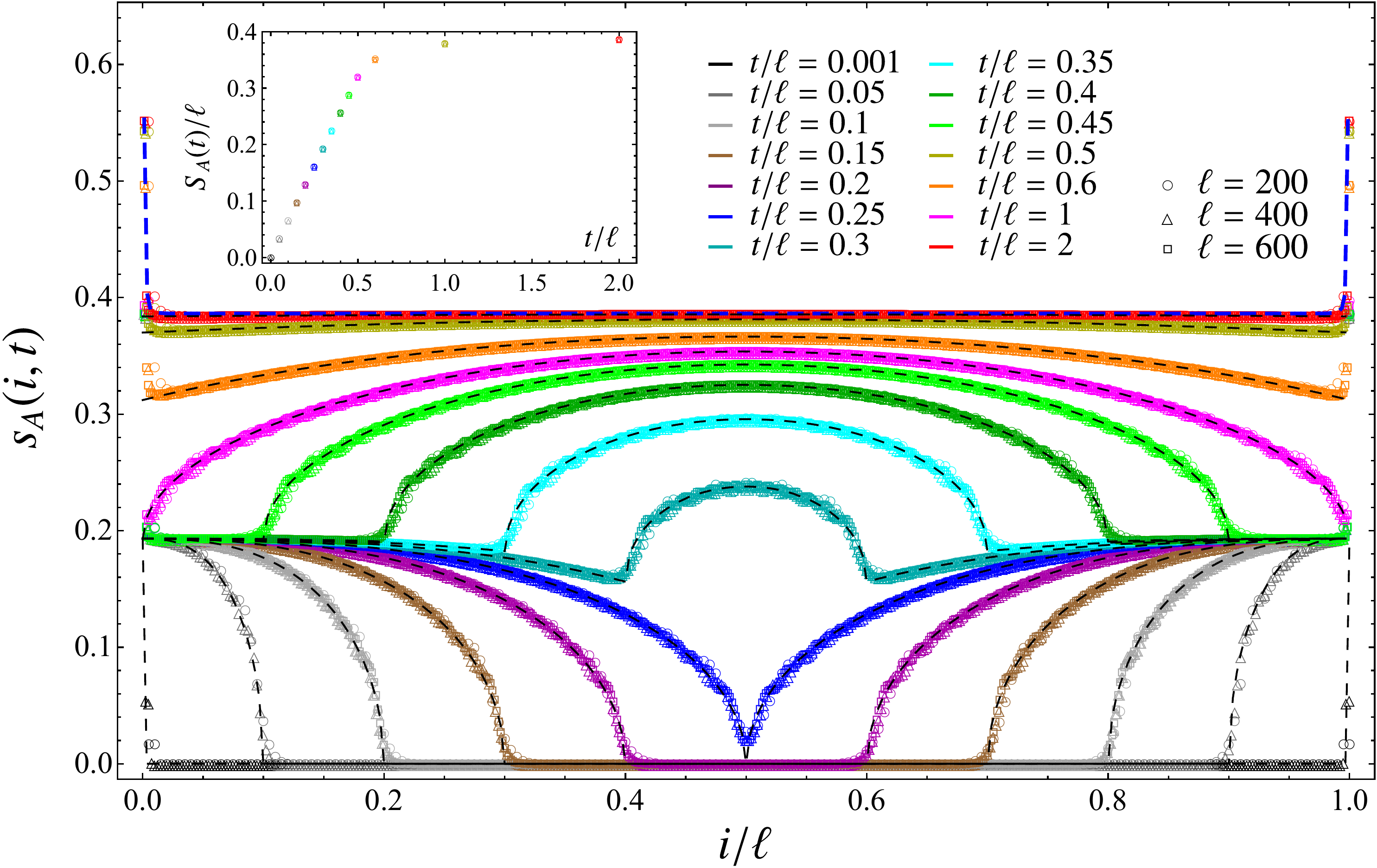}
\vspace{-.0cm}
\caption{
Chain of free fermions:
Temporal evolution after the global quench of the contour for the entanglement entropy 
of an interval made by $\ell$ sites,
evaluated from (\ref{contour from mpf}), (\ref{EE fermions}) and (\ref{mpf fermion A-case}).
The inset shows the entanglement entropy for certain values of $t$, according to the same colour code of the main plot. 
The black dashed lines are obtained from the quasi-particle picture, 
by using (\ref{QPcontourguess}) with $f_0(x)=0$ and other expressions discussed in \S\ref{sec:QPfermions}. 
The blue dashed curve is obtained from (\ref{contour large_t})
and it does not change significantly for the values of $\ell$ corresponding to these data. 
}
\label{fig:ContourFermion}
\end{figure}

 \begin{figure}[t!]
\vspace{.2cm}
\hspace{2.5cm}
\includegraphics[width=.68\textwidth]{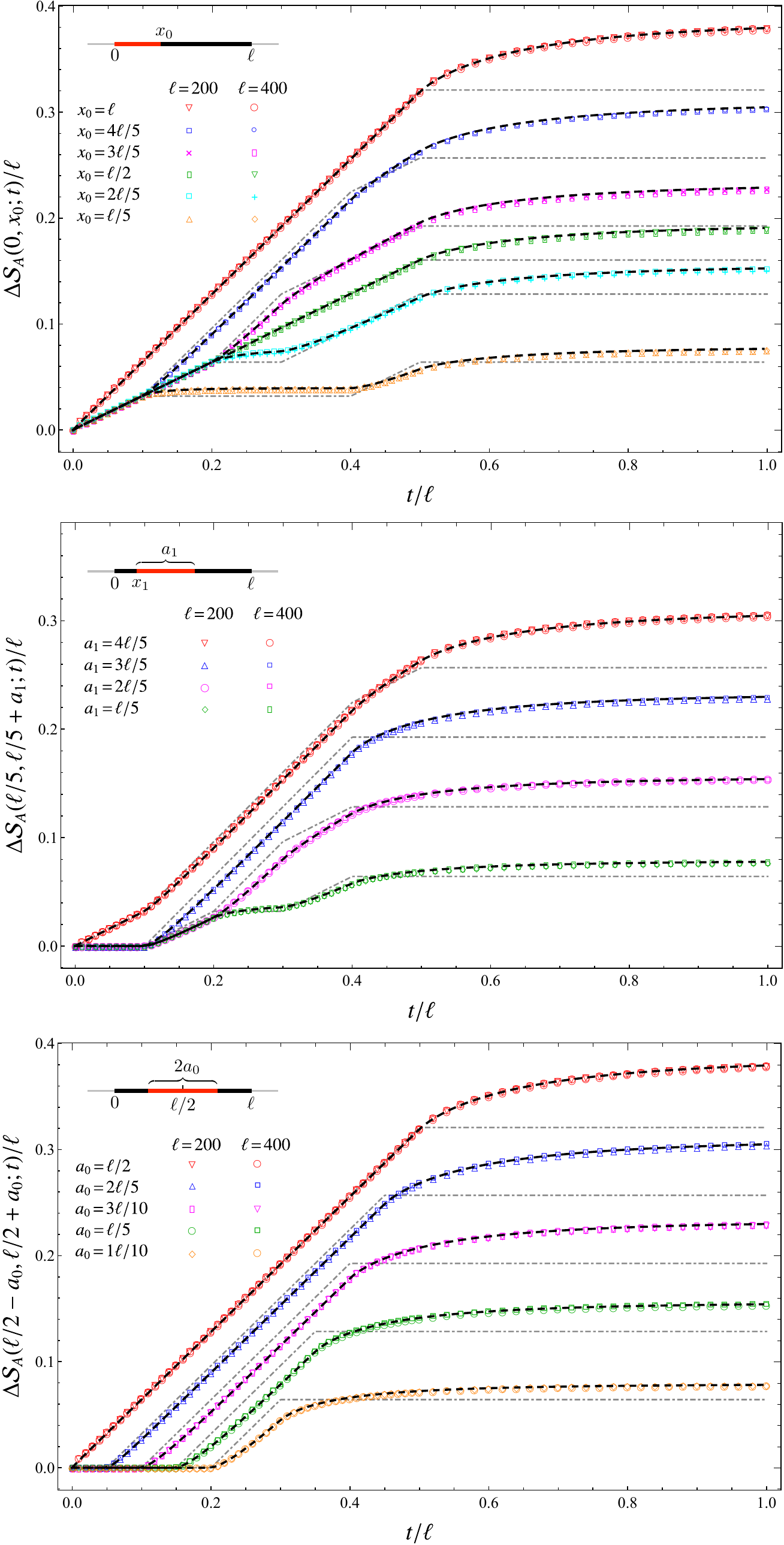}
\vspace{-.0cm}
\caption{
Chain of free fermions:
Temporal evolution of (\ref{integrated contour x1x2 discrete sub}) for $n=1$.
}
\label{fig:IC_fermions}
\end{figure}

\subsection{A contour function from the quasi-particle picture}
\label{sec:QPfermions}

In order to construct a contour function for the entanglement entropy by employing the quasi-particle picture,
we need to know the dispersion relation of the model after the quench and the entropy density in momentum space of the quasi-particles
(see \S\ref{sec:QP}), which are quantities corresponding to the entire system.
The dispersion relation of the hamiltonian (\ref{H_ff homo}), which determines the evolution of the system after the quench, 
is given by $\omega(\theta)=\cos\theta$, where $-\pi\leqslant \theta\leqslant \pi$ \cite{lieb-schulz-ExactlySolvModels}.
In order to compute the entropy density, let us consider the asymptotic state of the system after the quench. 
For $t\to\infty$, the correlators are given by (\ref{Cinf matrix def}) with $i$ and $j$ labelling two generic sites of the infinite chain. 
Considering a periodic chain made by an even number $L$ of sites,
a circulant matrix $\widetilde{C}^{(\infty)}_{i,j}$ with $i,j \in [ - L/2, L/2]$ is obtained, 
whose eigenvalues are $\zeta_k=\left[1+\cos( 2\pi k /L)\right]/2$, with $k\in [ - L/2, L/2]$. 
Plugging these eigenvalues into (\ref{EE fermions}), one finds the asymptotic entropy 
\be
\label{asymptotic entropy}
S^{(\infty)}
=
-\,2\!\! \sum_{k= -L/2}^{L/2} \!
\Bigg\{
\cos^2\left(\frac{\pi k}{L}\right)  \log\left[\cos\left(\frac{\pi k}{L}\right)\right]  +    \sin^2\left(\frac{\pi k}{L}\right) \log\left[\sin\left(\frac{\pi k}{L}\right)\right] 
\Bigg\}\,.
\ee 
After introducing $\theta= 2\pi k/L$, we take the limit $L\to\infty$ of $S^{(\infty)} / L$ with $S^{(\infty)}$ given by (\ref{asymptotic entropy}),
and the result provides the entropy density $s^{(\infty)}(\theta)$ as follows
\be
\label{asymptotic entr density FF}
-\frac{1}{\pi}
\int_{-\pi}^\pi\,
 \Big[ \big( \cos(\theta/2) \big)^2 \log[\cos(\theta/2)]  +   \big( \sin(\theta/2) \big)^2 \log[\sin(\theta/2)] \Big] d\theta\,
\equiv
\int_{-\pi}^\pi\,
s^{(\infty)}(\theta)\,
d\theta
\ee
These observations lead to construct a contour function from the quasi-particle picture
by employing $s^{(\infty)}(\theta)$ and $v(\theta)\equiv\omega'(\theta)= \sin \theta$ 
into (\ref{QPcontourguess}), being $\theta \in [-\pi, \pi]$.

In Fig.\;\ref{fig:ContourFermion} we show the temporal evolution of the contour 
for the entanglement entropy evaluated through the prescription of \cite{chen-vidal}
discussed in \S\ref{sec:fermions-special cases}, 
namely by employing (\ref{contour_lattice_def}), (\ref{pki condition}), 
(\ref{renyi-fermions}) and (\ref{mpf fermion A-case}).
Notice that the data corresponding to different values of $\ell$ collapse on the same curves. 
The black dashed curves are obtained from (\ref{QPcontourguess}) with $f_0(x)=0$ and (\ref{asymptotic entr density FF}) as explained above: 
they display perfect agreement with the data, except in the neighbourhoods of the endpoints. 
Similarly to the case of the harmonic chain described in \S\ref{sec:QP},
the divergencies of the contour function close to the endpoints can be reproduced by choosing a suitable expression for $f_0(x)$ 
(e.g. (\ref{contour CFT t=0}) with $n=1$ and the proper $\tau_0$, as done for the harmonic chain in the right panel of Fig.\;\ref{fig:ContourBosonQP}).
The qualitative behaviour of the curves in Fig.\;\ref{fig:ContourFermion} 
is the same observed in \cite{chen-vidal} for a different global quench
and it is roughly reprodiced also by the naive CFT formula (\ref{contour_cft_naive}) 
discussed in \S\ref{sec:cft-naive} (see the right panel in Fig.\;\ref{fig:CFTnaive}).
The blue dashed line in Fig.\;\ref{fig:ContourFermion} corresponds to the
asymptotic regime $t\to \infty$ discussed below in \S\ref{sec:long time}.
Also for this global quench the numerical data display linear divergencies 
close to the endpoints of the interval that are independent of time.

In Fig.\;\ref{fig:IC_fermions} we show the temporal evolution of (33) with $n = 1$ 
for the same choices of $(i_1,i_2)$ considered in Fig.\;\ref{fig:IC_bosons}. 
The qualitative behaviour of the curves is similar to the corresponding curves obtained 
for the global quench in the harmonic chain, hence
the qualitative arguments reported \S\ref{sec:QP} about the changes of the slopes holds also in this case. 
On the other hand, for $t/\ell>1/2$  the asymptotic values for long times are different from the ones observed
in the harmonic chain, as expected from the fact that the asymptotic value of the entanglement entropy depends
both on the model and on the initial state.
We find worth remarking that for the chain of free fermions we have access to larger values of $\ell$ with respect to the harmonic chain,
hence the curves corresponding to different values of $\ell$ display a better collapse than the ones shown in Fig.\;\ref{fig:IC_bosons} for the harmonic chain,
capturing the asymptotic curves for $\ell \to \infty$.
The black dashed lines,
obtained from the quasi-particle picture analysis by using (\ref{integrated contour x1x2 discrete}), (\ref{integrated contour x1x2 discrete sub}) and (\ref{QPcontourguess}),
display a very good agreement with the numerical data.
The data are also compared against the corresponding naive CFT expression (\ref{IntegContS2sub}) (grey dotted-dashed curves),
where $\tau_0 = 3.26 $ has been used.  
%


\subsection{Long time regime}
\label{sec:long time}

In the asymptotic regime $t \to \infty$,  
the imaginary part of the correlation matrix (\ref{correl}) vanishes,
as already remarked above.
In particular, $C_{i,j}(t) \to C^{(\infty)}_{i,j}$
being $C^{(\infty)}_{i,j}$
the real symmetric and tridiagonal matrix defined in (\ref{Cinf matrix def}),
which has the same value along a given diagonal.
Because of the simple structure of $C^{(\infty)}_{i,j}$,
analytic expressions for its eigenvalues and for the corresponding eigenvectors can be written.
The eigenvalues of $C^{(\infty)}$ read \cite{ep-07-local-quench}
\be
\label{fermions large time analytic e-val}
\zeta_k = \frac{1+ \cos \theta_k}{2} 
\;\;\;\; \qquad \;\;\;\;
\theta_k \equiv \frac{\pi \, k}{\ell+1}\,.
\ee

 \begin{figure}[t!]
\vspace{.2cm}
\hspace{-.8cm}
\includegraphics[width=1.05\textwidth]{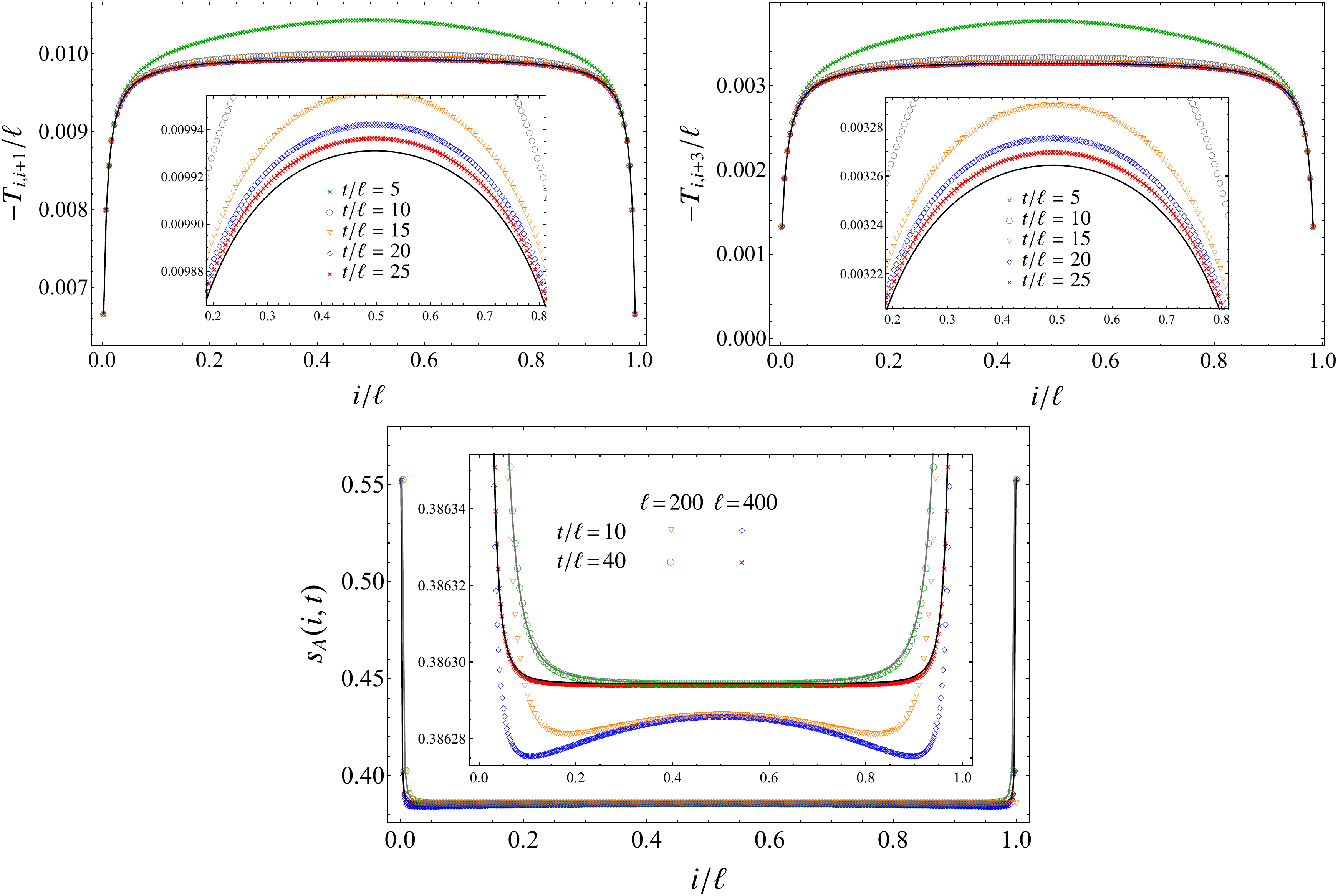}
\vspace{-.5cm}
\caption{
Chain of free fermions:
Long time regime after the global quench (see \S\ref{sec:long time}).
Top: First diagonal (left) and third diagonal (right) of the entanglement hamiltonian matrix 
of an interval with $\ell=200$.
The data points have been found by using (\ref{T-decomposition-peschel}), 
while the solid black curves correspond to (\ref{T large_t cos-sin step2}) with $p=1$ (left) and $p=3$ (right).
Bottom: Contour for the entanglement entropy for two values of $\ell$
and two (large) values of $t$ for each length. 
The data points, obtained from 
(\ref{contour from mpf}), (\ref{EE fermions}) and (\ref{mpf fermion A-case}),
agree with the solid lines found from (\ref{contour large_t}) with $n=1$.
}
\label{fig:H_fermion_large_time}
\end{figure}

\begin{figure}[t!]
\vspace{.2cm}
\hspace{-.8cm}
\includegraphics[width=1.05\textwidth]{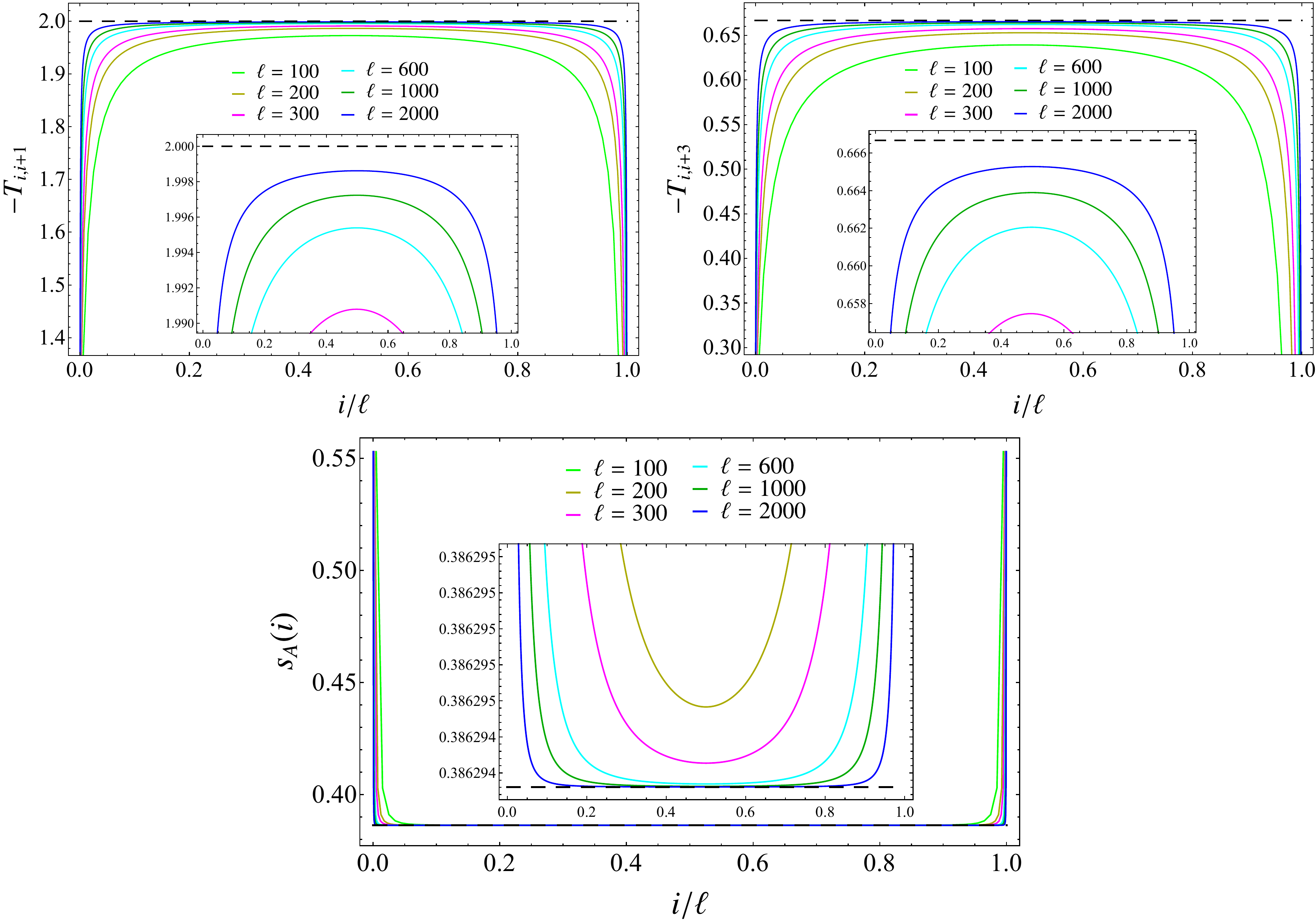}
\vspace{-.5cm}
\caption{
Chain of free fermions:
$\ell \to \infty$ in the long time regime after the global quench (see \S\ref{sec:long time}).
Top: First diagonal (left) and third diagonal (right) of the entanglement hamiltonian matrix for increasing values of $\ell$.
The solid curves are obtained from (\ref{T large_t cos-sin step2}) and the dashed lines correspond to (\ref{HA td diag}).
Bottom: Contour for the entanglement entropy.
The solid lines are obtained from (\ref{contour large_t}) with $n=1$, while
the horizontal dashed line corresponds to (\ref{EE large_t flat}).
}
\label{fig:H_fermion_TD}
\end{figure}

In order to study the matrix occurring in the entanglement hamiltonian, 
we also need the eigenvectors corresponding to the eigenvalues (\ref{fermions large time analytic e-val}), 
that are given by
\be
\label{fermions large time analytic e-vec}
u_k(r) = \frac{\sin( r\,\theta_k)}{\sqrt{(\ell+1)/2}} 
\; \qquad\;
\;\;\;\qquad \;\;\;
1\leqslant r \leqslant \ell
\ee
where the normalisation condition $\sum_r u_k(r)^2 =1$ has been imposed. 
Then, the orthogonal matrix $\widetilde{U}$ having the eigenvector $u_k(r)$ 
in (\ref{fermions large time analytic e-vec}) as $k$-th column can be constructed
and, by employing this matrix in (\ref{T-decomposition-peschel}), 
the generic element of the $\ell \times \ell$ entanglement hamiltonian matrix reads
\be
\label{Tlarge-t-sinsin}
T_{i,j} 
= \frac{2}{\ell+1}\sum_{k=1}^\ell \eta_k \, \sin(i \theta_k) \, \sin(j \theta_k)
\;\;\qquad\;\;
\eta_k = 2 \log\!\big[\! \tan(\theta_k/2)\big]
\ee
where $1\leqslant i,j \leqslant \ell$ and the single particle entanglement energies have been obtained  from (\ref{peschel02 H_A_gen}) and (\ref{fermions large time analytic e-val}). 
Notice that $\sin(r \theta_k)  =0 $ when $r=0$ or $r=L+1$.
The matrix $T$ in (\ref{Tlarge-t-sinsin}) is symmetric, as expected; 
hence we can focus on its $p$-th diagonal with $0 \leqslant p\leqslant \ell-1$, whose generic element is
\be
\label{T large_t cos-sin step2}
T_{i,i+p} 
\,=\,
\frac{1}{\ell+1}
 \sum_{k=1}^\ell \eta_k  \Big(\! \cos(p\,\theta_k) \big[ 1 - \cos(2 i \theta_k) \big] + \sin(p\,\theta_k) \,\sin(2 i \theta_k)  \Big) 
\qquad
 1\leqslant i \leqslant \ell - p\,.
\ee
Notice that $\theta_{\ell - k+1} = \pi - \theta_{k}$, 
and this implies $\eta_{\,\ell - k+1} =  - \eta_{\,k}$.
By using this property and splitting the sum in (\ref{T large_t cos-sin step2}) 
into a sum from $1$ to $\ell / 2$ and a sum from $\ell/2+1$ to $\ell$, 
one can show that the even diagonals vanish, namely $T_{i, i+2p} = 0 $ for non negative integers $p$. 
Furthermore, simple trigonometric identities allow to observe that the $p$-th diagonal
is symmetric with respect to its middle point,
namely $T_{\ell-i-p+1,\ell-i+1}=T_{i,i+p}$.

In the top panels of Fig.\;\ref{fig:H_fermion_large_time},
the expression (\ref{T large_t cos-sin step2}) has been checked against numerical data corresponding to large values of $t$ for $p=1$ (left) and $p=3$ (right).
The agreement is satisfactory, but the insets show that larger values of $t$ should be taken in order to improve it.

The contour for the entanglement entropies can be written by employing
(\ref{contour from mpf}), (\ref{mpf fermion A-case}) and (\ref{fermions large time analytic e-vec}),
together with the observations made above.
This gives
\be
\label{contour large_t}
s^{(n)}_A(i)
=
\sum_{k=1}^\ell s_n(\zeta_k) \,u_k(i)^2
=
\frac{2}{\ell+1}
\sum_{k=1}^\ell s_n(\zeta_k) \, \big[ \! \sin(i \theta_k) \big]^2
\ee
where $s_n(\zeta_k)$ is obtained by evaluating (\ref{renyi-fermions}) or (\ref{EE fermions}) 
for the $\zeta_k$'s given in (\ref{fermions large time analytic e-val}).
The expression (\ref{contour large_t}) corresponds to the black dashed curve in Fig.\,\ref{fig:ContourFermion} and to the solid lines in the bottom panel of Fig.\;\ref{fig:H_fermion_large_time}.
Agreement is observed with the data points obtained for large values of time.

It is useful to consider  the limit $\ell \to \infty$ of the above expressions
holding in the asymptotic regime of long time. 
As the entanglement hamiltonian matrix element (\ref{T large_t cos-sin step2}),
by neglecting the highly oscillating terms coming from $\cos(2 i \theta_k) $ and $\sin(2 i \theta_k) $, for $i$ far enough from $1$ and $\ell-p$,
one finds (see also section 7.1 of \cite{ep-rev})
\be
\label{HA td diag}
-T_{i,i+p} 
\;\longrightarrow\;\,
-\,\frac{1}{\pi} \int_0^\pi  \eta(\theta) \, \cos(p\,\theta) \, d\theta
\,=\,
\left\{ \begin{array}{ll}
0 \hspace{1cm} & \textrm{even $p$}
\\
\rule{0pt}{.5cm}
2/p & \textrm{odd $p$}
\end{array}\right.
\ee
where $\eta(\theta) \equiv 2 \log[\tan(\theta/2)]$ (see (\ref{Tlarge-t-sinsin})).

The limit (\ref{HA td diag}) is checked in the top panels of Fig.\;\ref{fig:H_fermion_TD} 
for $p=1$ (left) and $p=3$ (right).

As for the entanglement entropy, from (\ref{EE fermions}) and the eigenvalues in 
(\ref{fermions large time analytic e-val}), one obtains
\be
\label{EE large_t flat}
\frac{S_A}{\ell} \,\longrightarrow\, \frac{1}{\pi} \int_0^\pi s(\theta) \,d\theta =   -1 + \log 4 \simeq 0.386
\;\; \qquad\;\;
\ell \to \infty
\ee
that is the flat value reached for $t \to \infty$ both in the inset and in the main plot of Fig.\,\ref{fig:ContourFermion}. 
Also the horizontal dashed line in the bottom panel of Fig.\,\ref{fig:H_fermion_TD}
corresponds to (\ref{EE large_t flat}).
The consistency of this result can be checked by employing (\ref{contour large_t}), that for $\ell \to \infty$ gives
\be
\label{contour large-time td-limit}
-\frac{2}{\pi} \int_0^\pi \Big[ \big( \cos(\theta/2) \big)^2 \log[\cos(\theta/2)]  +   \big( \sin(\theta/2) \big)^2 \log[\sin(\theta/2)] \Big] d\theta
=
 -1 + \log 4 
\ee
where we used that $1/2$ is the mean value constant coming from  $[ \sin(i \theta_k)]^2$.
For the R\'enyi entropies, from (\ref{renyi-fermions}) one finds
\be
\frac{S^{(n)}_A}{\ell}  \,\longrightarrow\, 
\frac{1}{(1-n)\pi} \int_0^\pi \! 
\log \! \Big[ \big( \cos(\theta/2) \big)^{2n} +\big( \sin(\theta/2) \big)^{2n} \Big]\,d\theta
\ee
that can be evaluated for explicit values of $n$, finding e.g. 
$\log(24 - 16 \sqrt{2})$ for $n=2$ and $\log(4/3)$  for $n=3$.

\begin{figure}[t!]
\vspace{.2cm}
\hspace{-.8cm}
\includegraphics[width=1.05\textwidth]{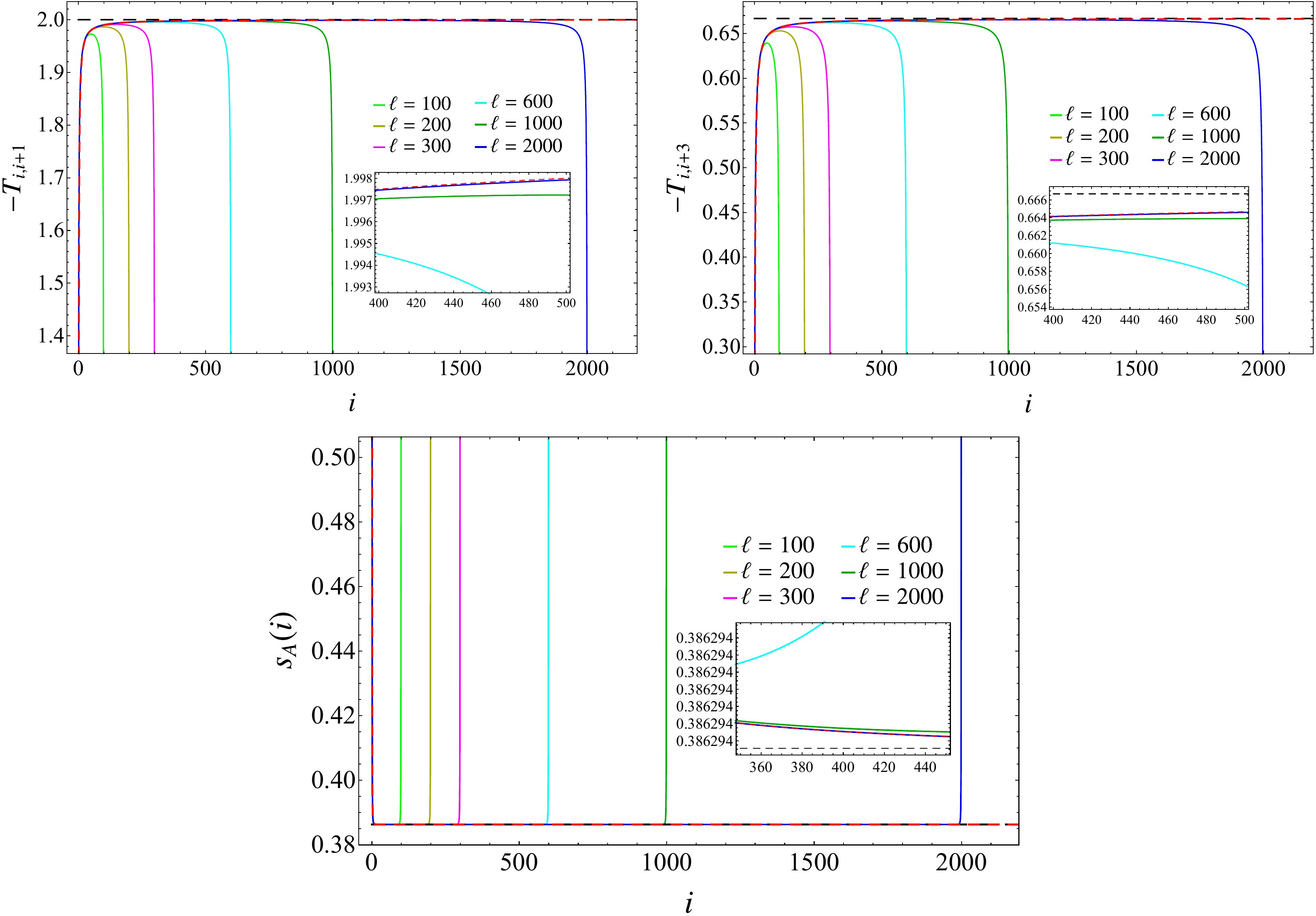}
\vspace{-.5cm}
\caption{
Chain of free fermions:
$\ell \to \infty$ in the analytic expressions for the long time regime after the global quench (see \S\ref{sec:long time}).
Top: First diagonal (left) and third diagonal (right) of the entanglement hamiltonian matrix. 
The solid curves are obtained from (\ref{T large_t cos-sin step2}), while the black and the red dashed lines 
correspond respectively to (\ref{HA td diag}) and (\ref{HA td anal}).
Bottom: Contour for the entanglement entropy.
The solid lines are obtained from (\ref{contour large_t}) with $n=1$, 
while the black and the red dashed lines correspond respectively  to (\ref{EE large_t flat}) and (\ref{contour large-time td-limit anal}).
}
\label{fig:H_fermion_TD_peschel}
\end{figure}

The previous analysis has provided the values of the entanglement hamiltonian and of the contour for the entanglement entropies
in the central part of the interval.
In the following we improve this analysis by capturing also the behaviour of these quantities close to one of the endpoints. 
Here we only mention the results, 
reporting the details of their derivation in the Appendix\;\ref{app:larget-largel-derivation}\footnote{These results and their derivation reported in the Appendix \ref{app:larget-largel-derivation} 
have been obtained by Ingo Peschel.
We are grateful to him for allowing us to include his analysis in this manuscript. 
}.

When $\ell \to \infty$, the expression in (\ref{Tlarge-t-sinsin}) becomes
\be
\label{HA td anal}
T_{i,j}
\,=\,
\left\{ \begin{array}{ll}
0
& \hspace{1.5cm}  \textrm{even $|i\pm j|$}
\\
\rule{0pt}{.5cm}
\displaystyle
2\left( \frac{1}{i+j}\,-\,\frac{1}{|i-j|} \right) 
& \hspace{1.5cm}  \textrm{odd $|i\pm j|$}
\end{array}\right. 
\ee
where $1\leqslant i \leqslant \ell$ and $1\leqslant j \leqslant \ell$.
This implies that  the generic element of the $p$-th diagonal 
is $T_{i,i+p} = 0$ when $p$ is even and
 $T_{i,i+p} = -4i /[p(2i+p)]$ when $p$ is odd.
 Taking $i \to \infty$ in these expressions,  
 one finds $T_{i,i+p}\approx-2/p + 1/i + O(1/i^2)$ for odd $p$;
 hence (\ref{HA td diag})  is recovered.
 A numerical check of (\ref{HA td anal}) is reported in the top panels of Fig.\,\ref{fig:H_fermion_TD_peschel}.

The contour for the entanglement entropy can be analysed in a similar way.
Taking $\ell\to\infty$ in (\ref{contour large_t}) with $n=1$, one obtains
\be
\label{contour large-time td-limit anal}
s_A(i)
\,=\,
\log 4 -1+ \frac{1}{2i (4 i^2-1)}
\ee
being $i = 1,2,3, \dots$ up to infinity. 
The limit $i\to \infty$ of this expression gives $s_A(i) \to \log 4 -1 + O(1/i^3)$,
which allows to recover (\ref{contour large-time td-limit})
and also to find the power law decay $O(1/i^{3})$ for the contour function $s_A(i)$ 
close to the endpoint. 
The expression (\ref{contour large-time td-limit anal}) has been checked 
numerically in the bottom panel of Fig.\,\ref{fig:H_fermion_TD_peschel}.

In this analysis the interval becomes a semi-infinite line; 
hence we can study the behaviour of the entanglement hamiltonian matrix 
and of the contour function only around one of the endpoints of the interval.

%
%

\section{Conclusions}
\label{sec:conclusions}

We studied the temporal evolution of the entanglement hamiltonian of an interval after a global quantum quench in two simple free lattice models: the harmonic chain and a chain of free fermions. 
In the harmonic chain (in the thermodynamic limit) we considered a quench of the frequency parameter such that the evolution hamiltonian is massless. In the chain of free fermions at half filling, 
we explored the global quench introduced in \cite{ep-07-local-quench}, 
where the initial state is the ground state of a dimerised chain and the evolution hamiltonian is determined by the homogenous hopping hamiltonian. 
In these free models, the time dependent entanglement hamiltonian is a quadratic operator; 
hence it is fully determined by a matrix that can be written explicitly as a function of the matrix whose elements are the two point correlators \cite{Peschel-Chung-oscillators, Peschel-Chung-00, ch-rev, banchi-pirandola-15, peschel-03-modham}.
The entanglement hamiltonian matrix can be decomposed in terms of a diagonal matrix containing the single particle entanglement spectrum and another matrix that is symplectic for the harmonic chain or orthogonal for the chain of free fermions. 
The single particle entanglement spectrum and the matrix occurring in this decomposition can be employed to construct also a contour for the entanglement entropies \cite{chen-vidal, cdt-17-contour}, that encodes information about the spatial structure of the bipartite entanglement. 
For static configurations, some entanglement hamiltonian matrices in the harmonic chain and in a chain of free fermions have been explored in \cite{Peschel-Chung-oscillators, Peschel-Chung-00, ch-rev, banchi-pirandola-15, peschel-03-modham, Arias-16, Arias-17, Arias-18, eisler-peschel-17, ep-18, etp-19}.

In this manuscript we explored  the temporal evolution of the entanglement hamiltonian matrices 
and of the contours for the entanglement entropy of an interval made by $\ell$ sites after the global quenches mentioned above. 
During the temporal evolution, the entanglement hamiltonian matrix is not block diagonal in the harmonic chain and it is complex in the chain of free fermions. 
All the diagonals of the $\ell \times \ell$ blocks composing the entanglement hamiltonian matrix in the harmonic chain and of the $\ell \times \ell$ real and imaginary parts of the entanglement hamiltonian matrix in the chain of free fermions (with the obvious exception of the ones that vanish at half filling)
display a non trivial temporal evolution.
Bands of diagonals around the main diagonal in these $\ell \times \ell$ matrices can be identified whose width growths linearly with velocity equal to one.
This is observed also in the covariance matrix for the harmonic chain 
and in the correlation matrix for the chain of free fermions, where all the elements of each diagonal are equal.

The analytic results obtained in CFT for the entanglement hamiltonian of a semi-infinite line  \cite{ct-16} 
have been exploited to write expressions for the finite interval in the continuum limit 
that reproduce qualitatively the behaviour of the numerical data of some quantities in some regimes.
In particular, the linear growth of the gaps in the entanglement spectrum before $t/\ell \simeq 1/2$ 
and the qualitative temporal evolution of the contour for the entanglement entropy have been obtained. 
In our numerical analysis we have also observed that the temporal evolution of the ratios of the gaps in the entanglement spectrum can be employed to read the ratios in the conformal spectrum of the underlying CFT when the evolution hamiltonian is critical. 
This observation holds both for $t/\ell < 1/2$ (as predicted in \cite{ct-16}) and in the asymptotic regime of long time. 
Furthermore, it seems true also for different initial states. 
It would be interesting to further explore this idea by considering 
global quenches governed by different critical hamiltonians.

The quasi-particle picture of \cite{cc-05-global quench} has been employed to obtain
semi-empirical analytic expressions of the contour for the entanglement entropy,
finding reasonable agreements with the numerical data, in particular for the chain of free fermions, 
where larger subsystems have been studied. 
For the harmonic chain, the formula for the contour is valid also for the global quenches 
where the mass in the evolution hamiltonian is non vanishing; hence it would be interesting to test it also for these protocols. 
For the global quench in the chain of free fermions, the entanglement hamiltonian matrix 
and the contour for the entanglement entropy have been studied in the asymptotic regime of $t \to \infty$.

The analysis of entanglement hamiltonians and 
the contours for the entanglement entropies can be extended in many interesting directions. 
For instance, it would be useful to find some analytic expressions describing 
the asymptotic regime of long time for the global quench considered here in the harmonic chain, 
like the ones discussed in \S\ref{sec:long time} in the chain of free fermions. 
Other kinds of global quenches can be considered 
or also quenches of different nature, like the local quenches \cite{cc-07-local quench}.
An interesting development is given by more complicated bipartitions,
like the one where  $A$ is made by disjoint intervals \cite{ch-09-eh-2int, ct-16, Arias-18}
whose entanglement entropies have been studied in CFT, lattice models 
\cite{2 disjoint intervals} and holography \cite{2 disjoint intervals - holog}.
Connections could be established also with some recent results in harmonic lattices
\cite{sierra-18-19}.

It is worth studying the temporal evolutions of 
the entanglement hamiltonians and 
of the contours for the entanglement entropies after global and local quenches
also in interacting models and in higher dimensions.
Indeed, in this manuscript we have highlighted the role of the 
entanglement gaps to obtain the conformal dimensions of the underlying CFT 
when the evolution hamiltonian is critical;
hence it would be interesting to test this idea also in interacting models. 
In the context of the gauge/gravity correspondence, the bit threads approach 
to the holographic entanglement entropy \cite{hf-bit-threads}
suggests a possible gravitational dual of the contour for the entanglement entropy 
\cite{contour-holog}.

The entanglement hamiltonians and their dynamics are very important 
also because they are the building blocks to explore other interesting quantities that capture different
aspects of entanglement. 
For instance, the bipartite entanglement of mixed states can be quantified through 
the partial transpose and the logarithmic negativity \cite{neg-def}, 
which have been studied also in quantum field theories \cite{neg-qft} 
and in some lattice models \cite{neg-lattice}.
Other interesting directions to explore concern excited states \cite{sierra-ee-excited}
and quantities involving them, like e.g. the relative entropy \cite{rel-ent}.



\section*{Acknowledgements}

It is our pleasure to thank Andrea Coser, Viktor Eisler, Ingo Peschel, Giulia Piccitto and Luca Tagliacozzo for important insights.
We are also grateful to Horacio Casini, Fabian Essler, Paul Fendley, Juan Maldacena, Sara Murciano, Giuseppe Mussardo, 
Mohammad Ali Rajabpour, Germ\'an Sierra, Hubert Saleur and Jacopo Surace for useful discussions. 
We thank an anonymous referee for observations
that have led us to the contour function from the quasi-particle picture discussed in \S\ref{sec:QPfermions}.
This work started at the Galileo Galilei Institute (GGI) in Florence during the programme {\it Entanglement in Quantum Systems}
in June and July 2018: we acknowledge GGI for financial support and the stimulating environment. 
RA thanks Associazione di Fondazioni e di Casse di Risparmio (ACRI) for financial support.


\begin{appendices}

\section*{Appendices}

\section{On the Williamson's decomposition of $H_A$}
\label{sec_app:EH derivation}

In the harmonic lattices described in \S\ref{sec:EHwilliamson_total},
the entanglement hamiltonian matrix (\ref{banchi-EH}) and its equivalent form (\ref{eh-cov-mat-B}) 
can be obtained from its Williamson's decomposition (\ref{H_A Williamson_dec}).
In this Appendix we discuss the details of this derivation.

By employing the Williamson's decomposition  (\ref{williamson th gammaA})
of the reduced covariance matrix $\gamma_A$, one finds
\be
\label{D2J diag}
\textrm{i}\mathcal{D}_\textrm{\tiny d} \,J = (J W) \, \textrm{i} J \gamma_A \,(J W)^{-1}
\;\; \qquad \;\;
\textrm{i}J\,\mathcal{D}_\textrm{\tiny d}  = (W^{\textrm t} J)^{-1} \,\textrm{i} \gamma_A J \,(W^{\textrm t} J)\,.
\ee
Then, since
$(\textrm{i}\mathcal{D}_\textrm{\tiny d} J)^2 = (\textrm{i}J\mathcal{D}_\textrm{\tiny d})^2 = \mathcal{D}_\textrm{\tiny d}^2$,
we have that
\be
\label{D2 powers}
\mathcal{D}_\textrm{\tiny d}^{2k} = (\textrm{i}\mathcal{D}_\textrm{\tiny d} J)^{2k} = (\textrm{i}J\mathcal{D}_\textrm{\tiny d})^{2k}
\;\; \qquad \;\;
\mathcal{D}_\textrm{\tiny d}^{2k+1} = (\textrm{i}\mathcal{D}_\textrm{\tiny d} J)^{2k+1} \,\textrm{i}J = \textrm{i}J  \,(\textrm{i}J\mathcal{D}_\textrm{\tiny d})^{2k+1}\,.
\ee
Considering a generic even function $f_{\textrm{\tiny e}}(x)$, 
the first expression in (\ref{D2 powers}) leads to
$f_{\textrm{\tiny e}}(\mathcal{D}_\textrm{\tiny d}) =
f_{\textrm{\tiny e}}(\textrm{i}\mathcal{D}_\textrm{\tiny d} J)= 
f_{\textrm{\tiny e}}(\textrm{i}J\mathcal{D}_\textrm{\tiny d})$.
By employing (\ref{D2J diag}) into this result, one obtains 
\be
\label{f_even_D2_1}
f_{\textrm{\tiny e}}(\mathcal{D}_\textrm{\tiny d})
=
(J W) \, f_{\textrm{\tiny e}}(\textrm{i}J \gamma_A) \, (J W)^{-1}
=
(W^{\textrm t} J)^{-1}  \, f_{\textrm{\tiny e}}(\textrm{i}\gamma_A J) \, (W^{\textrm t} J)\,.
\ee
Similarly, for a generic odd function $f_{\textrm{\tiny o}}(x)$,
the second expression in (\ref{D2 powers}) implies that
$f_{\textrm{\tiny o}}(\mathcal{D}_\textrm{\tiny d}) 
=f_{\textrm{\tiny o}}(\textrm{i}\mathcal{D}_\textrm{\tiny d} J) \,\textrm{i}J 
=\textrm{i}J  \, f_{\textrm{\tiny o}}(\textrm{i}J\mathcal{D}_\textrm{\tiny d}) $.
This result combined with (\ref{D2J diag}) leads to
\be
\label{f_odd 2}
f_{\textrm{\tiny o}}(\mathcal{D}_\textrm{\tiny d}) 
=
(J W) \, f_{\textrm{\tiny o}}(\textrm{i}J \gamma_A) \, (J W)^{-1} \,\textrm{i}J 
=
\textrm{i}J \,(W^{\textrm t} J)^{-1}  \, f_{\textrm{\tiny o}}(\textrm{i}\gamma_A J) \, (W^{\textrm t} J)\,.
\ee

Another useful relation can be derived by noticing that 
 $(\textrm{i} J \gamma_A)^n = J\,(\textrm{i}\gamma_A J)^n J^{-1}$ for any integer $n\geqslant 0$.
 This observation implies that, for a generic function $f(x)$, we have 
\be
\label{Jexchange-f}
f(\textrm{i} J \gamma_A) 
=
-\, J\, f(\textrm{i} \gamma_A J)\, J\,.
\ee

The relation (\ref{Ediag_Ddiag}) gives 
$\mathcal{E}_\textrm{\tiny d} = 2\, \textrm{arccoth}(2\mathcal{D}_\textrm{\tiny d} )$,  that is an odd function of $\mathcal{D}_\textrm{\tiny d}$;
hence, by specifying (\ref{f_odd 2}) to this case, we find that
\be
\mathcal{E}_\textrm{\tiny d} 
=
(J W^{\textrm t} )^{-1}
 \big[ 2\,\textrm{i} \, J  \,  \textrm{arccoth}(2\,\textrm{i}\gamma_A J) \big]
(W^{\textrm t} J)\,.
\ee
By isolating the matrix within the square brackets, one obtains
the first expression in (\ref{banchi-EH}) with $H_A$ written through its Williamson's decomposition 
(\ref{H_A Williamson_dec}) and $W_H$ given by (\ref{W_H is Wtilde}).
The second expression in (\ref{banchi-EH}) can be easily found from the first one by using (\ref{Jexchange-f}).

Let us observe that, from the relations (\ref{H_A Williamson_dec}) and (\ref{W widetildeW relation}),
we have $J^{\textrm t} H_A J  = W^{\textrm t}    \mathcal{E}_\textrm{\tiny d} W$,
which tells that the Williamson's decompositions of $\gamma_A$ and $J^{\textrm t} H_A \,J$ involve the same symplectic matrix $W$.

An equivalent way to express the entanglement hamiltonian matrix $H_A$ in terms of $\gamma_A$ is given by (\ref{eh-cov-mat-B}).
Indeed, from the Williamson's decomposition of $H_A$ in (\ref{H_A Williamson_dec}) 
with $W_H$ given by (\ref{W_H is Wtilde}) and (\ref{W widetildeW relation}),
by using that $\mathcal{E}_\textrm{\tiny d}=h(\mathcal{D}_\textrm{\tiny d}) \mathcal{D}_\textrm{\tiny d}$,
with $h(y)$ defined in the text below (\ref{eh-cov-mat-B}), one finds
\be
\label{H_A w_form_2}
H_A 
=
W^{-1} J^{-1}\, h(\mathcal{D}_\textrm{\tiny d}) \, \mathcal{D}_\textrm{\tiny d}  \,W J 
=
\big[ (JW)^{-1}\, h(\mathcal{D}_\textrm{\tiny d})\, (JW)\big]
 J^{\textrm t} (W^{\textrm t}  \, \mathcal{D}_\textrm{\tiny d}  \,W)\, J \,.
\ee
In the second step the identity matrix $\boldsymbol{1} = \widetilde{W}\, W^{\textrm t} $ has been properly inserted 
in order to recognise the Williamson's decomposition (\ref{williamson th gammaA}) for $\gamma_A$
and the relation (\ref{f_even_D2_1}) for the even function $h(y)$,
that allow to write the last expression of (\ref{H_A w_form_2}) in the form reported in the 
last step of (\ref{eh-cov-mat-B}).
Similarly, by inserting the identity matrix $\boldsymbol{1} = (W^{\textrm t}J)^{-1} (W^{\textrm t}J) $ 
into the first expression in (\ref{H_A w_form_2}) in the proper way, we find that it can written as follows 
\be
\label{H_A w_form_3}
H_A 
=
W^{-1}  \mathcal{D}_\textrm{\tiny d}  (W^{\textrm t})^{-1}
\big[ (W^{\textrm t}J)\, h(\mathcal{D}_\textrm{\tiny d})\, (W^{\textrm t}J)^{-1}\big]
\ee
where, by using  (\ref{williamson th gammaA}) and (\ref{f_even_D2_1}),
one recognises the first expression in (\ref{eh-cov-mat-B}).

Another equivalent expression for the entanglement hamiltonian matrix $H_A$ can be found by
observing that the relation (\ref{Ediag_Ddiag}) between $\mathcal{D}$ and $\mathcal{E}$ gives
\be
\label{D2andE2diag-app}
\mathcal{D}_\textrm{\tiny d} 
\,=\,
\frac{1}{2} \, \coth (  \mathcal{E}_\textrm{\tiny d}/2 )
\;\; \qquad \;\;
\mathcal{E}_\textrm{\tiny d} 
\,=\, 
\log\bigg[
\coth\bigg(
\frac{1}{2} \log(2\,\mathcal{D}_\textrm{\tiny d} )
\bigg)
\bigg]\,.
\ee
By employing these relations, (\ref{williamson th gammaA}), (\ref{W widetildeW relation}) and (\ref{H_A Williamson_dec}),
we arrive to
\be
\mathcal{D}_\textrm{\tiny d} 
=
(W^{\textrm t})^{-1}  \, \gamma_A  \,W^{-1}
=
 \widetilde{W}  \, \gamma_A \, \widetilde{W}^{\textrm t}
\;\; \qquad \;\;
\mathcal{E}_\textrm{\tiny d} 
=
(\widetilde{W}^{\textrm t})^{-1}  \, H_A  \,\widetilde{W}^{-1}
=
 W  \, H_A \, W^{\textrm t}\,.
\ee
Plugging these expressions into (\ref{D2andE2diag-app}), one easily finds 
\be
\gamma_A
\,=\,
\frac{1}{2} \; W^{\textrm t} \bigg[ \coth \bigg(  \frac{1}{2}\, W  \, H_A \, W^{\textrm t} \bigg) \bigg] W
\ee
and
\be
H_A
\,=\, 
\widetilde{W}^{\textrm t}
\, \bigg\{\!
\log\!\bigg[
\coth\bigg(
\frac{1}{2} \log(2\, \widetilde{W}   \gamma_A  \widetilde{W}^{\textrm t} )
\bigg)
\bigg]
\bigg\}
\,\widetilde{W}
\ee
which provides another form to express the relation between the reduced covariance matrix $\gamma_A$ 
and the entanglement hamiltonian matrix $H_A$.

\section{On the numerical determination of the symplectic matrix $W$}
\label{app:cholesky}

The symplectic matrix $W$ occurring in the Williamson's decomposition (\ref{williamson th gammaA}) 
of the reduced covariance matrix $\gamma_A$ plays a crucial role throughout this manuscript. 
In this Appendix we discuss the method employed to construct $W$ numerically.

The reduced covariance matrix $\gamma_A$ is symmetric and positive definite;
hence, according to the Cholesky decomposition, there exists a unique way to write $\gamma_A$ as follows 
\be
\label{cholesky gamma}
\gamma_A = L_\gamma L_\gamma^{\textrm{t}}
\ee
where $L_\gamma$ is a real lower triangular matrix. 
Plugging (\ref{cholesky gamma}) into (\ref{Jgamma relations}), one obtains
\be
\label{LJgammaJL}
(J L_\gamma )^{\textrm{t}} \,\gamma_A\, (J L_\gamma )
= 
(W L_\gamma^{-\textrm{t}})^{-1} \,\mathcal{D}_{\textrm{\tiny d}}^2  \,(W L_\gamma^{-\textrm{t}})
=
(\widetilde{W}   L_\gamma)^{-1} \, \mathcal{D}_{\textrm{\tiny d}}^2 \, (\widetilde{W}   L_\gamma)\,.
\ee
The first expression in this sequence of equalities is 
a real symmetric matrix that can be diagonalised by 
an orthogonal matrix $W_\gamma$, namely
$(J L_\gamma )^{\textrm{t}} \,\gamma_A (J L_\gamma ) = 
W_\gamma^{-1}  \Lambda \,W_\gamma$,
being $\Lambda$ a diagonal matrix. 
Comparing this formula with (\ref{LJgammaJL}), we find that 
$\Lambda =\mathcal{D}_{\textrm{\tiny d}}^2$
and $ W_\gamma =  \mathcal{N}\,W L_\gamma^{-\textrm{t}}$,
where  $\mathcal{N}$ is an invertible diagonal matrix.
By imposing that $W_\gamma$ is orthogonal, 
one obtains $ \mathcal{N} = \mathcal{D}_\textrm{\tiny d}^{1/2} $.
Thus, the matrix $W$ can be written as
\be
\label{W decomposition W_gamma}
W 
=  \mathcal{D}_\textrm{\tiny d}^{-1/2} \,W_\gamma \,L_\gamma^{\textrm{t}} \,.
\ee
By employing the fact that $W_\gamma$ is orthogonal and the Cholesky decomposition (\ref{cholesky gamma}), 
it is straightforward to verify that (\ref{W decomposition W_gamma})  satisfies (\ref{williamson th gammaA}), as expected. 
The fact that (\ref{W decomposition W_gamma}) is symplectic can be checked by following the procedure 
adopted to verify this property in the proof of the Williamson's theorem given in \cite{Simon99 proofWilliamson}.
As consistency check, one can repeat this analysis starting from the last expression in (\ref{LJgammaJL})
and this leads (\ref{W decomposition W_gamma}) again.

In our numerical analysis we have constructed $W$ by employing (\ref{W decomposition W_gamma})
and Wolfram Mathematica.
In particular, given the reduced covariance matrix $\gamma_A$, the software provides the unique 
matrix $L_\gamma$ and the orthogonal matrix $W_\gamma$ that diagonalises the symmetric matrix
$(J L_\gamma )^{\textrm{t}} \,\gamma_A\, (J L_\gamma )$.
Given these two matrices and the symplectic spectrum obtained from (\ref{Jgamma relations}), 
the symplectic matrix $W$ is constructed as in (\ref{W decomposition W_gamma}).

\section{Block diagonal entanglement hamiltonians in harmonic lattices}
\label{app:EH-block-diag}

In the harmonic lattices and for static configurations, 
both  the reduced covariance matrix $\gamma_A$ and the entanglement hamiltonian matrix $H_A$ are block diagonal
\cite{ch-rev, Arias-16, Arias-17, Arias-18}. 
In this Appendix we specify the results discussed in \S\ref{sec:EHwilliamson} to the simpler case.

When $\gamma_A = Q\oplus P$ is block diagonal, namely $R=\boldsymbol{0}$ in (\ref{gamma-W-block-dec}),
the symplectic matrix $W$ in the Williamson's decomposition (\ref{williamson th gammaA}) of $\gamma_A$  is block diagonal as well, 
namely $W= U \oplus V$;
hence the Williamson's decomposition of $\gamma_A$ can be written as 
\be
\label{UdU and VdV}
 Q = U^{\textrm{t}} \,\mathcal{D} \,U
\;\;\qquad\;\;
 P = V^{\textrm{t}} \, \mathcal{D} \,V\,.
\ee
The condition that $W$ and $W^{\textrm t}$ are symplectic leads respectively to 
\be
\label{symp-cond-UV}
U V^{\textrm{t}} = V U^{\textrm{t}} = \boldsymbol{1} 
\;\; \qquad \;\;
U^{\textrm{t}} V = V^{\textrm{t}} U = \boldsymbol{1} \
\ee
telling that $U^{-1}= V^{\textrm{t}} $ and $V^{-1}= U^{\textrm{t}} $.
In particular, $U$ and $V$ are not orthogonal matrices.
The relations (\ref{UdU and VdV}) do not provide the diagonalization 
of the real and symmetric matrices $Q$ and $P$ because 
$U$ and $V$ are invertible but not orthogonal. 
Moreover, being  $W= U \oplus V$, we have that (\ref{W widetildeW relation}) becomes $ \widetilde{W} = V \oplus U$.

The matrices (\ref{Jgamma relations})
to diagonalise in order to find the symplectic spectrum simplify to
$(\textrm{i} J \gamma_A)^2 = (PQ)\oplus (QP)$ 
and $(\textrm{i} \gamma_A J)^2 = (QP)\oplus (PQ)$
when $\gamma_A$ is block diagonal.
From  this observation and  (\ref{symp-cond-UV}), 
the first expression in (\ref{Jgamma relations}) gives
\be
\label{QP diag UV}
Q P = U^{\textrm{t}} \mathcal{D}^2 \,V = V^{-1} \mathcal{D}^2 \,V
\;\; \qquad \;\;
P Q = V^{\textrm{t}} \mathcal{D}^2 \,U = U^{-1} \mathcal{D}^2 \,U\,.
\ee

When $\gamma_A$ is block diagonal, the entanglement hamiltonian matrix $H_A$ is block diagonal as well. 
In particular, from the above remarks and $ J^{\textrm t}  \gamma_A  J  = P\oplus Q$,
one observes  that the expressions in (\ref{eh-cov-mat-B}) reduce to \cite{ch-rev}
\bea
\label{eh-block-ch-version}
H_A 
= 
M \oplus N
&=&
\Big( h\big(\sqrt{PQ}\,\big)   \oplus h\big(\sqrt{QP}\,\big) \Big) 
 \big( P \oplus Q \big)
 \\
 \label{eh-block-ch-version-2}
&=&
 \big( P \oplus Q \big)
 \Big( h\big(\sqrt{QP}\,\big)   \oplus h\big(\sqrt{PQ}\,\big) \Big) \,.
\eea
The equivalence of these two expressions can be verified by transposing one of them
and employing that $M$ and $N$ are symmetric combined with 
$(\sqrt{QP}\,)^{\textrm t} = \sqrt{PQ}$, that can be easily obtained from the fact 
that also $Q$ and $P$ are symmetric.
Another useful expression for the block diagonal entanglement hamiltonian matrix $H_A$ can be obtained 
by specifying (\ref{HA-block-dec-extended-final}) to this simpler case where 
$W= U \oplus V$ (i.e. $Y=Z=\boldsymbol{0}$ in (\ref{W Wtilde block form})).
This gives
\be
\label{HA-block-dec-extended-final-diag}
H_A 
=
\big( V^{\textrm t} \, \mathcal{E} \, V \big) \oplus \big( U^{\textrm t} \, \mathcal{E} \, U \big)\,.
\ee

Notice that, for a block diagonal $\gamma_A$, in (\ref{cholesky gamma}) we have  $L_\gamma = L_Q \oplus L_P$;
therefore also the expressions reported in the Appendix\;\ref{app:cholesky} simplify.

\section{On the regime of long time and large $\ell$ for the fermionic chain}
\label{app:larget-largel-derivation}


In this appendix we discuss the derivation of the expressions (\ref{HA td anal}) and (\ref{contour large-time td-limit anal}) 
about the limit $\ell \to \infty$ of the regime of long time for the quench of the fermionic chain (see \S\ref{sec:long time}).

For a given function $f$, the large $\ell$ limit of the sum $\frac{1}{\ell +1} \sum_{k=1}^\ell f(\theta_k)$ 
gives $\int_0^{\pi} \frac{d\theta}{\pi} f(\theta)$.
Thus, for (\ref{Tlarge-t-sinsin}) in this limit we get
\be
\label{T large_t large_l step2}
T_{i,j}
\,=\,
-\,\int_0^{\pi} \frac{d\theta}{\pi}\,\eta(\theta)\left[\,\cos \left(\theta (i+j)\right)- \cos \left(\theta (i-j)\right) \,\right]
\ee
where $\eta(\theta)$ has been defined in the text below (\ref{HA td diag}).
Splitting (\ref{T large_t large_l step2}) as a sum of two integrals and focussing on one of them, we have
\be
\label{integ_app_D}
-\int_0^{\pi} \frac{d\theta}{\pi}\,\eta(\theta)\cos \left(\theta\, (i\pm j) \right)
=
\frac{2}{\pi (i\pm j)}\int_0^\pi  \frac{\sin(\theta(i\pm j))}{\sin \theta}\; dq
\,=\,
\left\{ \begin{array}{ll}
\hspace{.2cm} 0 \hspace{.5cm} & \textrm{even $(i\pm j)$}
\\
\rule{0pt}{.7cm}
\displaystyle
\frac{2}{|i\pm j|} & \textrm{odd $(i\pm j)$}
\end{array}\right.
\ee
where an integration by parts has been performed, observing that the boundary terms vanish.
The expression (\ref{HA td anal}) is obtained by employing (\ref{integ_app_D}) into (\ref{T large_t large_l step2}).


As for the contour for the entanglement entropy, in the limit $\ell \to \infty$ we find that (\ref{contour large_t}) becomes
\be
\label{cont large_t large_l}
s_A(i)
\,=\,
\frac{1}{\pi}\int_0^\pi d\theta \, s(\theta) \left[1 - \cos(2\,\theta\, i) \right]
\,=\,
\log 4 -1- \frac{1}{\pi}\int_0^\pi d\theta \, s(\theta) \cos(2\,\theta\, i)
\ee
where the expression for $s(\theta)$ introduced in (\ref{contour large-time td-limit}) can be written as
\be
s(\theta)
\,=\,
-\log\!\big((\sin\theta)/2 \big) + \cos\theta\, \log\!\big((\tan\theta)/2 \big) .
\ee
Plugging this expression into (\ref{cont large_t large_l}), an integral to compute is given by 
\be
\label{integ1_appD}
\frac{1}{\pi} \int_0^\pi \! 
\log\!\big((\sin\theta)/2 \big) 
\cos(2\,\theta\, i)\, d\theta
\,=\,
-\frac{1}{2\pi\,i} \int_0^\pi  \frac{\sin(2\,\theta\, i)}{\tan \theta}\, d\theta
\,=\,
-\frac{1}{2\,i}
\ee
where an integration by parts has been performed (the corresponding boundary terms vanish).
The remaining integral becomes a sum of two simpler integrals once the product
$\cos(\theta) \, \cos(2\theta i)$ is written as a sum of two single cosine functions. 
These two integrals in (\ref{cont large_t large_l}) can be evaluated as follows
\be
\label{integ2_appD}
-\frac{1}{2\pi} \int_0^\pi \!  \cos(\theta\,\mu)\,  \log\!\big((\tan\theta)/2 \big) \, d\theta
\,=\,
\frac{1}{2\pi \, \mu} \int_0^\pi   \frac{\sin(\theta\, \mu)}{\sin \theta}\; d\theta
\,=\,
\frac{1}{2 \, \mu}
\ee
where $\mu=2\,i\pm 1$ is an odd positive integer.
By employing (\ref{integ1_appD}) and (\ref{integ2_appD}) into (\ref{cont large_t large_l}),
the expression (\ref{contour large-time td-limit anal}) is obtained.


As consistency check for (\ref{contour large-time td-limit anal}), 
let us recover the value of the  entanglement entropy found in (\ref{EE large_t flat}) 
by summing (\ref{contour large-time td-limit anal}) over the sites of the interval $A$.
Since $A$ has $\ell$ sites, the first term of this sum gives  $\ell\, (\,\log 4 -1\,)$. 
Then, by using that $\sum_{i=1}^\infty \frac{1}{2\,i\,(4\, i^2-1)}\,=\, \frac{1}{2} (\log 4 -1)$ 
and taking into account that the interval has two endpoints, 
we obtain
$
\sum_{i \in A} s_A(i)
=
(\ell + 1) (\log 4 -1)
$,
which is the result reported in (\ref{EE large_t flat}).


\end{appendices}


\end{document}